\documentclass[pre,letterpaper,twocolumn]{revtex4}
\usepackage{epsfig}
\usepackage{times}

\begin{document}
\title{ APPLICATIONS OF PHYSICS \\
TO FINANCE AND ECONOMICS:\\
RETURNS, TRADING ACTIVITY AND INCOME\footnote{This document is a
 reformatted version of my PhD thesis. Professor Theodore L.
 Einstein, Professor Steve L. Heston, Professor Dilip B. Madan, Professor
Rajarshi Roy, Professor Victor M. Yakovenko (Chair/Advisor). }}

\author{A. Christian Silva}
\email[Email:]{silvaac@evafunds.com} \affiliation{Department of
Physics, University of Maryland, College Park, MD, 20742}
\thanks{}

\begin{abstract}

{\bf Abstract:} This dissertation reports work where physics
methods are applied to financial and economical problems. Some
material in this thesis is based on $3$ published papers
\cite{SY,SPY,income} which divide this study into two parts. The
first part studies stock market data (chapter 1 to 5). The second
part is devoted to personal income in the USA (chapter 6).

We first study the probability distribution of stock returns at mesoscopic
time lags (return horizons) ranging from about an hour to about a
month.  While at shorter microscopic time lags the distribution has
power-law tails, for mesoscopic times the bulk of the distribution
(more than 99\% of the probability) follows an exponential law.  The
slope of the exponential function is determined by the variance of
returns, which increases proportionally to the time lag.  At longer
times, the exponential law continuously evolves into Gaussian
distribution.  The exponential-to-Gaussian crossover is well
described by the analytical solution of the Heston model with
stochastic volatility.

After characterizing the stock returns at mesoscopic time lags, we
study the subordination hypothesis with one year of intraday data.
We verify that the integrated volatility $V_t$ constructed from
the number of trades process can be used as a subordinator for a
driftless Brownian motion. This subordination will be able to
describe $\approx 85\%$ of the stock returns for intraday time
lags that start at $\approx 1$ hour but are shorter than one day
(upper time limit is restricted by the short data span of one
year). We also show that the Heston model can be constructed by
subordinating a Brownian motion with the CIR process. Finally, we
show that the CIR process describes well enough the empirical
$V_t$ process, such that the corresponding Heston model is able to
describe the log-returns $x_t$ process, with approximately the
maximum quality that the subordination allows ($80\% - 85\%$).

Finally, we study the time evolution of the personal income
distribution. We find that the personal income distribution in the
USA has a well-defined two-income-class structure. The majority of
population (97--99\%) belongs to the
 lower income class characterized by the exponential Boltzmann-Gibbs
  (``thermal'') distribution, whereas the higher income class (1--3\% of
  population) has a Pareto power-law (``superthermal'') distribution.
  By analyzing income data for 1983--2001, we show that the
  ``thermal'' part is stationary in time, save for a gradual increase
  of the effective temperature, whereas the ``superthermal'' tail
  swells and shrinks following the stock market.  We discuss the
  concept of equilibrium inequality in a society, based on the
  principle of maximal entropy, and quantitatively show that it
  applies to the majority of population.
\end{abstract}

\maketitle

\tableofcontents

\section*{Acknowledgements}
I want to thank Professor Victor M. Yakovenko for all help trough
this 3 years I have spend with him working on different projects.
His assistance was vital in finishing my PhD. I also thank for the
financial support he provided. I thank Professor Richard Prange
for long discussions where I learned a lot of the buy side of
finance. His critical questioning was essential in developing my
work and in teaching me the practical option pricing concepts.

I thank Professors Theodore L. Einstein, Steve L. Heston, Dilip B.
Madan and Rajarshi Roy for accepting my invitation to serve in my
dissertation committee. Their questions and comments were
insightful and essential.

Trough nearly 6 years I spend at UMD, I have met incredible people
which generosity and knowledge was fundamental in developing my
technical and personal skills. One of such persons is Professor
Dilip Madan. Now that I come to think about it, Professor Madan
was one of the first professors that I met at UMD. No wander he
kept asking me when I was going to graduate! Professor Madan is an
incredible teacher with an incredibly deep understanding of
finance and math finance. Professor Madan took me in his math
finance group without reservations and for such generosity I am
forever thankful. Many thanks also to the members of the ever
increasing math finance group which weekly meetings under
Professor Madan and Professor Fu where a lot of fun. In particular
I have to thank Samvit Prakash which has one of the most positive
personalities around me. Well let's just say that Samvit believed
in me when I myself did not. I thank for a variety of discussions
and a fruitful interaction: George Panayotov, Huaqiang Ma, Qing
Xia, Ju-Yi J Yen and Sunhee Kim. I thank Bing Zhang for long
discussions on programming and the Q-P trade.

Before I worked with Professor Yakovenko, I had the incredible
privilege to work in the highly active nonlinear optics laboratory
under the guidance of Professor Rajarshi Roy. In my 2 years of
work in the nonlinear optics lab I learned experimental optics and
as strange as it might sound, I actually learned to pick up the
phone and call people! Turn out that this is one of the most
important skills one can have. I have to thank Raj for the
opportunity of working with him. I thank him for trying to teach
me his insightful and positive approach to life and to research.
As in the Math finance group I made a lot of friends in the
nonlinear optics lab. I would like to thank them for this
friendship and for teaching me different things in optics, form
stripping an optical fiber to how to better do a computation. In
particular I thank David DeShazer, Wing-Shun Lam, Ryan McAllister,
Min-Young Kim, Elizabeth Rogers. In particular I thank Bhaskar
Khubchandani and Dr. Parvez Guzdar for the close interaction that
resulted into a nice paper.

I thank also some of the best teachers I had. Their dedication and
skill have been inspiring, especially because I keep bothering
them and they had the patience to answer my confusing questions! I
thank Professors Steve Heston, Jack Semura, Pavel Smejtek and P.
T. Leung.

Finally I thank my family for the patience and support. Here I
also need to thank Samir Garzon and Norio Nakagaito. Samir helped
us a lot and Norio, well, Norio is just incredible.

\section{Introduction}\label{Intro}

The interest of physicists in interdisciplinary research has been
constantly growing and the area of what is today named
socio-economical physics is 10 years old \cite{Farmer1999}. This
new area in physics has started as an exercise in statistical
mechanics, where complex behavior arises from relatively simple
rules due to the interaction of a large number components. The
pioneering work in the modern stream of economical physics was
initiated by Mantegna \cite{Mantegna} and Li \cite{Li} in the
early nineties followed most notably by Mantegna and Stanley
\cite{Stanley1995} and thereafter by a stream of papers
\cite{Network} that attempt to identify and characterize universal
and non-universal features in economical data in general. This
statistical mechanical mind frame arises in direct analogy with
statistical mechanics of phase transitions, where materials (such
as a ferromagnetic and a liquid), that are different in nature,
can belong to the same universality class due to their behavior
near the critical point (point at which abruptly the phase
changes, say from liquid to solid in water, for instance). These
universality classes are identified by critical exponents for
quantities that diverge at the critical point, for instance the
specific heat $C \approx \epsilon^{-\alpha}$, where $\epsilon$ is
the reduced temperature and $\alpha$ the critical exponent
\cite{Stanley}. Therefore, the area of economical physics has
grown from, and it is still in great part concerned with,
``power-law tails'' with universal exponents. This constitutes the
empirical stream of socio-economical physics, where modelling and
characterizing the empirical data with methods and tools borrowed
from traditional physical problems is attempted \cite{BP,MS,R,V}.

Soon after Mantegna and Li initiated the modern empirical stream
of economical physics, simulations appeared. Once again, as in the
case of empirical work, these were based into fundamental
statistical mechanical models such as the Ising model. This
literature attempted to construct from simple rules complex
behavior that could then mimic the market and explain the price
formation mechanism
\cite{Zhang1997,Lane2003,Zhang2001,Challet2003,Challet2005}.

This dissertation belongs to the empirical stream of
socio-economical physics. We study here two distinct problems.
First, we use daily and intraday stock data to describe the
essential nature of the stochastic process of price returns at
different time ranges. Second, we use yearly income data to study
the time evolution of the distribution of income in the USA.

\subsection{Stock returns}

The study of stock returns has a long history dating back to
Bachalier in $1900$, which was the first to model stock dynamics
with a Brownian motion \cite{Taqqu}. He proposed that the absolute
price change $\Delta S_{t} = S_{T}-S_{T-t}$, where $t$ is the
return horizon, should follow a Gaussian random walk. The clear
drawback of such a hypothesis is that the prices of stocks could
become negative. It was apparently Renery
\cite{Taqqu,Laurent,Osborne}, who introduced the geometrical
Brownian motion for the stock price by assuming that log-returns
($x_{t}=\ln(S_{T})-\ln(S_{T-t}) \approx \Delta S_{t}/S_{t}$), and
not absolute returns, should follow a Brownian motion. The geometric
Brownian motion became popular and accepted as a main stream idea
with the work of Osborne \cite{Osborne1959} (see also \cite{Taqqu} for
historical notes) and Samuelson (cited in \cite{Taqqu}).


It was not until the $1960$'s, that the hypothesis of Gaussian
random walks was challenged by Mandelbrot \cite{Mandelbrot1963}
and Fama \cite{Fama1963,Fama1965} with studies on daily cotton
prices. Since then, Brownian motion has been consistently
questioned for a variety of assets. Today asset log-returns that
follow Brownian motion for all return horizons $t$ are considered
an exception.

In his pioneering work, Mandelbrot introduced, as an alternative
model for stock returns, the stable L\'{e}vy distribution. This
distribution has the drawback that it can present infinite
variance. Despite the unwanted mathematical properties that such a
process presents, it was not founded into economical reasoning. In
$1973$ Clark \cite{Clark} proposed, as an alternative to
Mandelbrot's model, to use subordination \cite{FellerBook} to
construct the distribution of assets returns. Subordination has a
direct financial implication, it can be
liked with financial information arrival. Clark suggests
that prices react to financial information and that if this
financial information is taken into account, the gaussian random
walk is recovered. He showed that the information arrival can be
captured by volume of trades and that if one takes returns
conditional on the volume, these should be Gaussian.

Note that in fact, Mandelbrot and Clark do not contradict
themselves, as Clark first implied. Mandelbrot's L\'{e}vy stable
distribution can also be constructed by subordination, if one
chooses the right subordinator for the Brownian motion. Therefore,
the problem is reduced to finding the right subordinator if one
accepts the subordination hypothesis.

In physics, the concept of subordination can be found in the
construction of non-Shannon entropies, in the limit of the
continuous-time random walk, in interface growth models and other
statistical mechanical problems
\cite{Cohen2003,FGIS,Sokolov2001a,Sokolov2001b,Sokolov2002}. The
mathematical- ``physical'' idea of subordination is that if the
stochastic process is analyzed at the correct reference frame, it
will always look like a simple gaussian diffusion. But since we are
dealing with stochastic processes, the reference frame is moving
randomly as well; just enough for the actual process in
observation to be described by Brownian motion. For further
mathematical development of subordination, see section \ref{model}.

After Clark, the concept of subordination has been extensively
used to construct asset return models
\cite{LevyBook,Madan1990,NIG1995,CGMY}. Most recently a series of
studies have used high-frequency data to verify Clark's
subordination hypothesis by either assuming that the volume
\cite{Smith1994,Manganelli2000} or the trading activity (number of
trades) \cite{Ane2000,Stanley2000} is responsible for price
changes. Strong evidence is found for both; nonetheless number of
trades appears better suited, since it has been extensively tested
for a large number of companies \cite{Stanley2000}.

Contemporary to Clark, a series of empirical studies indicated
that the variance ($variance = volatility^2$) of stock returns is
not constant (see \cite{Johnson1987} and references therein). This
resulted in models for stock returns such as Engle's ARCH and
Bollerslev's GARCH that attempted to account for the changing
variance in the assets returns by modelling both in a discrete
framework \cite{Engle2003}. At the same time, models with
stochastic volatility were introduced. These models generally
assume a mean reverting continuous stochastic differential
equation for the volatility \cite{Sircar,Hull1987,Heston,options}.
Notice that stochastic volatility models, GARCH and subordination,
are not entirely orthogonal to each other. Stochastic volatility
models can also be constructed by subordination \cite{CGMYSA} (see
also section \ref{model}) or as limits of
discrete GARCH type models \cite{Heston2003}.

In $1993$ Heston \cite{Heston} introduced an exactly solvable
stochastic volatility model that is also a limit process for the
GARCH(1,1) model \cite{Heston2003}. The Heston model become widely
used for option pricing and in the study of asset returns. We use
a modified version of the Heston model as developed in Ref.
\cite{DY} to describe the general shape of probability density
distribution (PDF) for the log-returns and the time evolution of
such PDF.


\subsection{Outline of the dissertation}

The outline of this thesis is as follows. In chapter
\ref{modelCH}, we introduce the Heston model for stock returns as
developed in Refs. \cite{SPY,DY}. We summarize the procedure for
finding the closed form solution of the probability distribution
for the log-returns, starting from the correlated stochastic
differential equations as given in Ref. \cite{DY}. We also
introduce subordination and show how to construct the Heston model
using a Cox-Ingersoll-Ross (CIR) subordinator \cite{CIR}.

In chapter \ref{data}, we present the data we use in this thesis.
We show the typical features of the stock data and how we
constructed such data.

In chapter \ref{expD}, we study the time evolution of the empirical
distribution function (EDF) for the stock returns at mesoscopic
time lags $t$ ($1\, hour<t<20\,days$). We show that in the short-time
limit $t<<1/\gamma$, the EDF progressively tends to the
double exponential distribution and for the long-time limit
$t>>1/\gamma$, the EDFs progressively tends towards a Gaussian,
where $1/\gamma$ is the characteristic time for such limits.
Furthermore, we show that the Heston model introduced in chapter
\ref{modelCH} presents these fundamental features.

In chapter \ref{subCh}, we study the hypothesis of subordination.
We first start by pointing out the effect of the discrete nature
of absolute price changes in the log-returns. Thereafter, we
verify the subordination hypothesis using both tick-by-tick data
(this data records all trades in a given day, see chapter
\ref{data}) as well as $5$ minutes log-returns and number of
trades (ticks) data. We find that if we use the integrated
variance ($V_t$), which is proportional to the number of trades
($N_t$), as our subordinator, we are able to explain approximately
the central $85\%$ of the probability distribution for the
log-returns $x_t$ between $1$ hour and $1$ day. Finally, we show
the quality of modelling the subordinator $V_t$ with the CIR
process introduced in section \ref{model} and discuss the
implication of such model for the log-returns $x_t$.

The last chapter of this thesis presents work on the time
evolution of the distribution of income. We show the evolution of
the distribution of personal income in the United States from
$1983$ to $2001$. We show that the bulk of the distribution
(excluding very small income and very large income), is described
by the Exponential distribution with average income changing from
year to year in approximately the same rate as inflation. We
conclude that the inflation-discounted income of the majority of
the population is approximately the same throughout time and
therefore well approximated by a system in thermal equilibrium. We
also show that the top $3\%$ earners have income that changes over
time even when inflation is accounted for. This chapter is self
contained and does not require any other part of the thesis to be
read.

\section{Heston model for asset returns}\label{modelCH}

The Heston model was introduced by Heston \cite{Heston} and
belongs to the class of stochastic volatility models, which have
received a great deal of attention in the financial literature
specially in connection with option pricing \cite{Sircar}.

Empirical verification of the Heston model was done for both stocks
\cite{SY,SPY,DY,Pan,Vicente}
and options \cite{Hull1987,Bakshi,Duffie,options}, and good agreement
with the data has been found in these studies. The version of the
Heston model for stock returns used in \cite{SY,SPY,DY}, as well as in
this thesis, was modified
from the original solution by Heston and has
evolved into a different formula with $3$ parameters. One parameter
for the variance ($\theta$), one parameter representing the
characteristic relaxation time to the Gaussian distribution
($1/\gamma$) and another that gives the general shape of the curve
($\alpha$).

The outline of this chapter is as follows. First, we present the
modified Heston model used in this work by
showing its evolution from solving the related stochastic
differential equations (SDE). Thereafter, we introduce subordination
and we show the development of the modified Heston model through
subordination.

\subsection{Heston model-SDE and symmetrization}

The formal way of presenting the Heston model is given by two
stochastic differential equations (SDE), one for the stock price $S_t$
and another for the variance $v_t$.

\begin{equation}\label{eqS}
  dS_t = \mu S_t\, dt + \sigma_t S_t\, dW_t^{(1)},
\end{equation}
\begin{equation} \label{eqVar}
  dv_t = -\gamma(v_t - \theta)\,dt + \kappa\sqrt{v_t}\,dW_t^{(2)},
\end{equation}

\noindent where the subscript $t$ indicates time dependence, $\mu$ is the drift
parameter, $W_t^{(1)}$ and $W_t^{(2)}$ are standard random Wiener processes,
$\sigma_t$ is the time-dependent volatility and $v_t=\sigma_t^2$ is
the variance. In general, the Wiener process in
(\ref{eqVar}) may be correlated with the Wiener process in
(\ref{eqS}):
\begin{equation} \label{rho}
  dW_t^{(2)} = \rho\,dW_t^{(1)} + \sqrt{1-\rho^2}\,dZ_t,
\end{equation}
where $Z_t$ is a Wiener process independent of $W_t^{(1)}$, and
$\rho\in[-1,1]$ is the correlation coefficient. Note that
(\ref{eqS}) and (\ref{eqVar}) are well known in finance. These
represent, respectively, the log-normal geometric Brownian motion
stock process introduced by Renery, Osborne and Samuelson
\cite{Taqqu} (used by Black-Melton-Scholes (BMS) \cite{BS,Merton}
for option pricing. See Ref. \cite{BSinPhysics} for a practical
application of BMS to physics) and the Cox-Ingersoll-Ross (CIR)
mean-reverting SDE first introduced for interest rate models
\cite{CIR,Shreve}.

In order to solve (\ref{eqS}) and (\ref{eqVar}) together with
(\ref{rho}), we first change variables from stock price $S_t$ to
mean removed (demean) log-return $x_t=ln(S_{t}/S_{0})-\mu t$ (\ref{eqX}). All
further results and solutions are constructed for the demean
log-return $x_t$, which we will simply refer to as log-return or
return:

\begin{equation}\label{eqX}
   dx_t = - \frac{v_t}{2}\,dt + \sqrt{v_t}\,dW_t^{(1)}.
\end{equation}

After performing the change of variables from price to return, we
solve the Fokker-Planck equation (\ref{FP}) \cite{Gardiner}
implied by SDEs (\ref{eqVar}) and (\ref{eqX}), for the transition
probability $P_t(x,v\,|\,v_i)$ to find the return $x$ and the
volatility $v$ at time $t$ given the initial demean log-return
$x=0$ and variance $v_i$ at $t=0$. For simplicity, we drop the
explicit time dependence notation for the returns $x_t$ and call
them $x$.

\begin{eqnarray}
  \frac{\partial}{\partial t}P &=&
     \gamma\frac{\partial}{\partial v}\left[(v-\theta)P\right]
     + \frac12\frac{\partial}{\partial x}(vP)
\label{FP} \\
  && {} +\rho\kappa\frac{\partial^2}{\partial x\,\partial v}(vP)
     +\frac12\frac{\partial^2}{\partial x^2}(vP)
     +\frac{\kappa^2}{2}\frac{\partial^2}{\partial v^2}(vP).
\nonumber
\end{eqnarray}

The general analytical solution of (\ref{FP}) for $P_t(x,v\,|\,v_i)$
with initial condition $P_{t=0}(x,v|\,v_i)=\delta(x)\delta(v-v_{i})$
can be found by taking a Fourier transform $x->p_x$ and a Laplace
transform $v->p_v$ (see \cite{DY} for details),

\begin{equation} \label{P}
   P_t(x\,|\,v_i)=\int\limits_{0}^{+\infty}\!\!dv\,P_t(x,v\,|\,v_i)
   =\int\frac{dp_x}{2\pi}e^{i p_x x}\widetilde{P}_{t,p_x}(0\,|\,v_i),
\end{equation}

\noindent where the hidden variable $v$ is integrated out, so $p_v=0$.
Therefore we have

\begin{eqnarray} \label{finalex}
   && P_t(x\,|\,v_i) = \int_{-\infty}^{+\infty}
   \frac{dp_x}{2\pi}\, e^{i p_x x
   - v_i\frac{p_x^2 - ip_x}{\Gamma + \Omega\coth{(\Omega t/2)}} }
\nonumber \\
   && \times\, e^{- \frac{2\gamma\theta}{\kappa^2}\ln\left(
   \cosh\frac{\Omega t}{2} +\frac{\Gamma}{\Omega}\sinh\frac{\Omega t}{2}
   \right) + \frac{\gamma\Gamma \theta t}{\kappa^2}}.
\end{eqnarray}

\noindent where
\begin{equation} \label{Gamma}
    \Gamma = \gamma + i\rho\kappa p_x
\end{equation}
and
\begin{equation} \label{eqOmega}
   \Omega=\sqrt{\Gamma^2 + \kappa^2(p_x^2-ip_x)}.
\end{equation}

The marginal
probability density $P_t(x\,|\,v_i)$ could then be compared to
empirical stock returns directly. Nevertheless, $v_i$ has to be
treated as an extra parameter. In order to avoid this, we assume
that $v_i$ has the stationary distribution of the CIR stochastic
differential equation (\ref{eqVar}), $\Pi_\ast(v)$,

\begin{equation} \label{Pi_v}
   \Pi_\ast(v) = \frac{\alpha^\alpha}{\Gamma(\alpha)} \,
   \frac{v^{\alpha-1}}{\theta^\alpha} \,
   e^{-\alpha v/\theta}, \qquad
   \alpha=\frac{2\gamma\theta}{\kappa^2}.
\end{equation}

Using equation (\ref{Pi_v}) we arrive at the probability distribution of
the demean log-returns $P_t(x)$,

\begin{equation} \label{dv_i}
  P_t(x)= \int_0^\infty \!\!dv_i\,\Pi_\ast(v_i)\,P_t(x\,|\,v_i)
\end{equation}

\noindent where the final solution is

\begin{equation} \label{Pfinal}
   P_t(x) = \frac{1}{2\pi}\int_{-\infty}^{+\infty} \!\!dp_x\,
   e^{ip_x x + F_t(p_x)}
\end{equation}
with
\begin{eqnarray}
  && F_t(p_x)=\frac{\gamma\theta}{\kappa^2}\, \Gamma t
\label{phaseF} \\
  && {} - \frac{2\gamma\theta}{\kappa^2}
  \ln\left[\cosh\frac{\Omega t}{2} +
  \frac{\Omega^2 -\Gamma^2 + 2\gamma\Gamma}{2\gamma\Omega}
  \sinh\frac{\Omega t}{2}\right]
\nonumber
\end{eqnarray}
\noindent where as before
\begin{equation} \label{Gamma}
    \Gamma = \gamma + i\rho\kappa p_x
\end{equation}
and
\begin{equation} \label{eqOmega}
   \Omega=\sqrt{\Gamma^2 + \kappa^2(p_x^2-ip_x)}.
\end{equation}

The operation of removing the initial volatility dependence of the
marginal probability density $P_t(x\,|\,v_i)$ using equation
(\ref{dv_i}) was first introduced in Ref. \cite{DY}. This removes
an additional degree of freedom and therefore simplifies the final
marginal probability density.

In order to further simplify the original Heston model, we assume that
equations (\ref{eqS}) and (\ref{eqVar}) are uncorrelated. That
amounts in taking $\rho=0$ in expression (\ref{phaseF}). This approximation
was shown to be acceptable for some companies and indexes in the
US market \cite{SY,SPY,DY} but might not be good for different
markets \cite{Vicente} or for option pricing \cite{Sircar,Heston}.

In order to arrive at the probability density function used in this
work, we need to further simplify the equation for
$P_t(x,\rho=0)$ (\ref{Pfinal}) into a zero skew symmetrical
function.

We replace in (\ref{Pfinal}) $p_x\to p_x+i/2$ and $\rho=0$ to find

\begin{equation} \label{Pfinal'}
   P_t(x) = e^{-x/2}\int_{-\infty}^{+\infty} \frac{dp_x}{2\pi} \,
   e^{ip_x x + F_t(p_x)},
\end{equation}

\noindent where $\alpha=2\gamma\theta/\kappa^2$,

\begin{equation} \label{phaseF'}
   F_t(p_x)=\frac{\alpha\gamma t}{2}
   - \alpha\ln\left[\cosh\frac{\Omega t}{2} +
   \frac{\Omega^2+\gamma^2}{2\gamma\Omega}
   \sinh\frac{\Omega t}{2}\right],
\end{equation}
and
\begin{equation} \label{eqOmega'}
   \Omega=\sqrt{\gamma^2 + \kappa^2(p_x^2+1/4)}
   \approx\gamma\sqrt{1+p_x^2(\kappa^2/\gamma^2)}.
\end{equation}

Finally, we drop the $e^{-x/2}$ term in (\ref{Pfinal'}). Notice that
both taking $e^{-x/2} \approx 1$ and $p_{x}^{2}+1/4 \approx p_{x}^{2}$
are needed to produce a new characteristic
function $e^{F_{t}(p_x)}$ that correctly goes to unity when $p_x=0$. The final
functional form for $P_t(x)$ is

\begin{eqnarray}
  &&P_t(x) =
  \int\limits_{-\infty}^{+\infty} \frac{dp_x}{2\pi}\,
  e^{ip_xx + F_{\tilde t}(p_x)}, \label{eq:DY0} \\
  && F_{\tilde t}(p_x)=\frac{\alpha\tilde t}{2}
  - \alpha\ln\left[\cosh\frac{\Omega\tilde t}{2} +
  \frac{\Omega^2+1}{2\Omega}
  \sinh\frac{\Omega\tilde t}{2}\right],
\label{eq:DY}  \\
  && \tilde t=\gamma t, \quad \alpha=2\gamma\theta/\kappa^2,\nonumber \\
 && \Omega=\sqrt{1+(p_x\kappa/\gamma)^2}, \quad
  \sigma_t^2\equiv\langle x^2\rangle=\theta t.
\label{eq:DY2}
\end{eqnarray}

We have expressed the original Heston model for the probability
density of log-returns $x$, in a highly symmetrical form with
three parameters, $\theta$, $\alpha$ and $\gamma$. The parameter
$\theta$ can be found by calculating the variance of demean
log-returns $\sigma_t^2\equiv\langle x_t^2\rangle = \theta t$
(\ref{eq:DY2}) of $P_t(x)$ (\ref{eq:DY0}). The remaining two
parameters, $\alpha$ and $\gamma$, are responsible for the general
shape of the curve and the relaxation rate of $P_t(x)$ to a
Gaussian distribution \cite{SPY,DY}. The parameter $\alpha$ is
also responsible to define the analyticity at zero return. If
$\alpha=1$, value used in this thesis, the short-time-limit is a
double exponential distribution (see next subsection). This
distribution is not analytical at zero but becomes when time
progresses. For $\alpha>1$ the distribution is always analytical
with a center that is Gaussian and when $\alpha<1$ the
distribution starts non-analytic at zero (going to zero as a
power-law with exponent $2\alpha - 1$ \cite{DY}) and then evolves
into a analytic distribution with Gaussian center.

Notice that the average for the log-returns $x$ from equation
(\ref{eq:DY0}) is $\langle x\rangle=0$. This average is not
consistent with SDE (\ref{eqX}), but with the simplified
$dx_t=\sqrt{v_t}dW_{t}^{(1)}$, where the drift term $v_t/2$ is set
to zero. Therefore, $x$ in equation (\ref{eq:DY0}) does only
approximately represent demean log-returns $x=ln(S_{t}/S_{0})-\mu
t$. This difference arises because we took $e^{-x/2} \approx 1$
and $p_{x}^{2}+1/4 \approx p_{x}^{2}$ in equation (\ref{phaseF'})
in order to derive equation (\ref{eq:DY}).

The log-returns $x$ in equation (\ref{eq:DY0}) can be exactly
given by $x=ln(S_{t}/S_{0})-\mu t - \omega(t)$, where the extra
term, $\omega(t)$, removes the non zero average of
$x=ln(S_{t}/S_{0})-\mu t$.

The extra term $\omega(t)$ arises because the average of the stock
price at time $t$ needs to be given by $\mu$ only. Hence

\begin{equation}
\langle S_t \rangle = S_0 e^{\mu t} \langle e^{Y_t}  \rangle,\,
\langle e^{Y_t} \rangle \equiv 1,
\label{c1}
\end{equation}

where $Y_t$ is the stochastic process

\begin{eqnarray}
S_{t}=S_{0}\frac{e^{\mu t + X_t}}{<e^{X_t}>}=S_{0}e^{\mu
  t-ln(<e^{X_t}>)+X_t}\nonumber \\
  \Rightarrow \omega(t) = -ln(<e^{X_t}>)
\nonumber \\
x_{t}=ln(S_{t})-ln(S_{0})-\mu t = X_t+\omega(t)\nonumber \\
\Rightarrow Y_t = X_t+\omega(t). \label{Stxt}
\end{eqnarray}

Empirically, the correction represented by $\omega(t)$ or by
working with equation (\ref{Pfinal'}) instead of equation
(\ref{eq:DY0}) is small, and it can be safely neglected. We choose
to work with $x=ln(S_{t}/S_{0})-\mu t - \omega(t)$, and we call $x$
in (\ref{eq:DY0}) the log-return.

\subsubsection{Short and long time limits of the Heston model}

The short time lag limit of the modified Heston model (\ref{eq:DY0}) can be found by assuming
$\Omega t \ll 2$ in expression (\ref{finalex}). We also take $\rho=0$
and $ip_x \rightarrow 0$, since we interested in the short-time-limit of the symmetric modified
Heston model of equation (\ref{eq:DY0}). When taking the limit $\Omega
t \ll 2$ in (\ref{finalex}), the resulting PDF is the Fourier inverse of the characteristic function of a Gaussian with random variance
$v_i$ and zero drift. Since $v_i$ is
a Gamma random variable with distribution (\ref{Pi_v}), the final
characteristic function for the short-time-limit distribution of the modified Heston model is

\begin{equation}
\tilde{P}_{t}(p_x)=\int_{0}^{\infty} dv_{i}
e^{\frac{-v_{i}p_{x}t}{2}} \Pi_{*}(v_i)=(1+\frac{\theta t
p_{x}^{2}}{2 \alpha})^{-\alpha}. \label{cfShort}
\end{equation}

The probability distribution can be found analytically \cite{DY} as

\begin{equation} \label{PshortF}
   P_t(x)=\frac{2^{1-\alpha}}{\Gamma(\alpha)}\,
   \sqrt{\frac{\alpha}{\pi \theta t}}\,
   y^{\alpha-1/2}
   K_{\alpha-1/2}(y),
\end{equation}
where $K$ is the modified Bessel function and

\begin{equation} \label{y}
   y=\sqrt{\frac{2\alpha x^2}{\theta t}}.
\end{equation}

For $\alpha=1$, we recover the Laplace (symmetrical double exponential)
distribution

\begin{equation}
P_t(x) = \frac{e^{-y}}{\sqrt{2 \theta t}},\, y=\sqrt{\frac{2\alpha x^2}{\theta t}}.
\label{dExp}
\end{equation}

Notice that the short time limit is not a Gaussian with variance
$v_i$, only because of the assumed randomization of $v_i$
(\ref{cfShort}). Therefore, this randomization has substantial
effect in the limiting distributions, which can be checked
empirically \cite{SPY} (empirical results will be presented in
chapter \ref{expD}).

The long time lag $t$ limit for the modified Heston model can be found by
taking the limit $\Omega t \gg 2$ in the characteristic function
(\ref{eq:DY}). The resulting characteristic function is

\begin{equation}
\tilde{P}_t(p_x)= \langle e^{ip_{x}x} \rangle =
e^{\frac{\alpha \gamma t}{2}\left(1-\sqrt{1+
      x_{0}^{2} p_{x}^{2}}\right)}, \, x_0=\kappa^{2}/\gamma^{2}.
\label{cfNIG}
\end{equation}

The characteristic function in equation (\ref{cfNIG}) is the
characteristic function for the zero skew Normal Inverse Gaussian
(NIG) model. NIG was first introduced by Barndorff-Nielsen to
describe the distribution of sand particles sizes \cite{BN1977}
and was subsequently used in other physical problems such as
turbulence \cite{BN1979}. In $1995$, Barndorff-Nielsen also
introduced NIG for stock returns \cite{NIG1995}. NIG can also be
obtained as a limit of the Generalized Hyperbolic distribution
\cite{LevyBook,Eberlein2002}, as well as by subordinating a
Brownian motion to the inverse gaussian distribution
\cite{LevyBook} (next section will introduce the idea of
subordination).

NIG is part of the wide class of L\'{e}vy pure jump models
\cite{LevyBook}, and the fact that it is recovered as a limit of
the simplified Heston stochastic volatility model (\ref{eq:DY0}),
is another consequence of the randomization of $v_i$. Notice that if
we take the long time limit before the randomization of $v_i$ in the full Heston model given in
Eq. (\ref{finalex}), we will not find NIG as the long time limit.

The central limit theorem can be invoked for NIG and therefore for
Heston \cite{BP,FellerBook,NIG1995,DY}. That is, as time
progresses, the distribution $P_{t}(x)$ of returns $x$ will
become increasingly Gaussian. The characteristic time scale for
the central limit theorem to act is $t_{0}=2/(\alpha \gamma)$. For
$t \gg t_0$ the probability distribution is essentially Normal with
mean zero and variance $\theta t$.

Notice that for long time lags $t$, there are two characteristic
time limits. Heston tends to NIG for times $t \gg 1/\gamma$ and
then NIG tends to a Normal distribution for times $t \gg 1/\alpha
\gamma$. If $\alpha \geq 1$, NIG and Heston regimes can not be
effectively distinguished. It is only in the case $\alpha<1$, that
there will be a distinguished NIG regime.

In summary, the most important limits for $P_t(x)$ that we use in this study
are: Exponential (if $\alpha = 1$) at short time lags and Gaussian at long
time lags,

\begin{equation}
  P_{t}(x)\propto\left\{
    \begin{array}{ll}
    \exp(-|x|\sqrt{2/\theta t}),  & \quad \tilde t =\gamma t\ll1, \\
    \exp(-x^{2}/2 \theta t), &  \quad \tilde t=\gamma t\gg1.
    \end{array}
    \right.
\label{short-long}
\end{equation}


\subsection{Heston model and subordination} \label{subTh}

Subordination is a form of randomization in which one constructs a
new probability distribution, by assuming one or more parameters
of the original probability distribution to be random
\cite{FellerBook},

\begin{equation}
P_{New}(y,z) = \int_{-\infty}^{\infty}d\theta P(y,\theta)Q(\theta,z).
\label{mix}
\end{equation}

In the case of subordination, a Markov process $Y(N)$ is
randomized by introducing a non-negative process $N(t)$, called a
randomized operational time. The resulting process $Y(N(t))$ does
not need to be Markovian in general \cite{FellerBook}. We restrict
ourselves to subordination of a Brownian motion with drift
$\theta$ and standard deviation $\sigma$ (\ref{gaussSub}). We also
assume in what follows, that $t$ is time lag in usual units of
time, unless otherwise indicated. The probability density $P_t(y)$
for the time changed Brownian motion $Y(N)$ can be written

\begin{equation}
P_t(y)=\int_{0}^{\infty}dN \frac{1}{\sqrt{2\pi\sigma^2N}}
e^{\frac{-(y-\theta N)^{2}}{2\sigma^{2}N}} P_t(N).
\label{gaussSub}
\end{equation}

The moments of a Brownian subordinated process are related to the
moments of the subordinator. If we use $P_t(y)$ in
(\ref{gaussSub}), the first $4$ moments can be calculated as

\begin{equation}
\langle y \rangle  = \theta \langle N \rangle_{N} \label{M1}
\end{equation}

\begin{equation}
\langle (y- \langle y \rangle )^{2} \rangle  = \sigma^{2} \langle N
\rangle_{N}+\theta^{2} \langle (N- \langle N \rangle_{N})^{2}
\rangle_{N}\label{M2}
\end{equation}

\begin{equation}
\langle (y- \langle y \rangle)^{3} \rangle = 3\sigma^{2}\theta
\langle (N- \langle N \rangle_{N})^{2} \rangle_{N} + \theta^{3}
\langle (N- \langle N \rangle_{N})^{3} \rangle_{N} \label{M3}
\end{equation}

\begin{eqnarray}
\langle (y- \langle y \rangle)^{4} \rangle = 3\sigma^{4}(
\langle(N-\langle N \rangle_{N})^{2} \rangle_{N}+ \langle N
\rangle_{N}^{2})+
\nonumber \\
6\theta^{2}\sigma^{2}( \langle (N- \langle N
\rangle_{N})^{3}\rangle_{N}+
\nonumber \\
\langle N \rangle_{N}\langle(N- \langle N \rangle_{N})^{2}
\rangle_{N})+
\theta^{4} \langle (N- \langle N \rangle_{N})^{4}
\rangle_{N},\label{M4}
\end{eqnarray}


\noindent where $\langle \rangle$ refers to taking the expected value and
$\langle \rangle_{N}$ refers
to taking the expected value with respect to $N$. The time $t$
dependence of the moments of $Y$ are given by the moments of the
randomized operational time $N$. Furthermore, even though the
subordinator has odd moments, odd moments in the resulting process
$Y$ are only different from zero, if the Gaussian in equation
(\ref{gaussSub}) has a drift $\theta \neq 0$. For the present
work, we assume that the odd moments are all zero since the
empirical probability distribution of log-returns are quite well
described by zero skew probability distributions and because we work
with mean zero returns \cite{SPY}. By assuming zero odd moments probability
distribution, we simplify the even moments. The second and fourth
moments for $Y$ depend only on the first and second moments of the
subordinator $N$ (\ref{M2},\ref{M4}).

In the case of the modified Heston model (\ref{eq:DY0}), the
subordination takes the following terms. We assume that the
log-returns $x$ follow a Brownian motion with zero drift and
variance $V_t$. The variance $V_t$ is our ``random operational
time'', since it changes randomly. We will show in chapter
\ref{subCh} that the variance $V_t$ can be estimated (at least
partially) using the number of trades $N_t$ that occur in a the
time interval $t$. The variance $V_t$ is then a constant times
$N_t$, $V_{t}=\sigma^{2} N_{t}$.

The variance $V_t$ is given by $V_t=\int_{0}^{t}ds \, v_s$, where
the instantaneous variance $v_t$ appearing in the SDE
(\ref{eqVar}) is integrated in the interval $0 \rightarrow t$. For
this reason, $V_t$ is also know as integrated variance. The
Laplace transform for the conditional probability density
$P_t(V_t|\,v_i)$ is analytically known \cite{LevyBook,CIR}.
Therefore, subordination becomes a useful tool to construct asset
models with stochastic variance having the CIR process as a
subordinator \cite{CGMYSA}.

The Laplace transform of the subordinator of the modified Heston
model (\ref{eq:DY}) can be read off immediately,

\begin{equation}
\tilde{P}(p_x)=\langle e^{ip_{x}x} \rangle \Rightarrow \tilde{P}(p_x)=\int_{0}^{\infty}dV_{t}
e^{-\frac{p_{x}^{2} V_{t}}{2}}P(V_{t})
\label{pvt}
\end{equation}

\noindent where the integral with respect to $V_{t}$ defines a Laplace
transform of the probability density $P(V_t)$, for which the
Laplace conjugated variable is calculated at $p_{x}^{2}/2$.
Therefore we arrive at

\begin{eqnarray}
  && P_t(V_t) =
  \int\limits_{0}^{+\infty} dp_{V_t}\,
  e^{p_{V_t}x + F_{\tilde t}(p_{V_t})}, \label{eq:CIR0} \\
  && F_{\tilde t}(p_{V_t})=\frac{\alpha\tilde t}{2}
  - \alpha\ln\left[\cosh\frac{\Omega\tilde t}{2} +
  \frac{\Omega^2+1}{2\Omega}
  \sinh\frac{\Omega\tilde t}{2}\right],
\label{eq:CIR}  \\
  && \tilde t=\gamma t, \quad \alpha=2\gamma\theta/\kappa^2,
  \quad \Omega=\sqrt{1+2(\kappa/\gamma)^2 p_{V_{t}}}.
\label{eq:CIR2}
\end{eqnarray}

The only difference between the characteristic exponent
(\ref{eq:CIR}) and the characteristic exponent for the Heston
model (\ref{eq:DY}) is in $\Omega$, where $p_{V_{t}}$ replaces
$p_{x}^{2}/2$ as the Laplace variable for $V_{t}$.


The first and second moments for the integrated CIR process (\ref{eq:CIR})
are

\begin{eqnarray}
&& \langle V_{t} \rangle = \theta t \label{M1cir} \\
&& \langle (V_{t}-\langle V_{t} \rangle)^{2} \rangle = \frac{2
  \theta^{2}}{\alpha \gamma^{2}}(e^{-\gamma t}-1+\gamma t). \label{M2cir}
\end{eqnarray}

The time dependence of the variance (\ref{M2cir}) shows that the CIR
process is not independent and identically distributed (IID). That is
expected since we have a mean reverting SDE (\ref{eqVar})
for the instantaneous variance $v_t$ with exponential relaxation to
the mean \cite{Gardiner,CIR,Shreve}.

We have shown that subordinating a zero drift gaussian to the
integrated $V_t$, given by equation (\ref{eq:CIR0}) is equivalent to solving
for the transition probability densities for the uncorrelated
($\rho = 0$ in equation (\ref{rho})) system of SDEs $dx_t =
\sqrt{v_t}dW_t^{(1)}$ and $dv_t = -\gamma(v_t-\theta)dt+\kappa
\sqrt{v_{t}}dW_{t}^{(2)}$(\ref{eqVar}). However, it is not clear
how to use subordination in order to produce a stochastic process
that is equivalent to the correlated ($\rho \neq 0 $) system of
SDEs \cite{CGMYSA}.

\section{General characteristics of the data and methods}\label{data}

We use 2 databases for this study. Daily closing prices are
downloaded from Yahoo \cite{Yahoo} and intraday data is
constructed using the TAQ database from the NYSE \cite{TAQ}. The
TAQ database records every transaction that occurred in the market
(tick-by-tick data), where the average number of
transactions in a day for a highly traded stock, such as Intel, is
$20000$ (from 1993 to 2001). That is equivalent, in terms of data
quantity, to approximately $77$ years of daily data.

Our data has the time that the transaction occurred, the price the
transaction was realized and the volume of the transaction (number
of shares that exchanged hands). The TAQ database does not account
for splits or dividends whereas Yahoo gives the prices corrected
for splits and dividends. However we do need to correct for splits
and dividends because the TAQ database is used only when
constructing intraday returns. The splits and dividends are
realized overnight and therefore will not show up if we calculate
intraday returns.

After downloading the TAQ data, we remove any trade that is
recorded as an error and also restrict the data to trades that
took place inside the conventional $6.5$ hours trading day from
$9\colon 30$ AM to $4\colon00$ PM. Any trade that happen before
$9\colon 30$ AM and after $4\colon 00$ PM is ignored. We choose to
restrict to business hours because we want our data set to agree
with Yahoo daily data in the limit of one day that is defined from
the open bell ($9\colon30$ AM) to close bell ($4\colon00$ PM).

\begin{figure}
\centerline{\epsfig{file=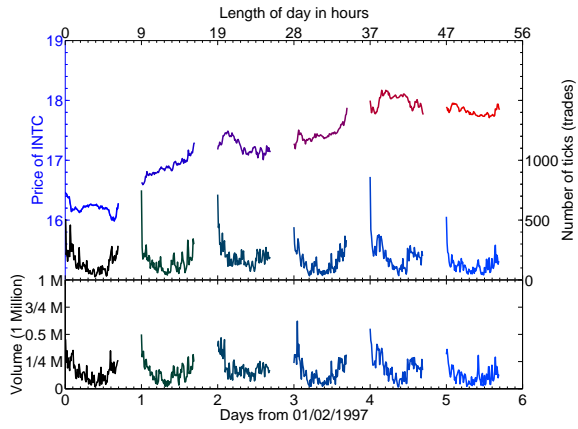,width=0.65\linewidth,angle=-90}}
\caption{Intraday stock price and number of trades constructed from the TAQ database at
  each $5$ minute interval from Thursday, 2nd of January 1997 to Thursday
  9th of January 1997 for Intel (upper panel). Volume of trades
  during each day is shown in the lower panel. Days are separated by
  an effective overnight time interval that is constructed from the
  data, such that the open-to-close variance and the close-to-close
  variance of the log-returns follow the same $ \propto t$ line
  (see Fig. \ref{fig:Var}).}
\label{fig:TS}
\end{figure}

We define as the daily open price, the price of the first trade
that happened after or at $9\colon30$ AM. We also define the daily
close price, the price of the trade that happened right before or
at $4\colon00$ PM. A typical time series for intraday prices,
number of trades and volumes for $1$ particular week is shown in
Fig. \ref{fig:TS}.

Notice that the intraday volume and trading activity (number of
trades) can be well described by a parabola (Fig. \ref{fig:Pat}).
This typical intraday pattern \cite{Osborne1962,Ord1985} has also
been found for high-frequency volatility proxies, such as the root
mean square return for all ticks that happen in a certain interval
of time
\cite{StanleyVol1999,Engle2000,Bollerslev1996,Bollerslev2001,Granger1996a,Granger1996b}.
The statistics for such a pattern for the number of trades of
Intel in the year $1997$ is shown in Fig. \ref{fig:Pat}. Notice
that the probability density for different parts of the day will
clearly have different widths and averages. Therefore, mixing all
parts of the day will result in a wider probability density for
number of trades and other intraday quantities \cite{Ord1985}. We
do not study the consequences of such a mixture, we only are
careful to work with intraday time lags that divide equally all
day \cite{SPY}. In such a way, all parts of the daily trend are
equally represented. Since we are working with prices quoted at every
$5$ minutes (Five minutes close prices)
and the day from open to close has only $78$ such
intervals, we work with returns that are
$t=5,10,15,30,65,130,195,390$ minutes long.

\begin{figure}
\centerline{\epsfig{file=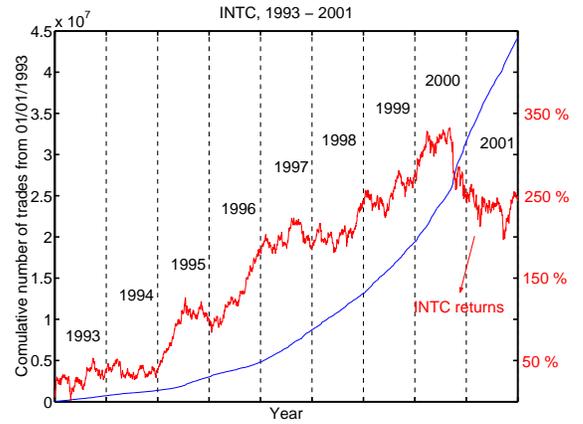,width=0.65\linewidth,angle=-90}}
\caption{Cumulative number of trades and return from $1993$ to
  $2001$ for Intel. The increase of the cumulative number of trades indicate that
  the parameters describing the stock are changing.}
\label{fig:CumN}
\end{figure}


Another important characteristic of daily and intraday  data is
shown in Fig. \ref{fig:CumN}. The cumulative number of trades from
$1993$ to $2001$ ($\sum_{i=01/01/1993}^{i=12/31/2001}N_i$)
increase almost exponentially. The behavior of the commutative
number of trades shows that the average number of trades change
from year to year. The same type of behavior is found for the
square of the demean log-returns (the variance of the returns).
Therefore, the probability density for the returns, volume and
number of trades is only approximately stationary throughout the
years. When studying returns (chapter \ref{expD}), we assume the
data as stationary, and we take data from $1993$ to $1999$. When
studying subordination using the number of trades (chapter
\ref{subCh}), we reduce the non-stationary effect of the data by
working with one year of data.

\begin{figure}
\centerline{\epsfig{file=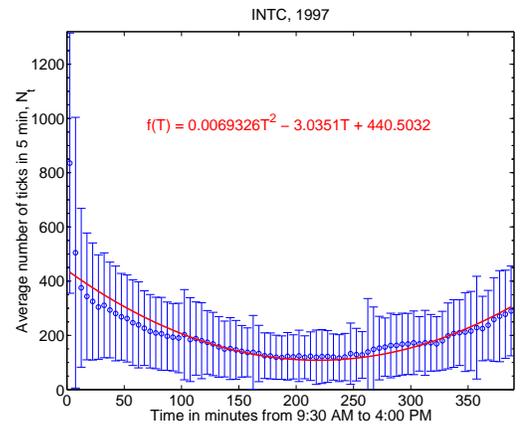,width=0.65\linewidth,angle=-90}}
\caption{Average number of trades (ticks) in a given period of the day. The
  error bars represent the volatility. The red solid line gives the best fit parabola to the average number of trades. Same type of pattern is found for
  absolute returns \cite{StanleyVol1999} and volume.}
\label{fig:Pat}
\end{figure}

In order to study intraday returns, we construct from the tick-by-tick
data, $5$ minutes close prices. The $5$ minute close price is defined
in analogy with the day close price. The 5 minutes volume ( or number
of trades (ticks)) is the sum of all traded volume ( or number of
trades (ticks)) in a $5$ minutes interval.

\begin{figure}
\centerline{\epsfig{file=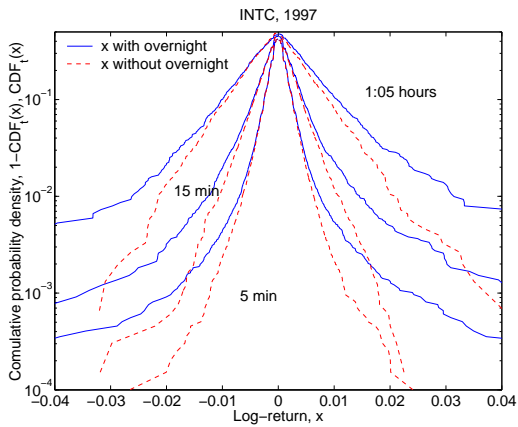,width=0.65\linewidth,angle=-90}}
\caption{Cumulative density function for the positive and negative
  log-returns of Intel. Log-returns constructed including overnight time
  lags (solid lines) show higher probability of large returns than
  log-returns that do
  not include overnight time lags (dashed lines). We choose not to
  include overnight time lags in our intraday return time series.}
\label{fig:OverN}
\end{figure}

When constructing intraday returns time series, we do not include
nights or weekends. Effectively our largest intraday return is
from open to close (time lag of 390 min = 6.5 hours). A common
procedure, not adopted here, is to assume the open of the next day
as the close of the present day \cite{Stanley1999a,Stanley1999b}.
This will include returns that are effectively overnight, where no
trades are present. The result of such practice is illustrated in
Fig. \ref{fig:OverN}. Clearly, the tails of the distribution of
returns including overnight time lags are considerably enhanced,
if compared with the distribution of intraday returns that do not
include overnight time lags.

When working with high-frequency (intraday) data recording errors
are inevitable. In order to remove errors in the tick-by-tick data
as well as our $5$ minutes close time series, created from the
tick-by-tick data, we use Yahoo database as our benchmark. We
assume that the daily Yahoo database does not have errors. Our
filtering technique consists of two parts. First, we calculate the
log-return between the maximum and minimum price of a given day
for the Yahoo data ($r_{HL}$). We then calculate the log-return
($r_{5min}=ln(S_T)-ln(S_{T-5min})$) for the $5$ minutes price data
in the same day and compare to $r_{HL}$. We replace any log-return
$|r_{t}|>r_{HL}$ with the return immediately preceding it. We also
replace the number of trades and volume of the ``corrupted'' 5
minute interval by the immediately preceding ones. The second
filtering procedure consists of requiring that the largest and
smallest 5 minutes log-return ($r_{5min}$) in a given day, be
between the maximum and the minimum of all the time series formed
by the yahoo open to close return data
($min(r_{OC})<r_{5min}<max(r_{OC})$). Once again, if the condition
$min(r_{OC})<r_{5min}<max(r_{OC})$ is not satisfied, we replace
the ``corrupted'' log-return, volume and number of trades by the
immediately preceding one.

The typical effect of such a simple error removal algorithm is to
change less than $1\%$ (on the order of $0.1\%$) of the data.

The same filtering procedure is used for tick-by-tick data, except
that instead of replacing the ``corrupted'' log-return and volume,
we just ignore it. In fact ignoring or replacing by the nearest
value is found to be equivalent (for tick-by-tick or 5 minutes
data) for the purpose of this work: the probability density and
moments are the same.


\section{Mesoscopic returns} \label{expD}

The actual observed empirical probability distribution functions (EDFs) for different
assets have been extensively studied in
recent years
\cite{SY,BP,DY,Stanley1999a,Stanley1999b,Vicente,India,Japan,Germany,Miranda,USA}. We
focus here on the EDFs of the returns of individual large American
companies from 1993 to 1999, a period without major market
disturbances.  By `return' we always mean `log-return', the difference
of the logarithms of prices at two times separated by a time lag $t$.

The time lag $t$ is an important parameter: the EDFs evolve with this
parameter. At micro lags (typically shorter than one hour), effects
such as the discreteness of prices and transaction times, correlations
between successive transactions, and fluctuations in trading rates
become important (for discreteness effects see chapter
\ref{subCh})\cite{BP,MS}.  Power-law tails of EDFs in this regime have
been
much discussed in the literature before
\cite{Stanley1999a,Stanley1999b}.  At `meso' time lags (typically from
an hour to a month), continuum approximations can be made, and some
sort of diffusion process is plausible, eventually leading to a normal
Gaussian distribution.  On the other hand, at `macro' time lags, the
changes in the mean market drifts and macroeconomic `convection'
effects can become important, so simple results are less likely to be
obtained.  The boundaries between these domains to an extent depend on
the stock, the market where it is traded, and the epoch. The
micro-meso boundary can be defined as the time lag above which
power-law tails constitute a very small part of the EDF. The
meso-macro boundary is more tentative, since statistical data at long
time lags become sparse.

The first result is that we extend to meso time lags a stylized
fact\footnote{Stylized facts is a term that comes from the economical
  literature. It refers to facts that can not be proved right. For instance,
the variance of returns is proportional to $t$ for a good quantity of
stocks but there might be stocks where this is not a fact.}
known since the 19th century \cite{Regnault} (quoted in \cite{Taqqu}):
with a careful definition of time lag $t$, the variance of returns is
proportional to $t$.

The second result is that log-linear plots of the EDFs show prominent
straight-line (tent-shape) character, i.e.\ the bulk (about 99\%) of
the probability distribution of log-return follows an exponential law.
The exponential law applies to the central part of EDFs, i.e.\ not too
big log-returns.  For the far tails of EDFs, usually associated with
power laws at micro time lags, we do not have enough statistically
reliable data points at meso lags to make a definite conclusion.
Exponential distributions have been reported for some world markets
\cite{SY,DY,Vicente,India,Japan,Germany,Miranda,USA} and briefly mentioned
in the book \cite{BP} (see Fig.\ 2.12).  However, the
exponential law has not yet achieved the status of a stylized fact.
Perhaps this is because influential work
\cite{Stanley1999a,Stanley1999b} has been interpreted as finding that
the individual returns of all the major US stocks for micro to macro
time lags have the same power law EDFs, if they are rescaled by the
volatility.

\begin{figure}[t]
\centerline{
\epsfig{file=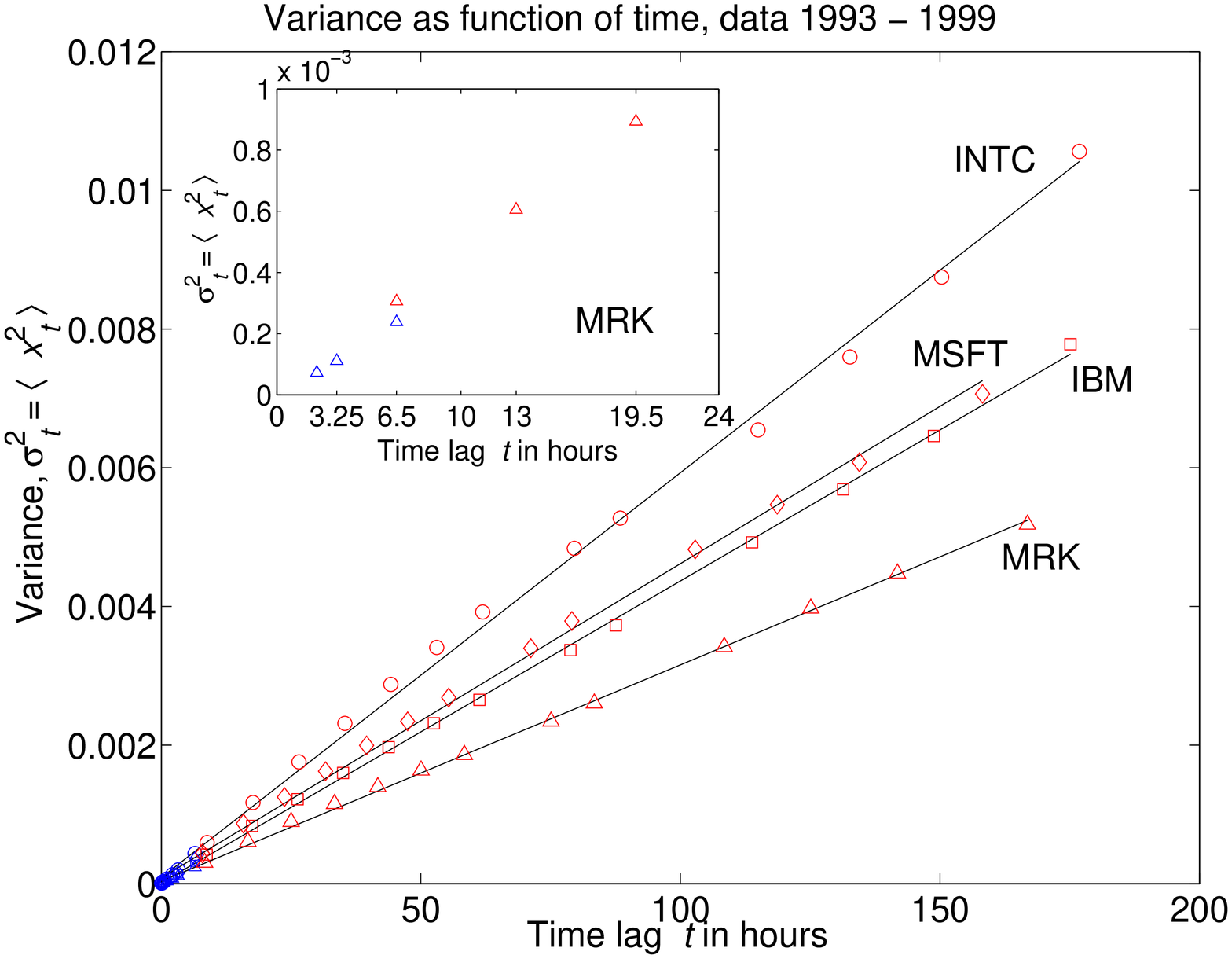,width=0.5\linewidth,angle=-90}}
\centerline{
\epsfig{file=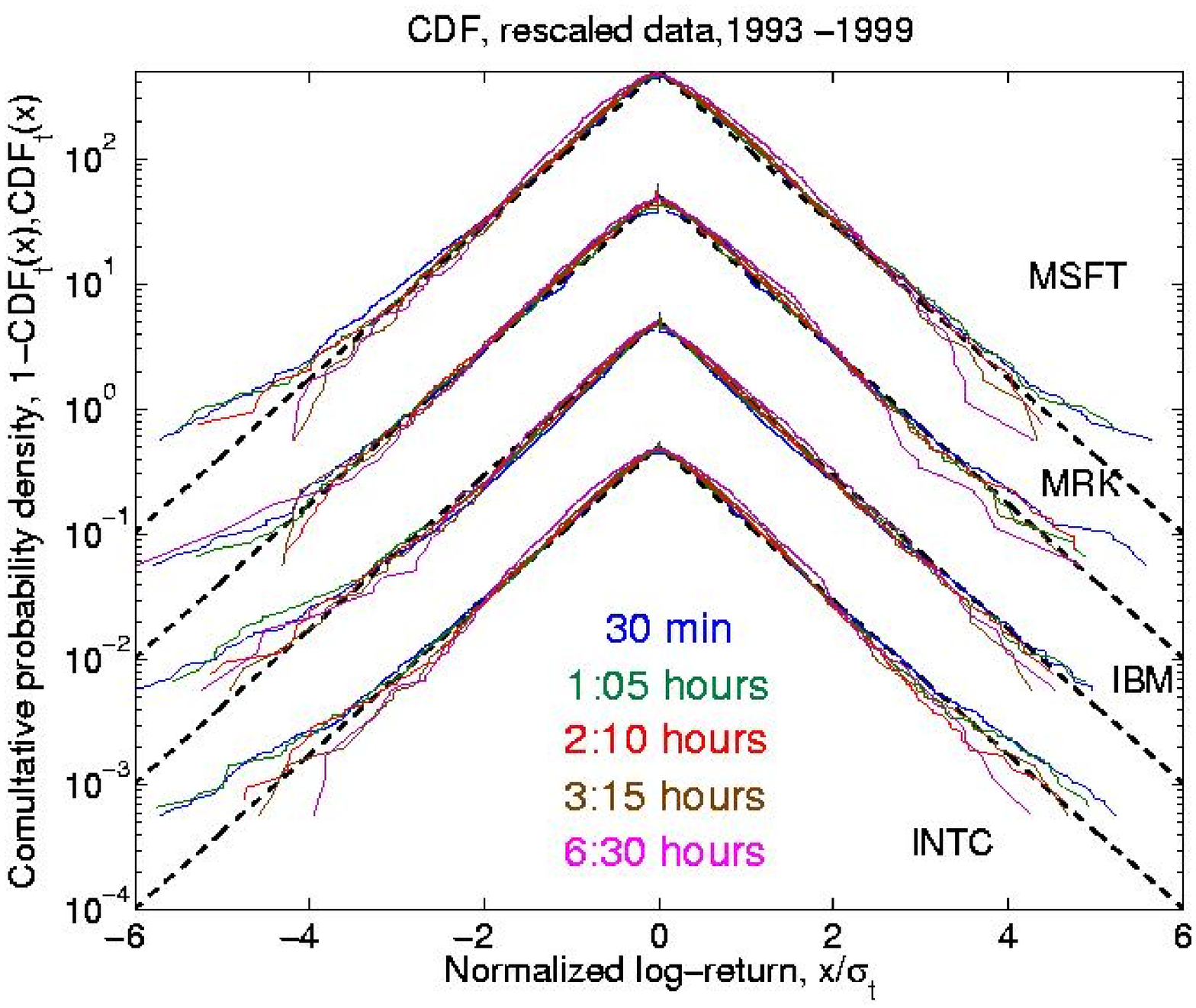,width=0.5\linewidth,angle=-90}}
\caption{Top panel: Variance
  $\langle x_t^2\rangle$ vs.\ time lag $t$. Solid lines: Linear fits
  $\langle x_t^2\rangle=\theta t$. Inset: Variances for MRK before
  adjustment for the effective overnight time $T_n$. Bottom panel:
  Log-linear plots of CDFs vs.\ $x/\sqrt{\theta t}$.  Straight dashed
  lines $-|x|\sqrt{2/\theta t}$ are predicted by the DY formula
  (\ref{short-long}) in the short-time limit.  The curves are offset
  by a factor of 10.}
\label{fig:Var}
\end{figure}

\begin{figure}[h]
\centerline{
  \epsfig{file=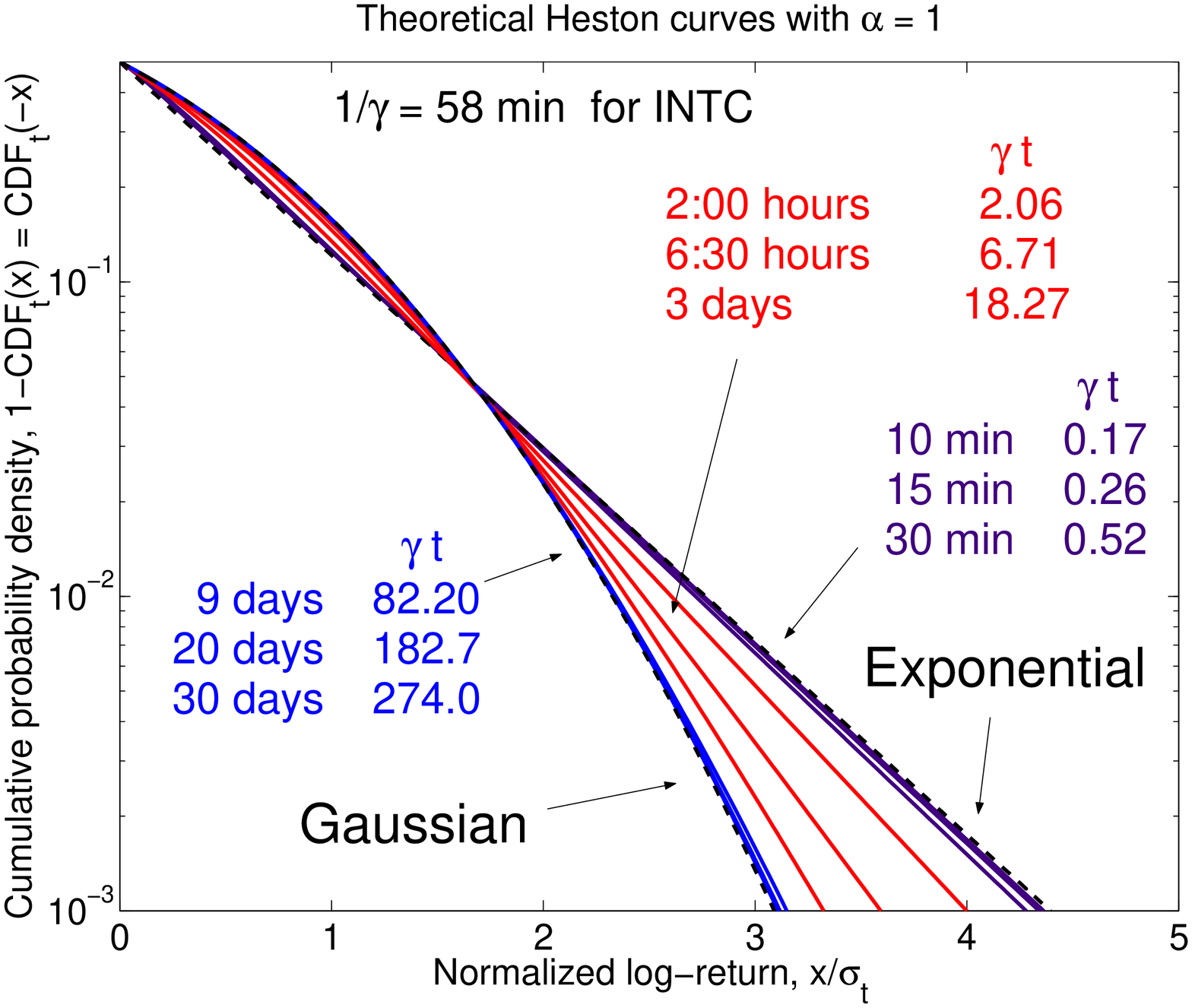,width=0.5\linewidth,angle=-90}}
\centerline{
  \epsfig{file=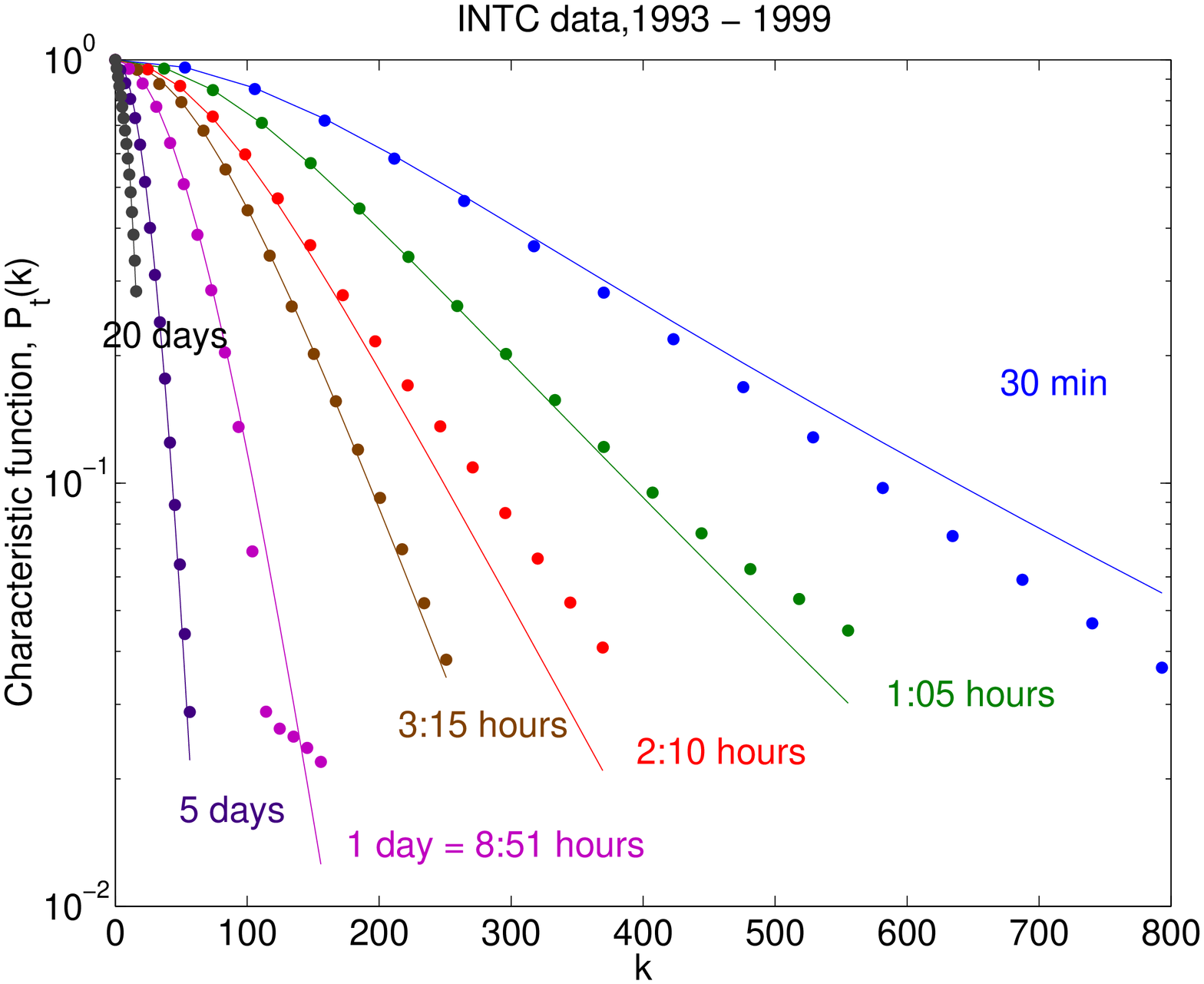,width=0.5\linewidth,angle=-90}}
\caption{Top panel: Theoretical CDFs for the Heston
  model plotted vs.\ $x/\sqrt{\theta t}$.  The curves interpolate
  between the short-time exponential and long-time Gaussian scalings.
  Bottom panel: Comparison between empirical (points) and the DY
  theoretical (curves) characteristic functions $\tilde P_t(k)$.}
\label{fig:ThPk}
\end{figure}

The Heston model is a plausible diffusion model with stochastic
volatility, which reproduces the timelag-variance proportionality and
the cross\-over from exponential distribution to Gaussian.  This model
was first introduced by Heston, who studied option prices
\cite{Heston}. Later Dr\u{a}gulescu and Yako\-ven\-ko (DY) derived a
convenient closed-form expression for the probability distribution of
returns in this model and applied it to stock indexes from 1 day to 1
year \cite{DY}.  The third result is that the DY formula with three
lag-independent parameters reasonably fits the time evolution of EDFs
at meso lags.

\subsection{Data analysis and discussion}

We analyzed the data from Jan/1993 to Jan/2000 for $27$ Dow companies,
but show results only for four large cap companies: Intel (INTC) and
Microsoft (MSFT) traded at NASDAQ, and IBM and Merck (MRK) traded at
NYSE (please see the appendix for more companies).  We use two databases, TAQ to construct the intraday returns and
Yahoo database for the interday returns (see Chapter \ref{data}). The intraday time lags were
chosen at multiples of 5 minutes, which divide exactly the 6.5 hours
(390 minutes) of the trading day. The interday returns are as
described in \cite{SY,DY} for time lags from 1 day to 1 month = 20
trading days.

In order to connect the interday and intraday data, we have to
introduce an effective overnight time lag $T_n$.  Without this
correction, the open-to-close and close-to-close variances would have
a discontinuous jump at 1 day, as shown in the inset of the left panel
of Fig.\ \ref{fig:Var}.  By taking the open-to-close time to be 6.5
hours, and the close-to-close time to be 6.5 hours + $T_n$, we find
that variance $\langle x_t^2\rangle$ is proportional to time $t$, as
shown in the left panel of Fig.\ \ref{fig:Var}.  The slope gives us
the Heston parameter $\theta$ in Eq.\ (\ref{eq:DY2}).  $T_n$ is about
2 hours (see Table \ref{Parameters}).

In the right panel of Fig.\ \ref{fig:Var}, we show the log-linear
plots of the cumulative distribution functions (CDFs) vs.\ normalized
return $x/\sqrt{\theta t}$.  The $\makebox{CDF}_t(x)$ is defined as
$\int_{-\infty}^xP_t(x')\,dx'$, and we show $\makebox{CDF}_t(x)$ for
$x<0$ and $1-\makebox{CDF}_t(x)$ for $x>0$.  We observe that CDFs for
different time lags $t$ collapse on a single straight line without any
further fitting (the parameter $\theta$ is taken from the fit in the
left panel).  More than 99\% of the probability in the central part of
the tent-shape distribution function is well described by the
exponential function.  Moreover, the collapsed CDF curves agree with
the DY formula (\ref{short-long})
$P_{t}(x)\propto\exp(-|x|\sqrt{2/\theta t})$ in the short-time limit
for $\alpha=1$ \cite{DY}, which is shown by the dashed lines.

\begin{table}[h]
\caption{Fitting parameters of the Heston model
  with $\alpha=1$ for the 1993--1999 data.
  \label{Parameters}}
\centerline{
\begin{tabular}{c|cccccccc}
\hline
& $\gamma$ & $1/\gamma$ & $\theta$ & $\mu$ & $T_{n}$ \\
& ${1\over{\rm hour}}$ & hour & ${1\over{\rm year}}$
& ${1\over{\rm year}}$ & hour \\
\hline
INTC & $1.029$ & $0 \colon 58$ & $13.04\%$ & $39.8\%$ & $2 \colon 21$ \\
\hline
IBM & $0.096$ & $10 \colon 25$ & $9.63\%$ & $35.3\%$ & $2 \colon 16$  \\
\hline
MRK & $0.554$ & $1 \colon 48$ & $6.57\%$ & $29.4\%$ & $1 \colon 51$ \\
\hline
MSFT & $1.284$ & $0 \colon 47$ & $9.06\%$ & $48.3\%$ & $1 \colon 25$ \\
\hline
\end{tabular}
}
\end{table}

Because the parameter $\gamma$ drops out of the asymptotic Eq.\
(\ref{short-long}), it can be determined only from the crossover
regime between short and long times, which is illustrated in the left
panel of Fig.\ \ref{fig:ThPk}.  We determine $\gamma$ by fitting the
characteristic function $\tilde P_t(k)$, a Fourier transform of
$P_t(x)$ with respect to $x$.  The theoretical characteristic function
of the Heston model is $\tilde P_t(k)=e^{F_{\tilde t}(k)}$
(\ref{eq:DY}).  The empirical characteristic functions (ECFs) can be
constructed from the data series by taking the sum $\tilde P_t(k)={\rm
  Re}\sum_{x_t}\exp(-ikx_t)$ over all returns $x_t$ for a given $t$
\cite{BookCh}.  Fits of ECFs to the DY formula (\ref{eq:DY}) are shown
in the right panel of Fig.\ \ref{fig:ThPk}.  The parameters determined
from the fits are given in Table \ref{Parameters}.

\begin{figure}
\centerline{
  \epsfig{file=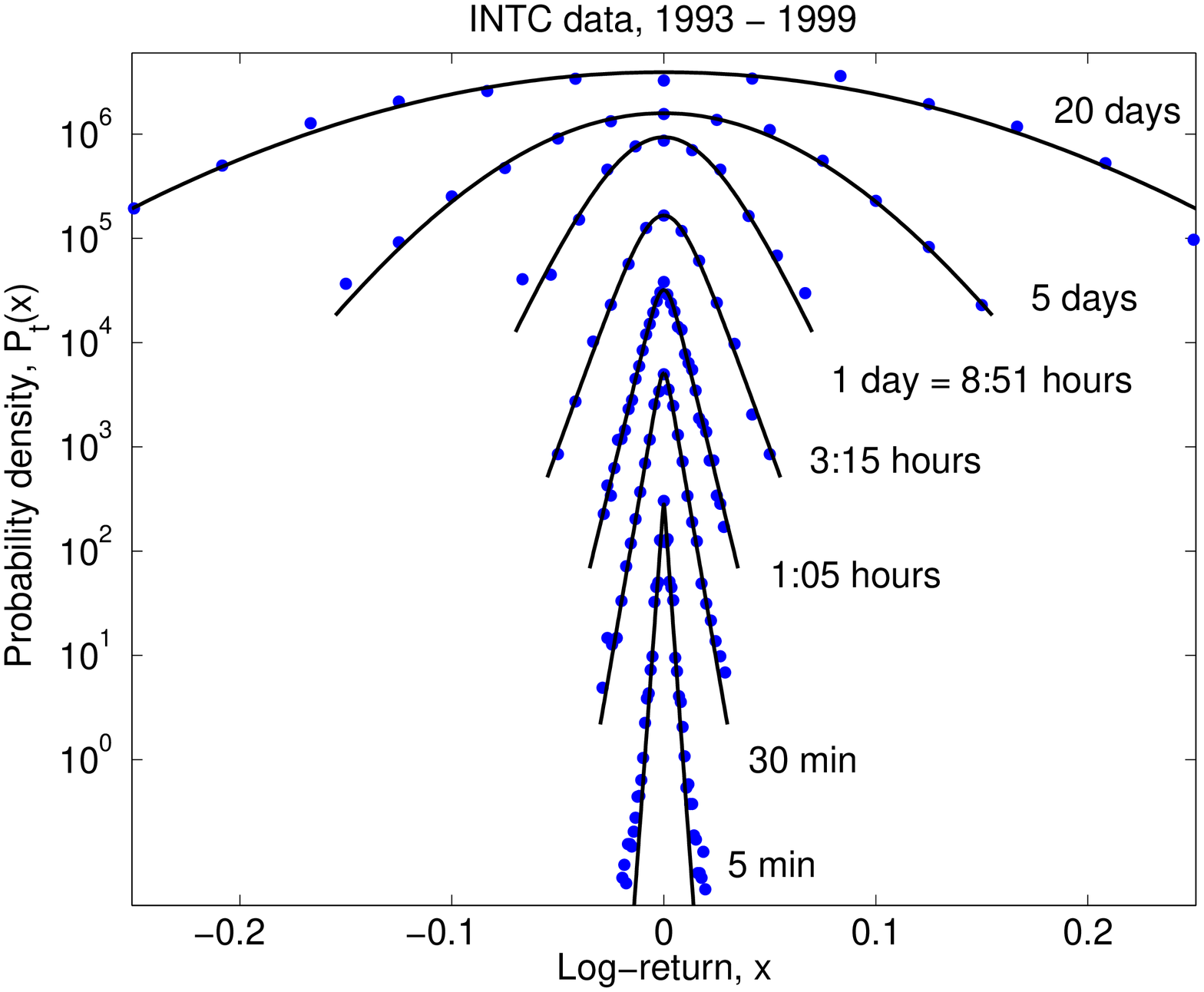,width=0.5\linewidth,angle=-90}}
\centerline{
  \epsfig{file=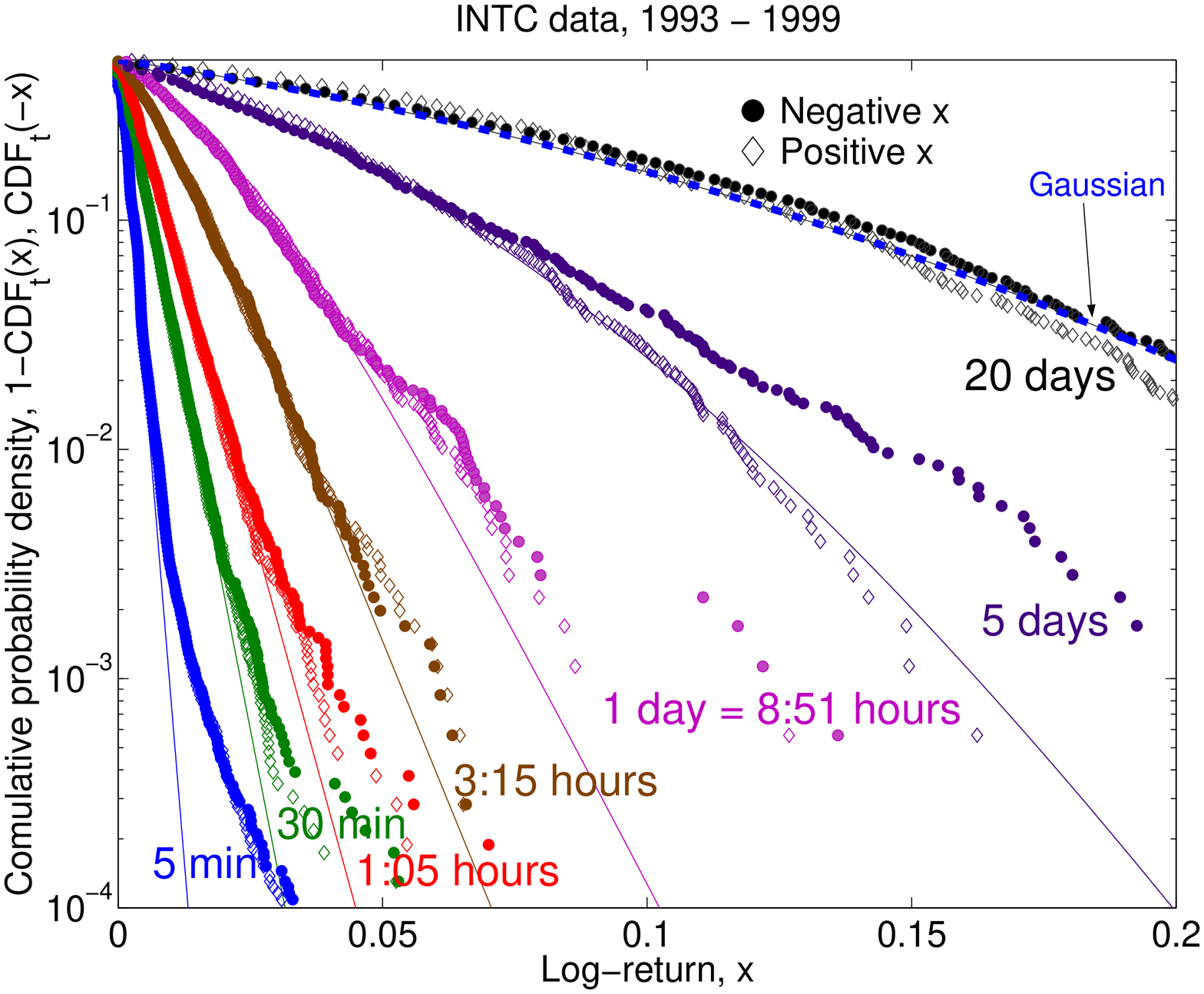,width=0.5\linewidth,angle=-90}}
\caption{Comparison between the 1993--1999 Intel
  data (points) and the DY formula (\ref{eq:DY}) (curves) for PDF
  (top panel) and CDF (bottom panel).}
\label{fig:PdfCdf}
\end{figure}

\begin{figure}[b]
\centerline{
  \epsfig{file=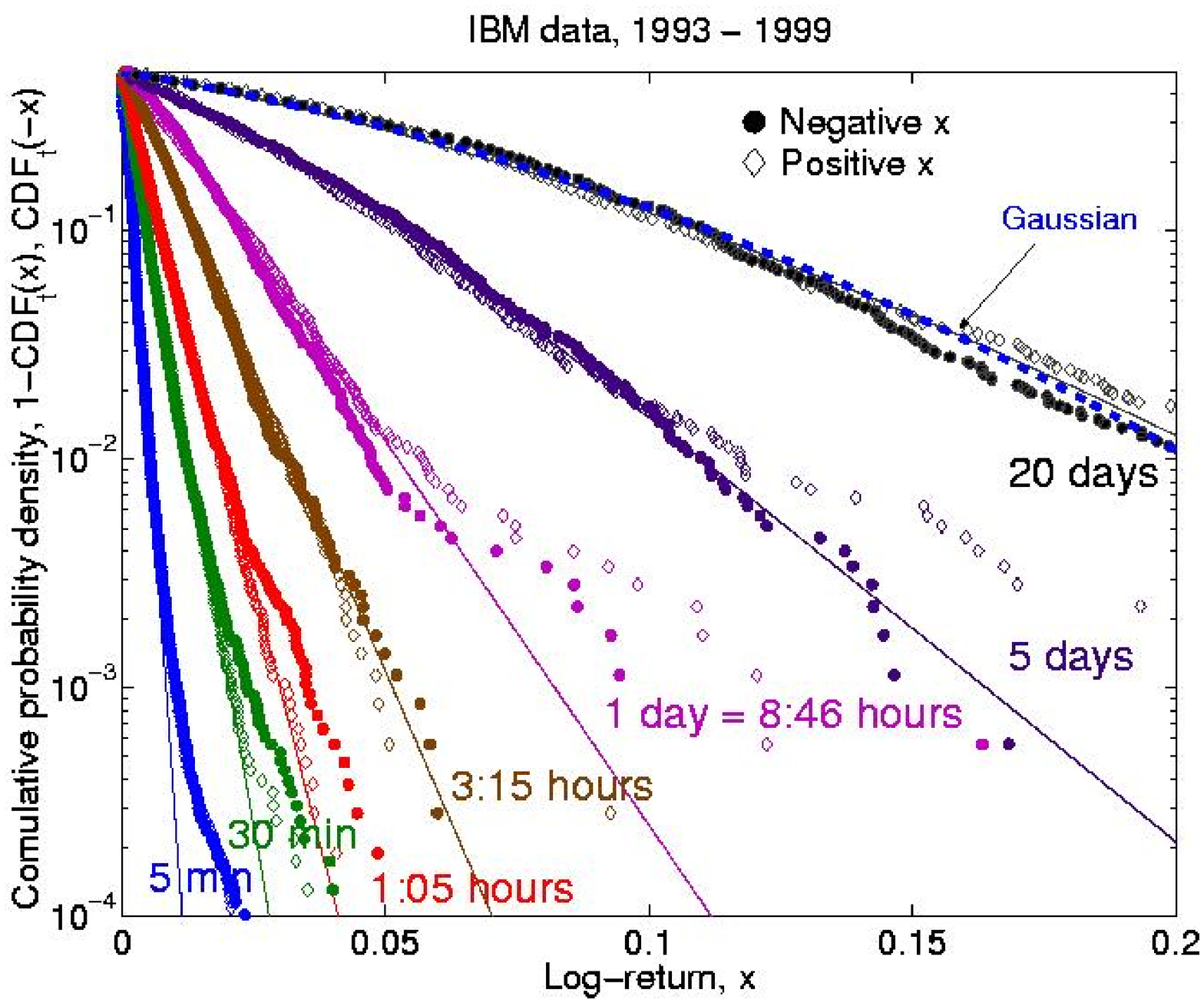,width=0.41\linewidth,angle=-90}
  \epsfig{file=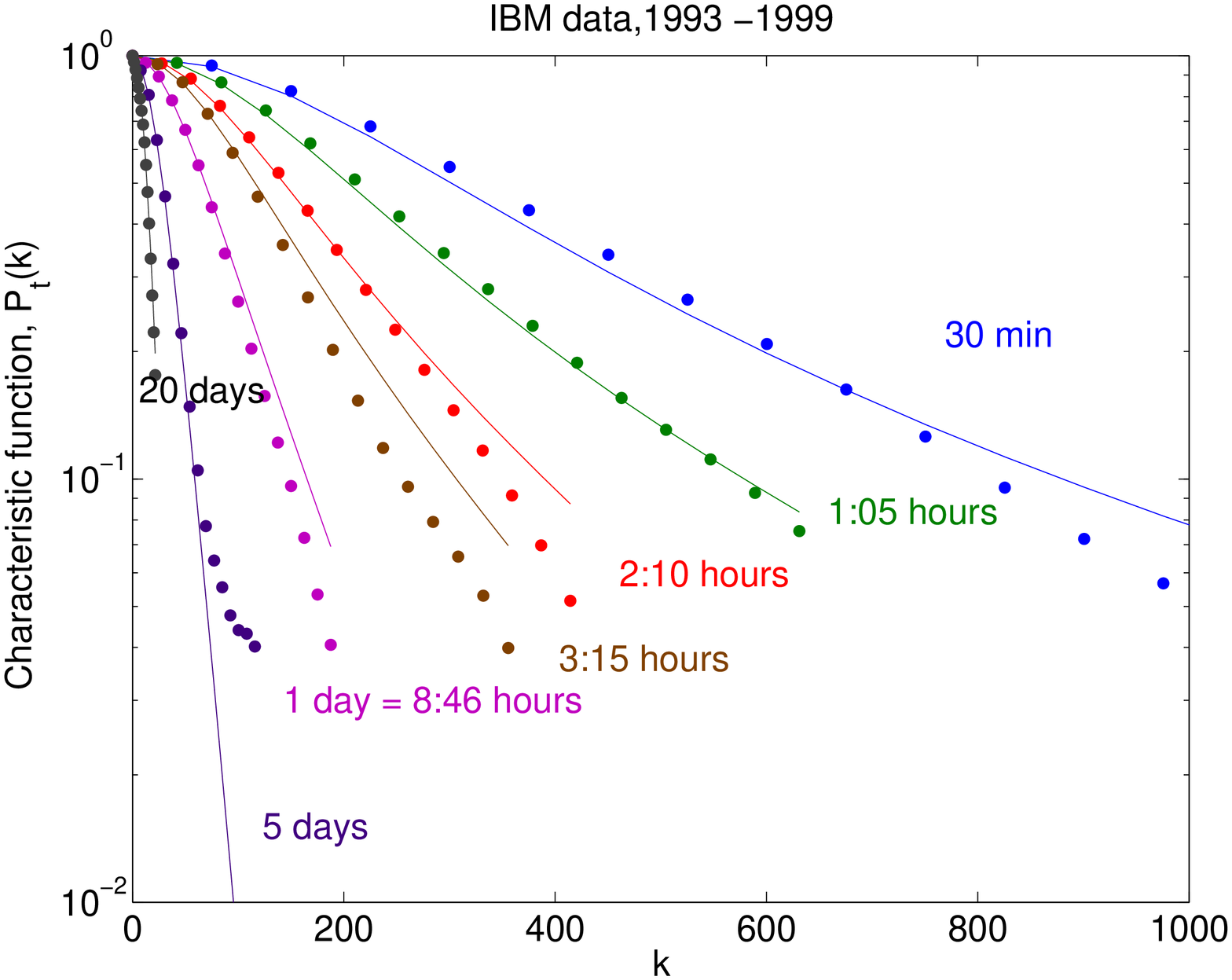,width=0.41\linewidth,angle=-90}}
\centerline{
  \epsfig{file=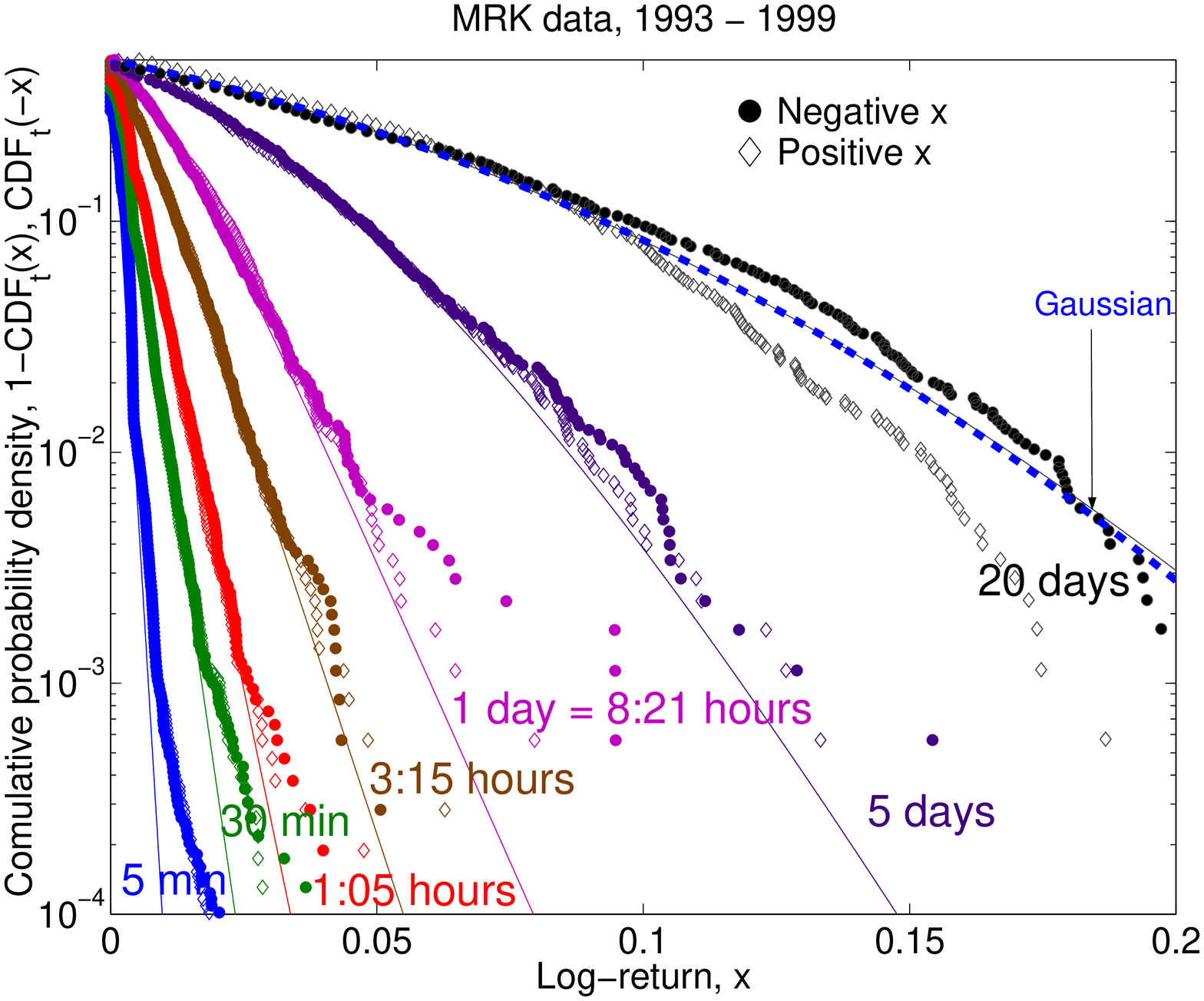,width=0.41\linewidth,angle=-90}
  \epsfig{file=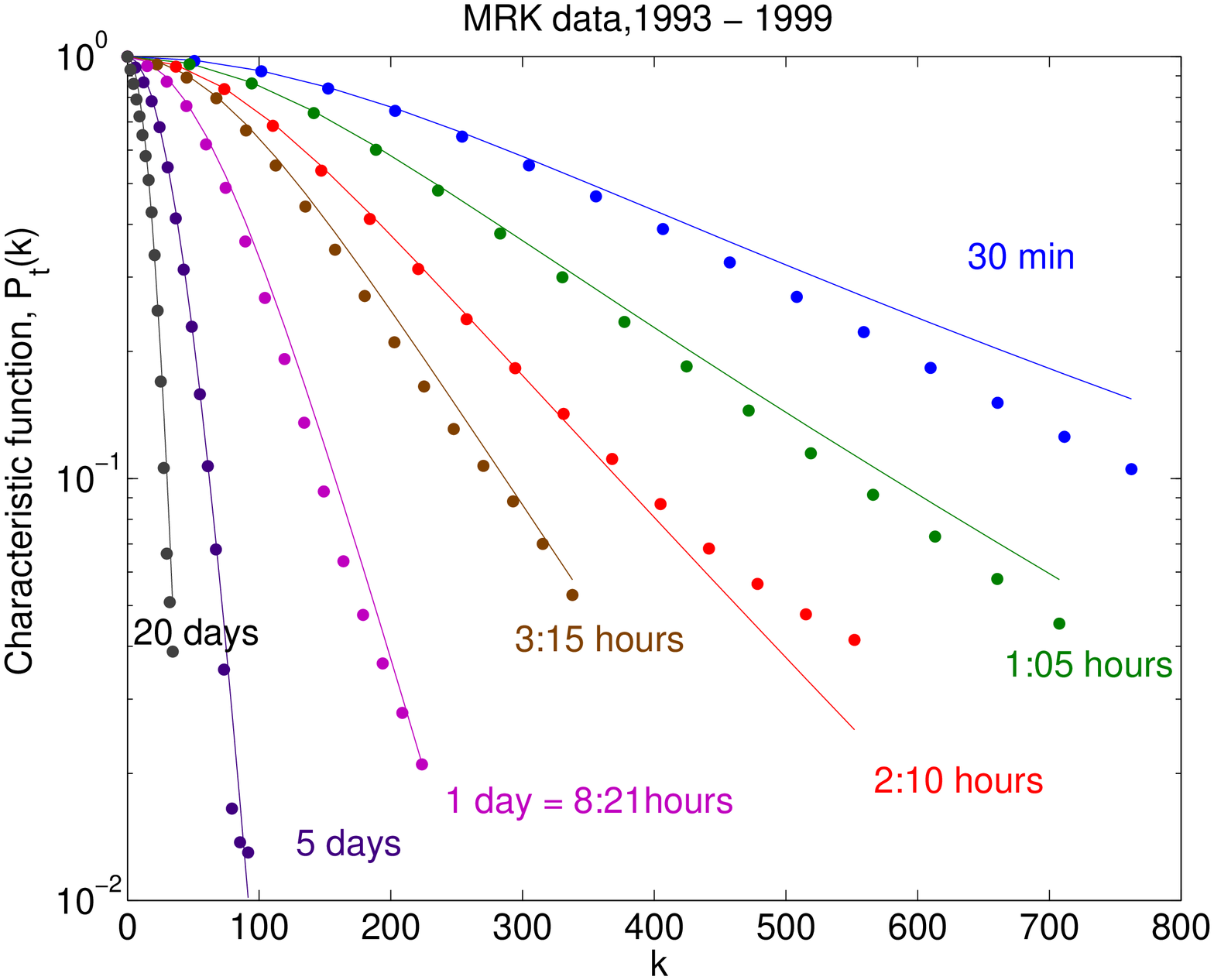,width=0.41\linewidth,angle=-90}}
\centerline{
  \epsfig{file=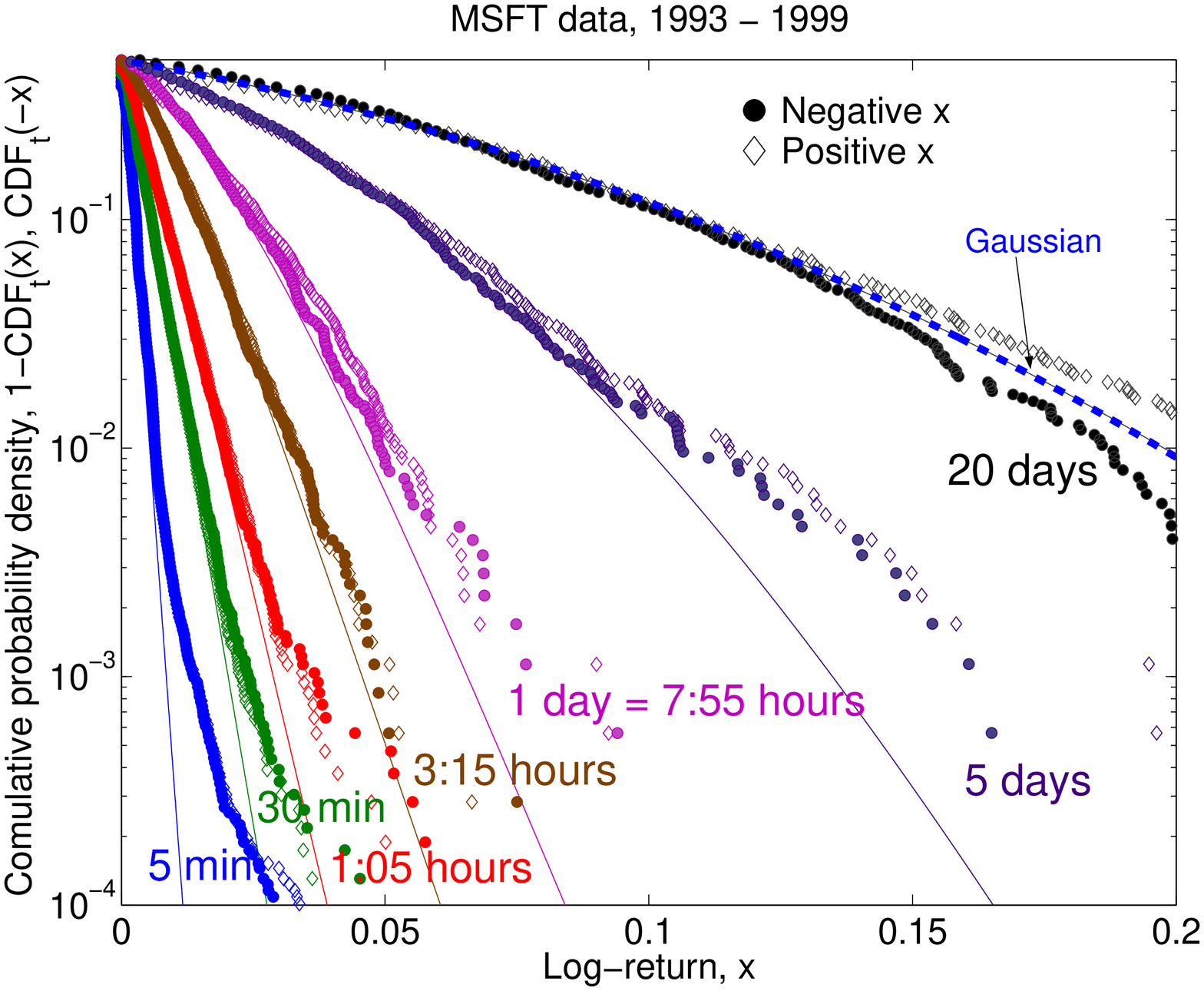,width=0.41\linewidth,angle=-90}
  \epsfig{file=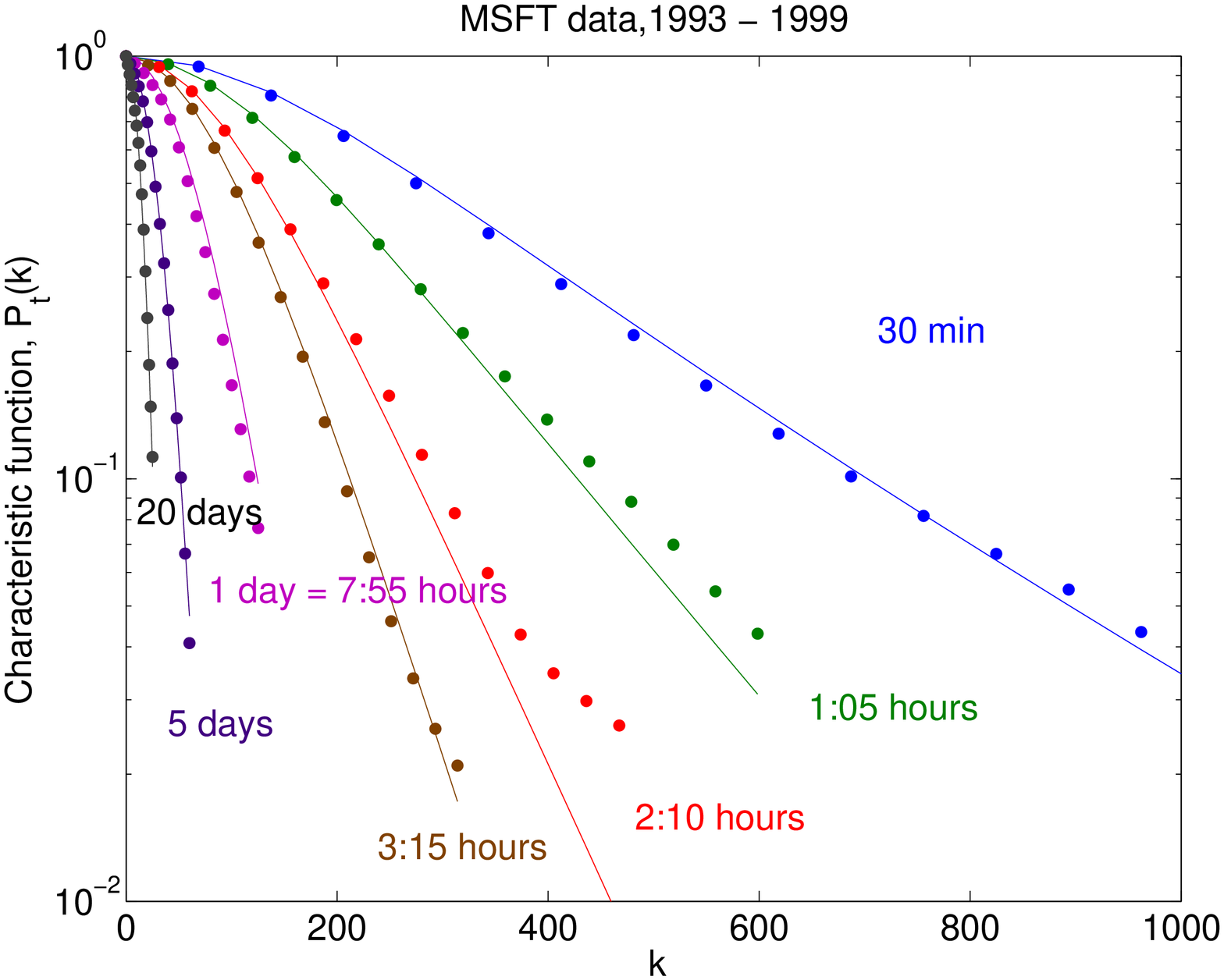,width=0.41\linewidth,angle=-90}}
\caption{Comparison between empirical data
  (symbols) and the DY formula (\ref{eq:DY}) (lines) for CDF (left
  panels) and characteristic function (right panels).}
  \label{fig:many}
\end{figure}

In the left panel of Fig.\ \ref{fig:PdfCdf} we compare the empirical
PDF $P_t(x)$ with the DY formula (\ref{eq:DY}).  The agreement is
quite good, except for the very short time lag of 5 minutes, where the
tails are visibly fatter than exponential.  In order to make a more
detailed comparison, we show the empirical CDFs (points) with the
theoretical DY formula (lines) in the right panel of Fig.\
\ref{fig:PdfCdf}.  We see that, for micro time lags of the order of 5
minutes, the power-law tails are significant.  However, for meso time
lags, the CDFs fall onto straight lines in the log-linear plot,
indicating exponential law.  For even longer time lags, they evolve
into the Gaussian distribution in agreement with the DY formula
(\ref{eq:DY}) for the Heston model.  To illustrate the point further,
we compare empirical and theoretical data for several other companies
in Fig.\ \ref{fig:many}.

In the empirical CDF plots, we actually show the ranking plots of
log-returns $x_t$ for a given $t$.  So, each point in the plot
represents a single instance of price change.  Thus, the last one or
two dozens of the points at the far tail of each plot constitute a
statistically small group and show large amount of noise.
Statistically reliable conclusions can be made only about the central
part of the distribution, where the points are dense, but not about
the far tails.

\subsection{Conclusions} \label{sec:conclusions}

We have shown that in the mesoscopic range of time lags, the
probability distribution of financial returns interpolates between
exponential and Gaussian law.  The time range where the distribution
is exponential depends on a particular company, but it is typically
between an hour and few days. Similar exponential distributions have
been reported for the Indian \cite{India}, Japanese \cite{Japan},
German \cite{Germany}, and Brazilian markets \cite{Vicente,Miranda}, as
well as for the US market \cite{SY,DY,USA} (see also Fig. 2.12 in
\cite{BP}).  The DY formula \cite{DY} for the Heston model
\cite{Heston} captures the main features of the probability
distribution of returns from an hour to a month with a single set of
parameters.

\section{Number of trades and subordination}\label{subCh}

The concept of subordination has important fundamental and
practical implications. From a fundamental point of view, it gives
a relation between microstructure of the market and price
formation that can be exploited in simulations and modelling
\cite{Farmer2004,Engle2000,OHara,Pohlmeier2003}. From a practical
point of view, the subordinator can be identified with the
integrated variance $V_t$ \cite{Bollerslev1996,Geman}. This would
imply a direct measure of
the mean square return which could impact pricing and hedging both
of options on a particular stock as well as variance swaps and
options on the variance.

In this chapter we verify and model the subordination hypothesis
as given by Eq. (\ref{pvt}). We will restrict our study to
intraday Intel data in the year $1997$. We restrict to a year of
data because of the nonlinear drift of the number of trades: we
would like to minimize this effect (see Fig. \ref{fig:cunN}). We
chose Intel because it has been studied by us in Ref. \cite{SPY}
(chapter \ref{expD}) and it can be modelled well with the Heston
model introduced in chapter \ref{expD}. It is true that it is a
highly traded stock, and that is an advantage, since that are a
lot of trades in a day and therefore the statistics is better.
Therefore smaller stocks should be also checked in the future. The
year of $1997$ represents most of what one finds for other years,
except perhaps $2000$ and $2001$ which we did not verified because
of technical problems (to large data set requires especial
computing techniques that should  be implemented in the future).

We begin by showing the influence of the discrete nature of the
absolute price change in the intraday log-return data. This is
rarely pointed out, even though there is a vast literature on
intraday log-returns
\cite{BP,Stanley1999a,Stanley1999b,BS,MullerBook}. This
discreteness has to be accounted for when considering
subordination, or even when studying intraday returns. It implies
that a continuous probability density is only a convenient
approximation for some return horizons.

In section \ref{subCheck}, we verify when and for what range of
data does subordination apply. We assume that the integrated
volatility $V_t$ is the random subordinator of a driftless
Brownian motion and that $V_t$ is proportional to the number of
trades $N_t$ in an interval of time $t$. We also use tick-by-tick
data to check for subordination by constructing the probability
density of the log-returns $x_N$ after $N$ trades (\ref{pvt}).

In section \ref{model}, we model the integrated variance $V_t$ with
the CIR process introduced in Eq. (\ref{eq:CIR}). We present the level
of agreement between the data and the theoretical CIR model and we
link these results to the distribution of log-returns $x_t$.

In the last section, we present a summary of our findings.

\subsection{Discrete nature of stock returns}\label{disc}

\begin{figure}
\centerline{\epsfig{file=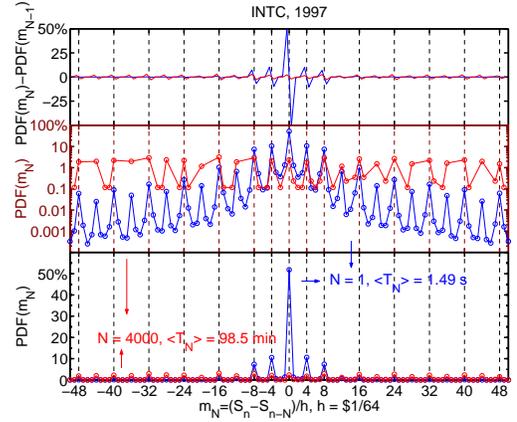,width=0.65\linewidth,angle=-90}}
\caption{Dimensionless absolute returns $m_{N}=(S_{n}-S_{n-N})/h$
for $N$ trades in log linear and linear scale (center and bottom
panels respectively). In the top panel we show the difference of
the PDFs for $m_{N}$ and $m_{N-1}$ to illustrate the oscillatory
nature of the discrete PDF for absolute returns: it evolves from a
``pulse'' like shape for $N=1$ to a ``constant wave'' for
$N=4000$.} \label{fig:dS}
\end{figure}

On a tick-by-tick level, price changes are discrete. There is a
minimal price change for bid and offers that is set by internal
rules of the stock exchange. In the case of Intel in the year of
$1997$, the minimal price change was $\$1/8$ for the first part of
the year and after June, 24th it became $\$1/16$
\cite{Goldstein2000,Pruitt2000}. Nevertheless, empirically we find
that the smallest price change on realized transactions is
$h=\$1/64$ (Fig. \ref{fig:dS}). This difference is a direct
consequence of the mechanism of trading, and we will not study it
here (see Ref. \cite{Farmer2003Prl, Farmer2005})\footnote{One of the
  possible reasons for the different between empirical $h$ and quoted
  price $h$ is the bid and ask spread. That is the difference in
  price between the buy and sell quote. Since we work with transaction
prices, these prices will tend to jump between the bid and ask.
And this gap is not quantized by law. Another point to remember is
that this quantum set by law only make sense for limit orders
(where the buyer of seller quotes his preference price) and not
market orders (the buyer or seller buys at the first available
price). TAQ does not distinguish between order types.}. We note
that the minimal price change set by law is clear in Fig.
\ref{fig:dS}, since the most probable price changes are indeed
$0$, $\pm 4h =\$ 1/16$ and $\pm 8h =\$1/8$, according to the rules
of the NASDAQ exchange in $1997$.

Our goal in this section is to identify the discrete nature of
absolute price changes \footnote{Absolute price change is used
here as an opposite to relative price changes. We do not refer to
the absolute value. What we refer as absolute price changes are
also known as the P\&L of the trade.} after $N$ trades ($m_{N}h =
S_{n}-S_{n-N}$) in the log-returns after $N$ trades
($x_N=ln(S_{n})-ln(S_{n-N})$) and in log-returns after a time-lag
$t$ ($x_t=ln(S_{T})-ln(S_{T-t})$), since these log-returns are the
quantities that we ultimately want to model. We want to point out
that the discrete nature of the log-returns for intraday work is
generally overlooked but it can influence in the analysis of short
returns.

We will refer to minimal price change $h=\$1/64$ as ``quantum of
price'' or simply ``quantum'' in analogy with quantum mechanics.

The discrete nature of the price change can be used to model the
price dynamics starting from a microscopic approach as recently
suggested in
\cite{Engle2000,Bollerslev2001,Engle1998,Pohlmeier2001}. We are
interested in the limit where the quantum effect is not noticeable
and therefore quantities such as number of trades and returns can
be treated as continuous random variables.

\begin{figure}
\centerline{\epsfig{file=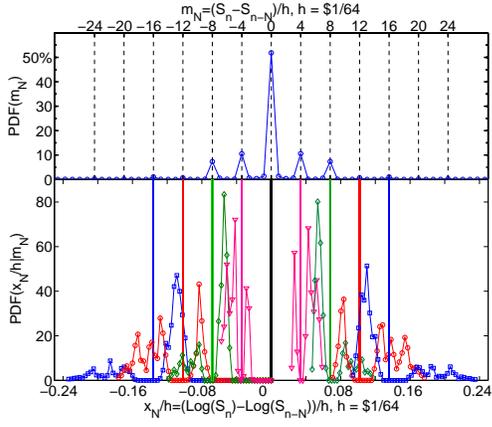,width=0.65\linewidth,angle=-90}}
\caption{Effect of taking log-returns instead of taking absolute
  returns. Lower panel shows the probability density of the dimensionless log-returns
  $x_{N}/h$ conditioned on $m_{N}$, $P(x_{N} / h | m_{N})$.
  The values concentrated about a multiple of $h$ (upper panel), spread about their respective $h$ value.
   The vertical color coded lines (lower panel) indicate the $h$ value from which each, equally color coded,
   $P(x_{N} / h | m_{N})$ originated. The discreteness of $m_N$ is removed by taking log-returns since the spread of
   $P(x_{N}/ h | m_{N})$ is larger than $h$.}
\label{fig:MixN1}
\end{figure}

Fig. \ref{fig:dS} shows the probability density for the
dimensionless absolute price return $m_{N} = (S_{n}-S_{n-N})/h$
after $N$ trades in steps of one quantum $h$. The nature of the
tick-by-tick distribution ($N=1$) is considerably different from
$N=4000$. More than $50\%$ of the returns are zero for $N=1$, and
most of the other returns have a probability of less than $1\%$
except $\pm 4h$ and $\pm 8h$. The probability has a clearly
oscillatory nature where multiples of $4h$ are maxima (Fig.
\ref{fig:dS}, top panel). After $4000$ trades the probability
distribution for $m_{N}$ has changed into a two level system (Fig.
\ref{fig:dS}). The probability of the most probable $m_N$ in $N=1$
have now approximately the same probability. Therefore, the zero
return has (after $4000$ trades) a comparable probability to the
other probability maxima.

The quantum nature of the price changes is removed by working with
log-returns, except for the zero return. Notice that intraday
log-returns can be approximated by the ratio \cite{Montero}

\begin{equation}
x_{N}=\ln{S_{n}}-\ln{S_{n-N}} \approx
\frac{S_{n}-S_{n-N}}{S_{n-N}}=\frac{m_{N}}{S_{n-N}/h}.
\label{xNratio}
\end{equation}

The log-returns can also be written

\begin{eqnarray}
&& m_{0,N}h=0 \nonumber \\
&& m_{i,N}h=S_{iN}-S_{(i-1)N}, i=1,2,3... \nonumber \\
&& x_{i,N}=\frac{m_{N}}{\sum_{j=0}^{j=i-1}m_{j,N}+ C},C=S_{0}/h,i=1,2,3,...,
\label{xm}
\end{eqnarray}

\noindent where $S_{0}$ is the first open of the year (in the case of Intel 1997,
$S_{0}=\$131.75$).

The effect of taking log-returns is illustrated in Fig.
\ref{fig:MixN1}. For each absolute return $m_N$, there is a
potentially different denominator $S_{n-N}/h$
(\ref{xNratio}) composed by a random walk with integer valued
steps about a level $C$ (\ref{xm}). Clearly the values of the
ratio $x_N$ will not be integer. Therefore, the ratio of $m_{N}$
in Eq. (\ref{xm}) spreads the concentrated discrete absolute
returns multiple of $h$, around the multiple.

The lower panel of
Fig. \ref{fig:MixN1} shows the probability density of $x_{N}/h$
conditioned on $m_{N}$. The conditional probability density
$P(x_{N}/h | m_{N})$ illustrates a spread for each $m_N$ that is
larger than $h$. This spread is enough to mix the discreteness with
exception of $m_N=0$.

\begin{figure}
\centerline{\epsfig{file=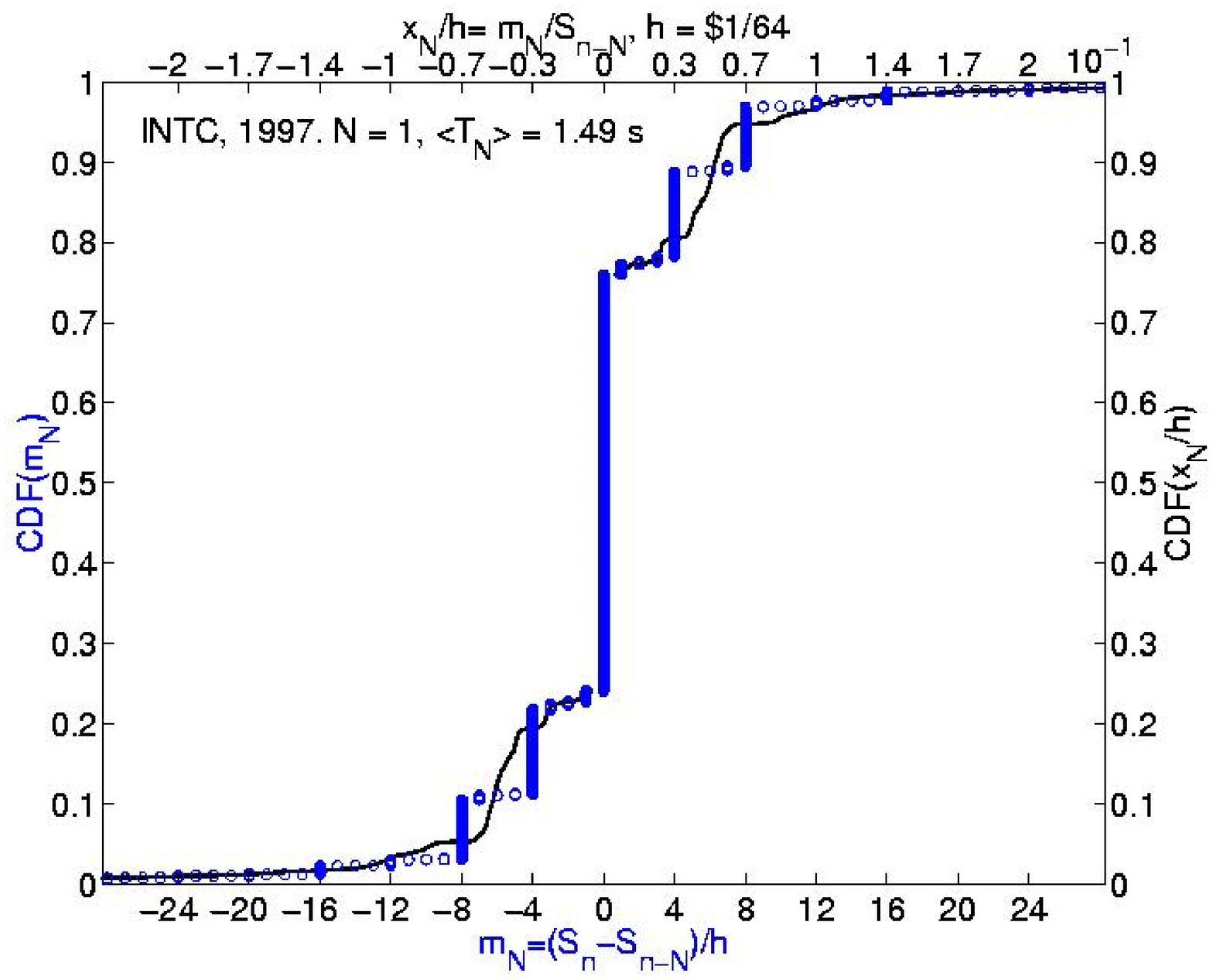,width=0.5\linewidth,angle=-90}}
\centerline{
\epsfig{file=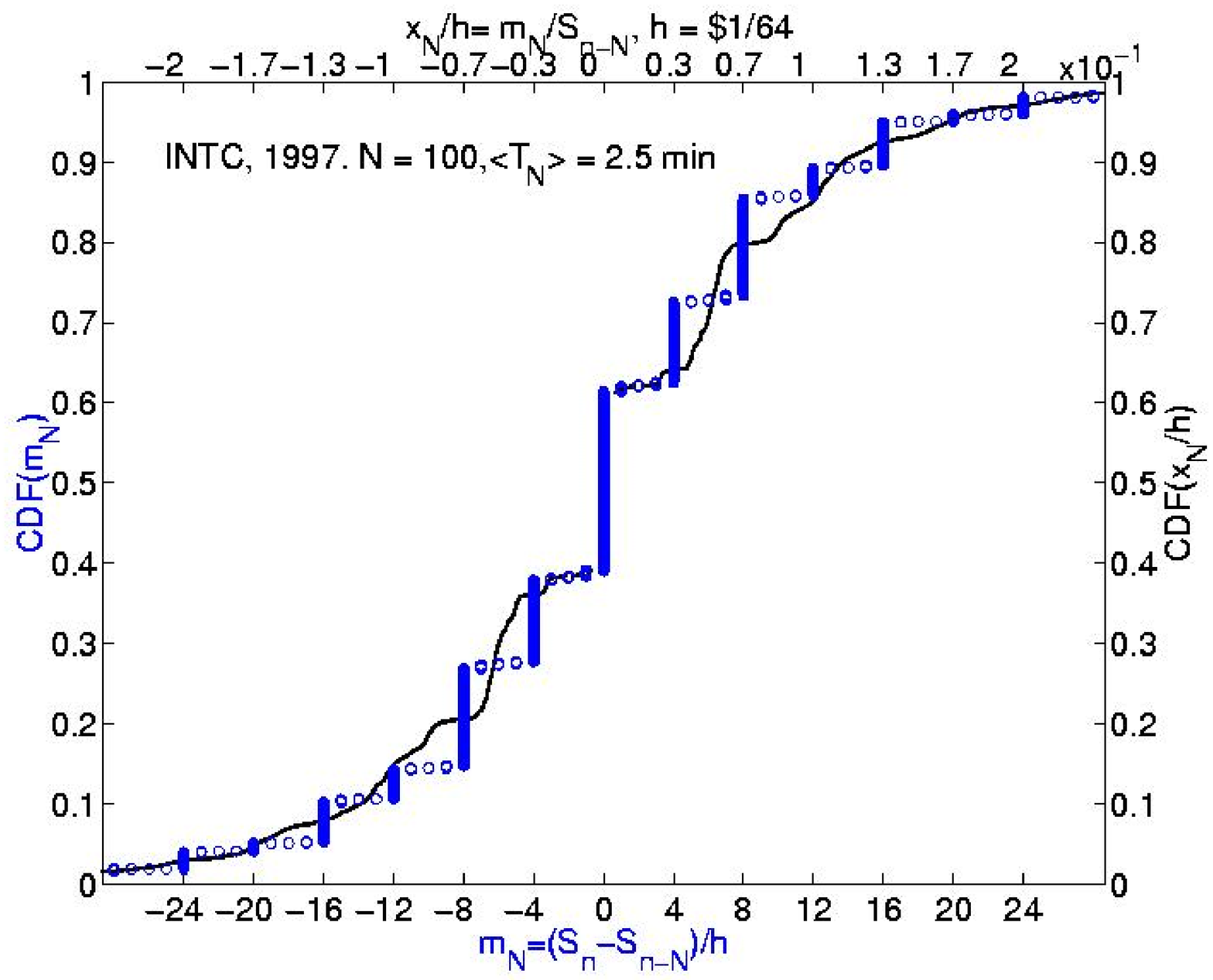,width=0.5\linewidth,angle=-90}}
\caption{Cumulative probability density for both dimensionless
log-returns, $x_N/h$ (black line), and dimensionless absolute
returns, $m_N$ (blue symbols). Even though the discreteness of
$m_N$ is removed with exception of $x_N=0$, the signature of such
discreteness is still visible. Notice the stepwise nature of the
black line.} \label{fig:MixCdf1}
\end{figure}

\begin{figure}
\centerline{\epsfig{file=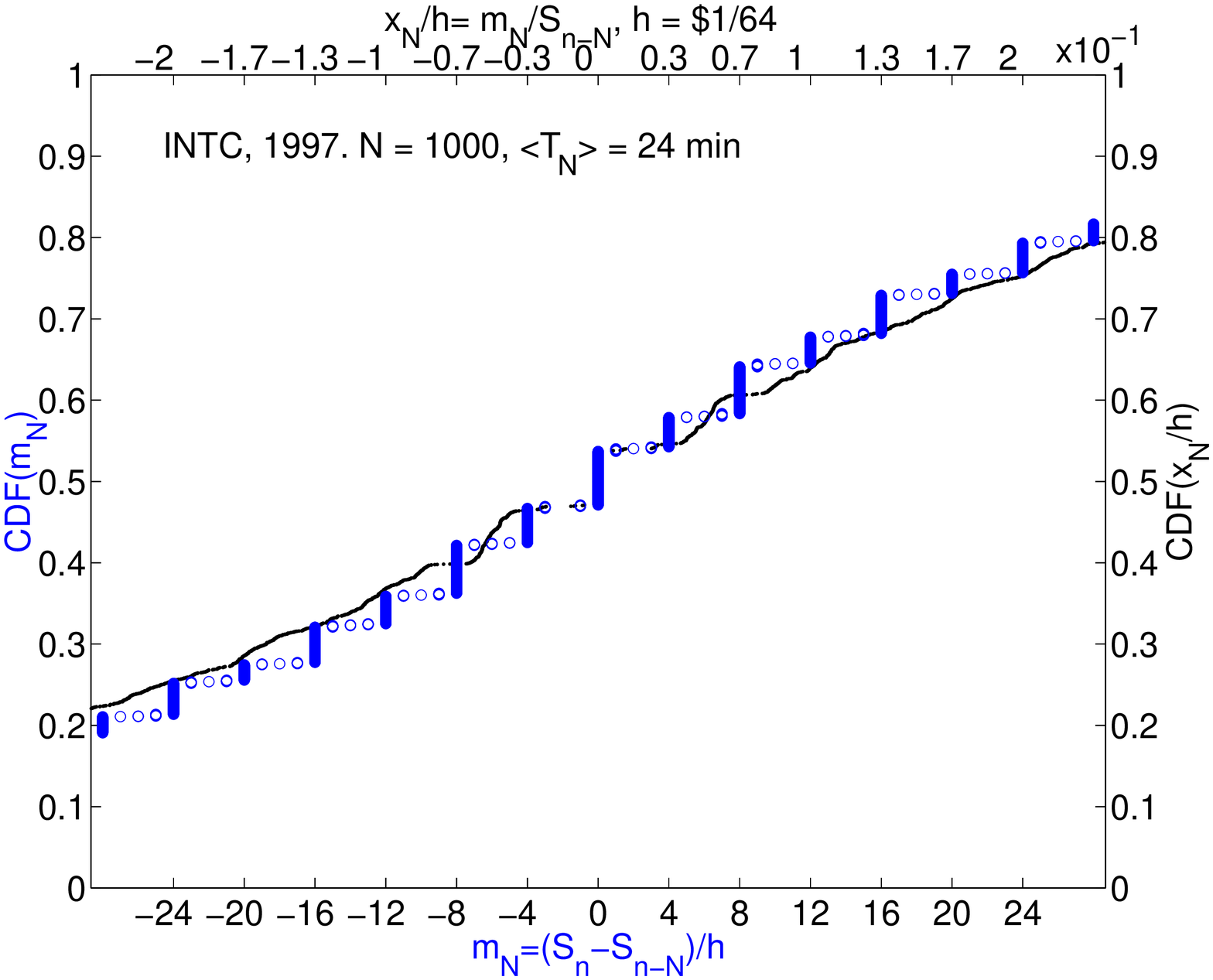,width=0.5\linewidth,angle=-90}}
\centerline{
\epsfig{file=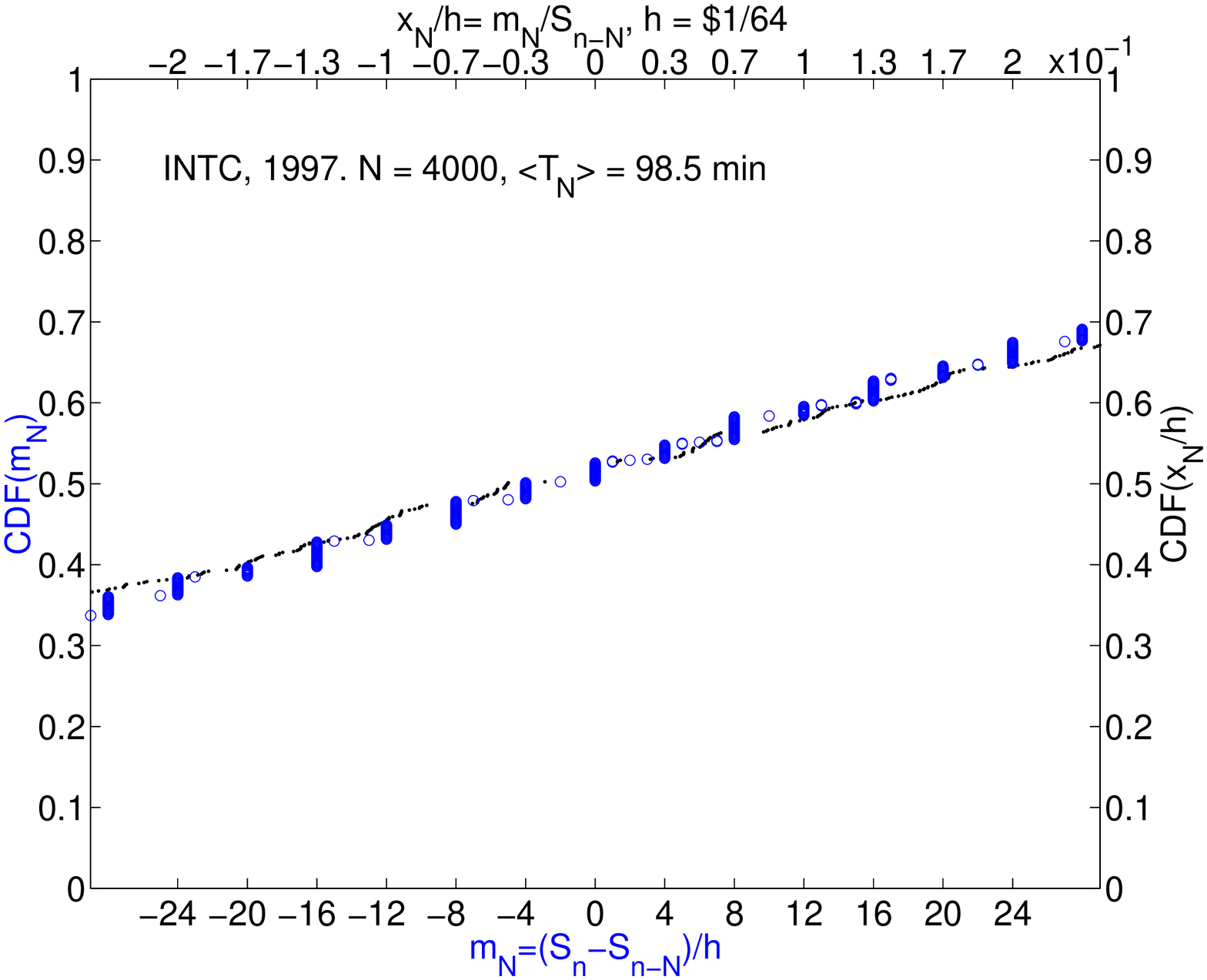,width=0.5\linewidth,angle=-90}}
\caption{Cumulative probability density for both dimensionless
log-returns, $x_N/h$, and dimensionless absolute returns, $m_N$.
When $N$ increases the CDF becomes progressively less oscillatory and the
discrete nature of the underlying absolute returns becomes less
clear.} \label{fig:MixCdf2}
\end{figure}

The quality of such a mixture can be seen in Fig.
\ref{fig:MixCdf1} and Fig. \ref{fig:MixCdf2}. Even though the
cumulative density function for $x_N$ is practically continuous
(even for $N=1$) with exception of $x_N=0$, the stepwise nature of
$m_N$ can be easily recognized up to $N=1000$ (Fig.
\ref{fig:MixCdf2}). The oscillations in the cumulative density
functions for $x_N$ are centered about the discrete steps of the
cumulative density function of $m_N$.

The discrete quantum effect at $m_N=0$ is quite persistent, but it
can be neglected for returns $x_N$ with large number of trades $N$
(for instance $N=4000$). Empirically, it appears that the criteria
for neglecting the $m_N=0$ effect is that the probability of
having $m_N=0$ is of the same order of magnitude as the
probability of having any other $m_N$ (Fig.\ref{fig:dS}). For
Intel $1997$ this transition starts approximately at $N=1000$.

\begin{figure}
\centerline{\epsfig{file=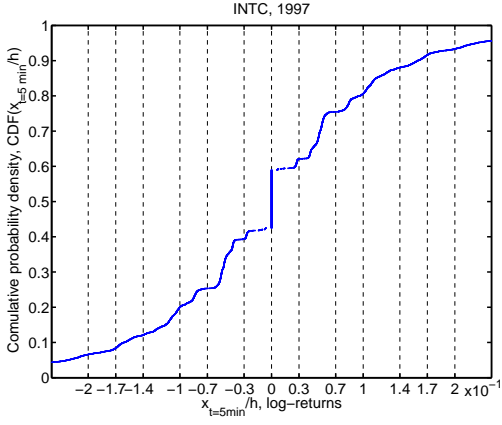,width=0.65\linewidth,angle=-90}}
\caption{Cumulative probability density for $x_t/h$ with $t=5$
minutes. The discreteness at zero persists from $x_N/h$ as well as
the oscillation (stepwise nature) of the CDF.} \label{fig:MixXt}
\end{figure}

The effect of data discreteness is also present in the log-return
$x_t$ of time lag $t$. From the log-return $x_t$, we can construct
$x_N$ by conditioning on the number of trades $N$ present in $t$
($N_t$). The opposite is also true, by conditioning on $t$ we can
construct $x_t$ from $x_N$. Therefore some of the discrete effects
that are present in $x_N$ will be present in $x_t$. As an example
consider $5$ minute log-returns. The average number of trades is
$\langle N_{t=5min} \rangle =  200 \pm 184$. Because of the
reciprocity in constructing the PDF for $x_t$ from $x_N$ (and
vice-versa) by conditioning, this shows that in the composition of
$x_{t=5min}$, there is a wide range of $x_{N}$ for which the
discrete features can not be ignored (clear oscillations and large
probability for $x_t=0$). If we approximate the PDF of
$N_{t=5min}$ by a Gaussian distribution, we would have in
$x_{t=5min}$, with the highest probability, $N_{t=5min}=200$.
Therefore some fraction of $x_N=200$ will be sampled when we
construct the probability of $x_{t=5min}$ by conditioning, these
returns clearly have a lot of discrete features (Fig.
\ref{fig:MixCdf2}) and these features will pass to $x_{t=5min}$.

Fig. \ref{fig:MixXt} shows the oscillatory stepwise cumulative
probability density and also the special nature of $x_{t=5min} =
0$ for the cumulative probability density of $x_{t=5min}$. Compare
this figure with Fig. \ref{fig:MixCdf1} and Fig.
\ref{fig:MixCdf2}. These features originate from $x_N$ and
represent small flat portions in the probability density function.

Finally, from the sequence of Figs. \ref{fig:MixCdf1} and
\ref{fig:MixCdf2} and the correspondence between $x_N$ and $x_t$,
we can conclude that the discrete effects become negligible for a
time lag $t>1$ hour.

\subsection{Verifying subordination with intraday
data}\label{subCheck}

The hypothesis of subordination introduced by Clark \cite{Clark}
has had a strong economical implication, and following his work
there is a vast body of theoretical and empirical work which
addresses the issue
\cite{Smith1994,Manganelli2000,Ane2000,Stanley2000,Farmer2004}.
Similar to the work of Refs. \cite{Ane2000,Stanley2000}, we verify
for subordination considering integrated variance $V_t$,
constructed from the number of trades $N_t$, to be the subordinator
of a Brownian motion.

Due to the discrete nature of the distribution of intraday returns
presented in section (\ref{disc}), we can only talk about
subordination as formulated in equation (\ref{pvt}) after the discrete
effects become small. In what follows, we will take all time lags
even those where the discrete effects are large. Nevertheless, we
will see that the best subordination will take place for time lags
for which discrete effects can be ignored.


\begin{figure}
\centerline{\epsfig{file=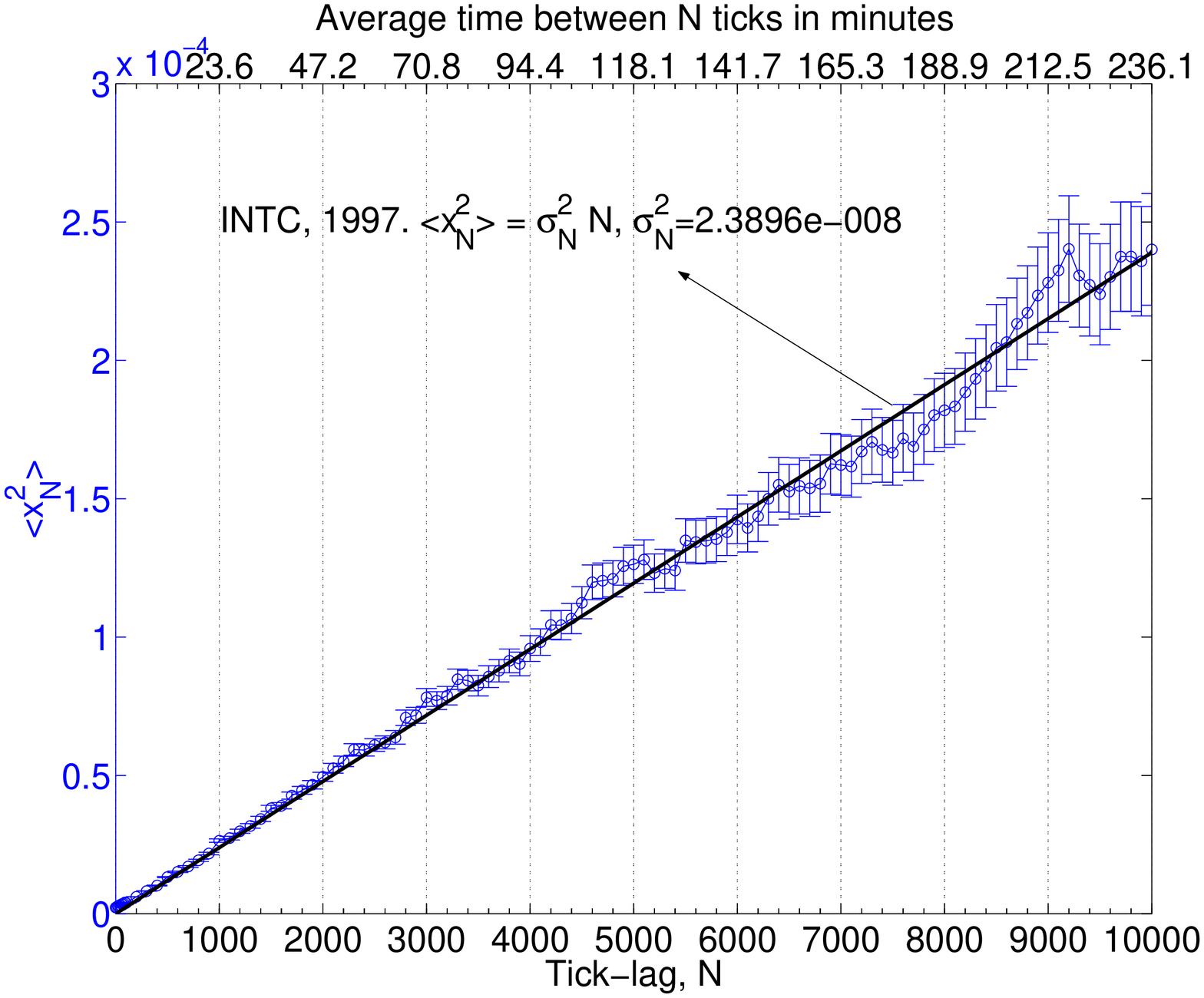,width=0.4\linewidth,angle=-90}}
\caption{Variance of the log-return $x_N$ for $N=1$ to $N=10000$.}
\label{fig:xN2}

\centerline{\epsfig{file=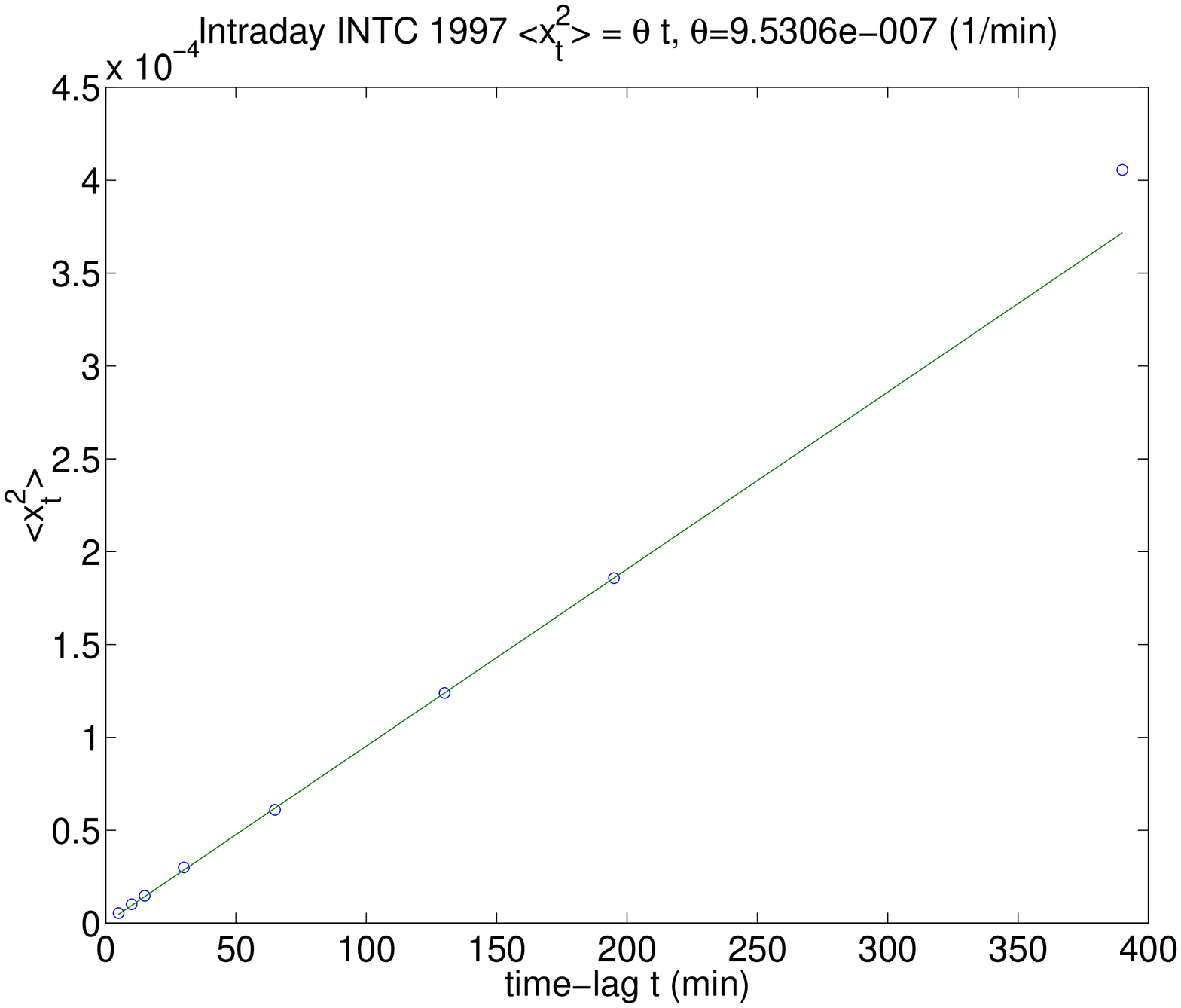,width=0.4\linewidth,angle=-90}}
\caption{Variance of the demean log-return $x_t$ for intraday time
lags $t$.} \label{fig:xt2}

\centerline{\epsfig{file=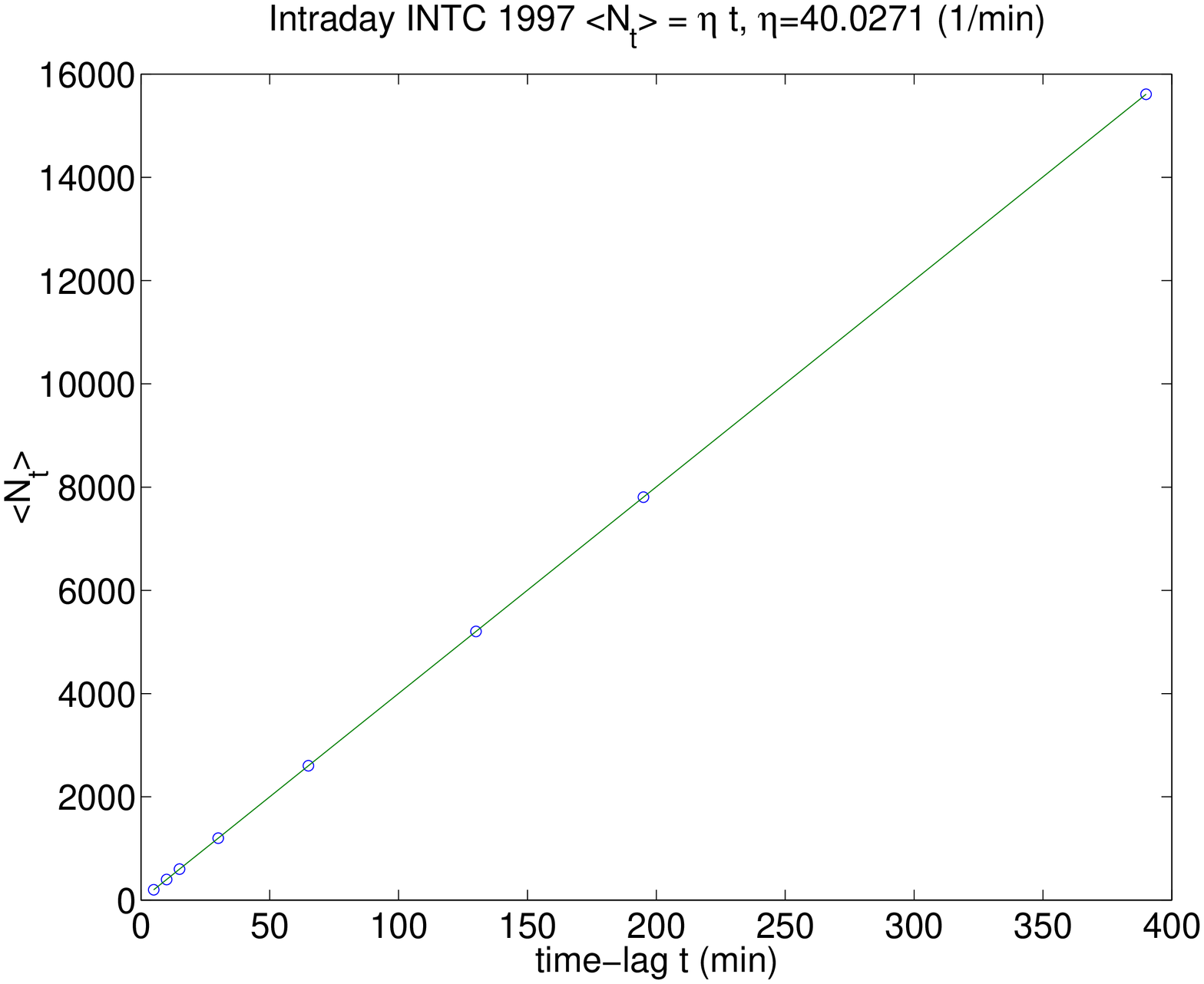,width=0.4\linewidth,angle=-90}}
\caption{Average number of trades in an intraday interval $t$.}
\label{fig:Nt}
\end{figure}

The first implication of subordination can be verified with the
use of moments given by equations (\ref{M2}) and (\ref{M4}). Figs.
\ref{fig:xt2} and \ref{fig:Nt} show the linear time relation for
both the variance of $x_t$ and the mean of $N_t$ as expected from
equation (\ref{M2}). Furthermore, since we are assuming a Brownian
motion with stochastic variance given by the number of trades, log-returns $x_N$ after $N$ trades should
be Gaussian distributed with variance $\langle x_{N}^{2} \rangle =
\sigma_{N}^{2} N$. Fig. \ref{fig:xN2} shows the linear relation of
$\langle x_{N}^{2} \rangle$ vs. $N$. The implied consistency
between the slope values in Figs. \ref{fig:xN2}, \ref{fig:xt2} and
\ref{fig:Nt} required by subordination is

\begin{equation}
\langle x_{t}^{2} \rangle = \theta t = \sigma_{N}^{2} \langle N_{t}
\rangle = \sigma_{N}^{2}\eta t \, \Rightarrow \, \theta =
\sigma_{N}^{2} \eta.
\label{slope}
\end{equation}

Using expression (\ref{slope}), the difference between $\theta$
measured (Fig.\ref{fig:xt2}) and $\theta = \eta \sigma_{N}^{2}$
from Fig. \ref{fig:xN2} and Fig. \ref{fig:Nt} is less than $1\%$.

\begin{figure}
\centerline{\epsfig{file=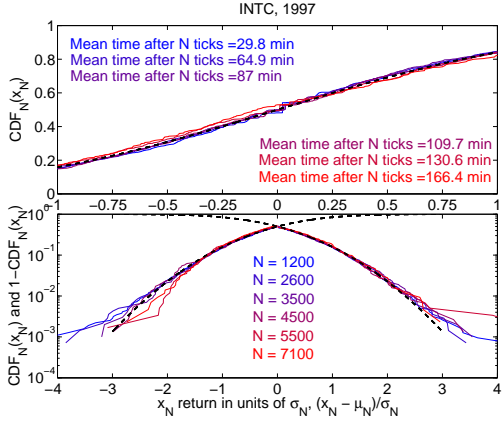,width=0.65\linewidth,angle=-90}}
\caption{Cumulative probability density for the demean and
standard deviation (STD) normalized $x_N$ log-returns (color coded solid
curves), compared to the Gaussian distribution of mean zero and STD
one (dashed curve). From small $N$ to large $N$, there is a
progressive agreement with the Gaussian with best agreement
between $N=3500$ and $N=4500$. While smaller values of $N$ have
CDFs above the Gaussian, larger values are below the Gaussian.}
\label{fig:subxN}
\end{figure}
\begin{figure}
\centerline{\epsfig{file=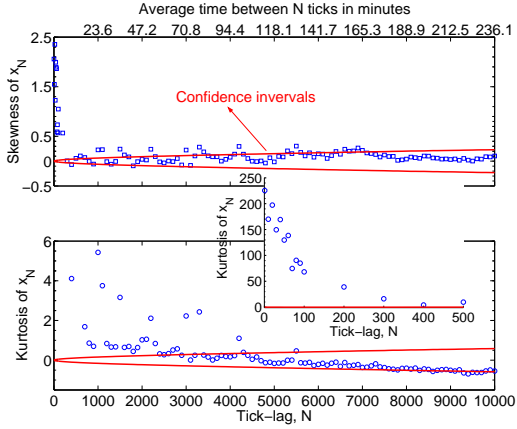,width=0.65\linewidth,angle=-90}}
\caption{Skewness and excess kurtosis (labelled as "kurtosis" in
the
  figure) as a function of $N$ for the normalized log-returns $x_N$ in
  Fig. \ref{fig:subxN}. For a Gaussian distribution the skewness is
  zero and the excess kurtosis is also zero. As the number of trades
  (ticks) $N$ increase the skewness and excess kurtosis become
  zero. The probability density for $x_N$ can be well approximated by
  a Gaussian for $N>2500$, since both skewness and excess kurtosis are small.}
\label{fig:xNskK}
\end{figure}

In order to find a time and a return range where subordination
takes place, we look at the data in $3$ different ways. First,
using tick-by-tick data, we construct the distribution of the
log-return $x_N$ after $N$ trades. $x_N$ should be Normal
distributed with mean zero and standard deviation
$\sigma_{N}\sqrt{N}$. We also present the $N$ dependence of the
skewness ($\langle x_{N}^{3} \rangle/(\langle x_{N}^{2}
\rangle^{3/2})$) and excess kurtosis ($\langle x_{N}^{4}
\rangle/(\langle x_{N}^{2} \rangle^{2})-3$) of $x_N$ in Fig.
\ref{fig:xNskK}.

Second, using $t$ minute
returns $x_t$ and the number of trades $N_t$ in the same $t$ interval,
we construct the time series

\begin{equation}
\epsilon_{t} = \frac{x_t}{\sqrt{V_{t}}}, \, V_{t}=\sigma_{N}^{2} N_{t},
\label{epsilon}
\end{equation}

\noindent where $V_t$ is the integrated variance in an interval $t$ and
$\sigma_N$ is the proportionality constant that converts number of
trades $N_t$ into variance. If indeed subordination holds,
$\epsilon_{t}$ is Normal distributed with mean zero and standard
deviation one, due to the central limit theorem
\cite{FellerBook,Stanley2000}.

\begin{figure}
\centerline{\epsfig{file=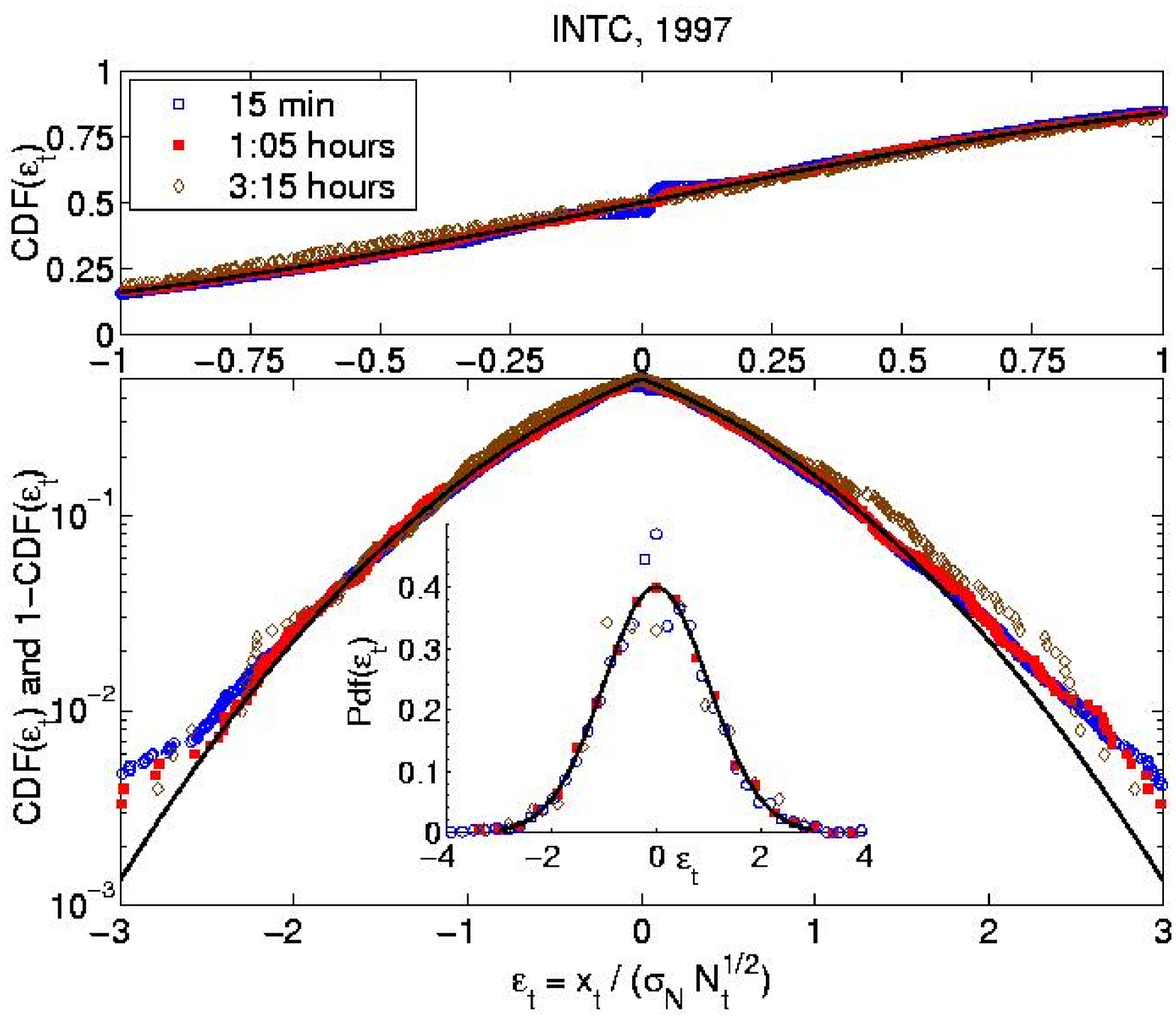,width=0.5\linewidth,angle=-90}}
\caption{Cumulative density function (CDF) for $\epsilon_{t}$ as
defined in equation (\ref{epsilon}) for three different $t$ compared
to the Gaussian (solid line). The parameters $\sigma_{N}$ in
(\ref{epsilon}) is chosen  for the best agreement between the
Gaussian and the data.} \label{fig:subRatio}

\centerline{\epsfig{file=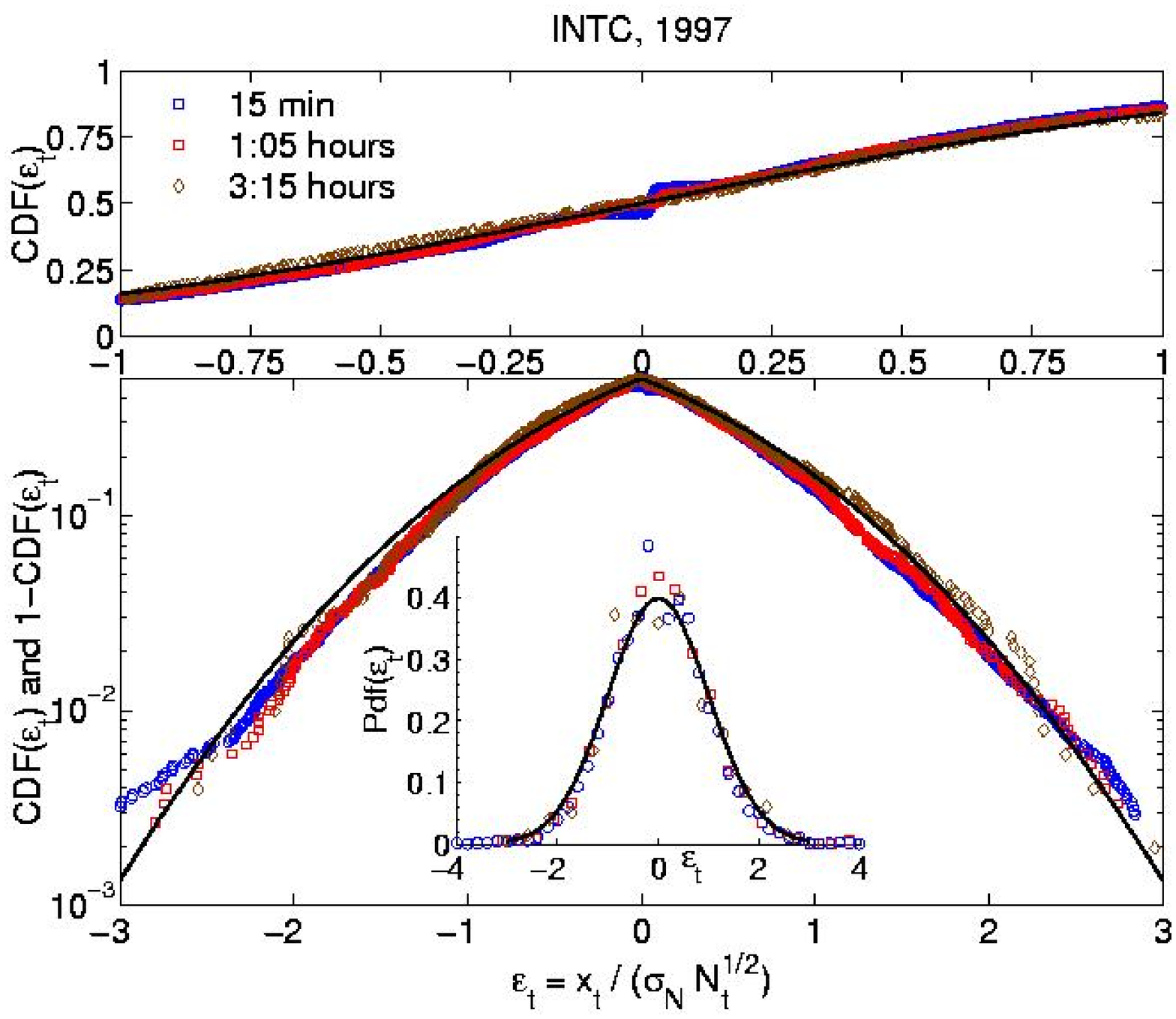,width=0.5\linewidth,angle=-90}}
\caption{Cumulative density function (CDF) for $\epsilon_{t}$ as
defined in equation (\ref{epsilon}) for three different $t$ compared
to the Gaussian (solid line). Contrary to Fig. \ref{fig:subRatio},
the parameter $\sigma_{N}$ in equation (\ref{epsilon}) is found
using Fig. \ref{fig:xN2}. Notice that the Gaussian lies above the
data in the tails.} \label{fig:subRatio2}
\end{figure}

\begin{figure}
\centerline{\epsfig{file=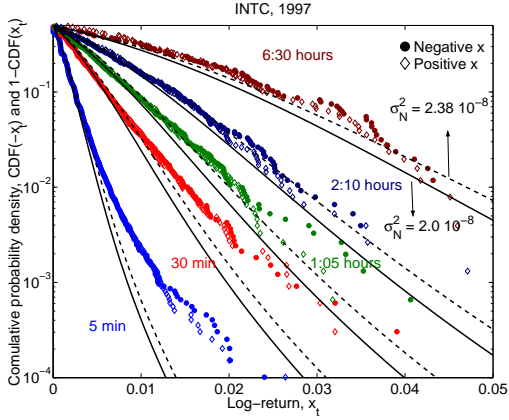,width=0.65\linewidth,angle=-90}}
\caption{Cumulative distribution of the stock returns $x_t$
compared to the reconstructed cumulative distribution function
(black lines) by randomizing the variance $V_t$ of a Gaussian
distribution. The probability of $V_t$ is constructed by binning
the number of trades, and this probability is used
non-parametrically in the integral (\ref{pvt}). The solid lines
have parameter $\sigma_N$ chosen in order to minimize the least
square error between the empirical $x_t$ distribution and
reconstructed variance changed Brownian motion (\ref{pvt}). The
dashed line has $\sigma_N$ found from Fig. \ref{fig:xN2}.}
\label{fig:subInt}
\end{figure}

Finally, we check subordination by numerically calculating the
probability mixture equation (\ref{pvt}). We construct the
probability density function of the number of trades $N_t$ inside
a time interval $t$ by binning the time series of $N_t$. The
choice for binwidth is according to Ref. \cite{Scott}. However,
the result appears independent of binwidth as long as the binwidth
chosen is not too large. The cumulative
probability density function for the measured $x_t$ and the
non-parametric reconstructed $x_t'$ are shown in Fig.
\ref{fig:subInt}.

The distributions in Fig. \ref{fig:subxN}, Fig. \ref{fig:subRatio}
and Fig. \ref{fig:subInt}(solid line) show an agreement of
approximately $85\%$ of the data with the subordination hypothesis
for time lags above $t>1$ hour or $N>2500$ (Fig. \ref{fig:xNskK}).
However, the subordination is clearly bad for times close to one
day ($t=6.5$ hours), where we do not have enough data (253 points)
to draw meaningful conclusions.

Notice the clear disagreement above $2$ standard deviations (STD)
as well as at zero in Fig. \ref{fig:subxN} and Fig.
\ref{fig:subRatio}. The deviations at zero are due to the discrete
nature of the data (section \ref{disc}) while the deviations above
$2$ STD show that the subordination hypothesis can not explain the
large changes in returns \cite{Farmer2004}.

For Fig. \ref{fig:subRatio} and Fig. \ref{fig:subInt}(solid
line), $\sigma_{N}^{2} = 2 \times 10^{-8}$ is found to give the best
agreement between the measured data and the
reconstructed data. For Fig. \ref{fig:subxN}, Fig.
\ref{fig:subRatio2} and Fig. \ref{fig:subInt}(dashed line),
$\sigma_{N}^{2} = 2.39 \times 10^{-8}$ is found from Fig.
\ref{fig:xN2}. Notice that the higher $\sigma_N$ in Fig.
\ref{fig:subRatio2} and Fig. \ref{fig:subInt} (dashed lines) seems
to indicate an overestimation of $\sigma_N$, since the curves
constructed by subordination are
generally above the data.

The lower value of $\sigma_{N}^{2}$ for Fig. \ref{fig:subRatio}
and Fig. \ref{fig:subInt} (solid line) leads to a violation of
relation (\ref{slope}). The difference between measured $\theta$
in Fig. \ref{fig:xt2} and the one calculated from $\eta
\sigma_{N}^{2}$ is now of approximately $16\%$. In order to verify
the origin of such difference, we remove $8\%$ of the largest
log-return $x_t$ data on both tails (ignore $8\%$ of the largest
$x_t$ on the positive and negative tail for all time lags $t$
used), a total of $16\%$ of the data. We find now a $\theta
\approx 8.01 \times 10^{-7}$. This new $\theta$ does not violate
relation (\ref{slope}) with $\sigma_{N}^{2} = 2 \times 10^{-8}$
and reconfirms that subordination with $V_t=\sigma_{N}^{2} N_t$ is
unable to explain large changes ($>85\%$) in the log-returns
$x_t$. This reconfirmation arises because we had to ignore $16\%$
of the data in the tails to reduce $\theta$. Dropping $16\%$ of
the tails is equivalent to looking only at the center $\approx
85\%$ of the data and saying that subordination is only valid of
it.

\subsection{Models for the subordinator}\label{model}

Having verified that a Brownian motion subordinated to the number
of trades $N_t$ via $V_t$ can describe approximately $85\%$ of the return
data for time lags larger than $1$ hour (or, if one ignores
discreetness effects such as the zero return effect, larger than
$30$ minutes), we can model $V_t$ instead of modelling $x_t$.

In this section, we verify the quality of modelling $V_t$ with a
CIR process as given in section (\ref{subTh}). We present the quality
of the CIR fit for Intel in the year $1997$. We also show that the
quality of the Heston fit to $x_t$ with parameters from the $V_t$
CIR fit is consistent with the quality of the subordination: we
are able to model most of the central $85\%$ of the $x_t$ distribution.

\begin{figure}[b]
\centerline{\epsfig{file=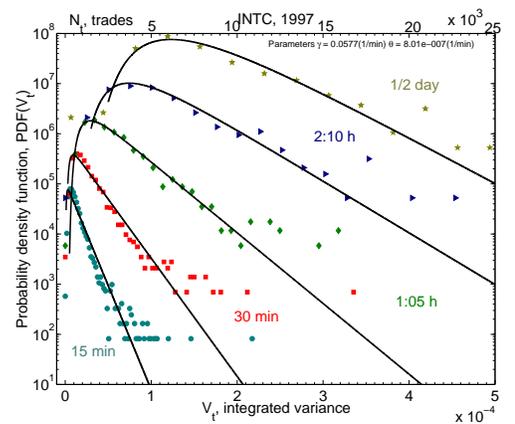,width=0.65\linewidth,angle=-90}}
\caption{Empirical probability density function for the number of
  trades (ticks) $N_t$ or integrated variance $V_t = \sigma_{N}^{2}
  N_{t}$, compared to the least square fit with the CIR formula
  (\ref{eq:CIR}). Curves are offset by a factors of 10.}
\label{fig:pdfNt}
\end{figure}

\begin{figure}[b]
\centerline{\epsfig{file=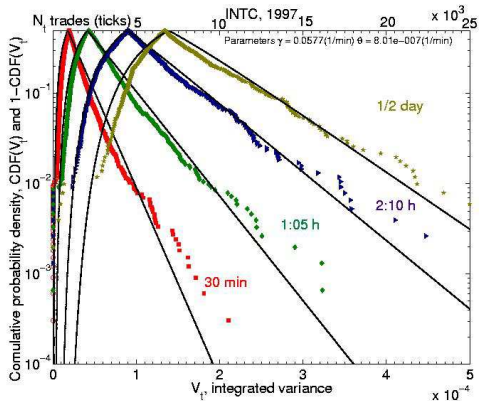,width=0.65\linewidth,angle=-90}}
\caption{Cumulative distribution function (CDF) for the number of
trades $N_t$ and integrated variance $V_t$ compared to the CIR fit
(solid lines). The $CDF(V_t)$ goes from $0$ to $0.5$.
$1-CDF(V_t)$ goes from $0.5$ to $0$. The lower tail ($V_t: 0->0.5$) of
the $CDF$ is to the left and the upper tail ($V_t: 0.5->0$) to the right of
$0.5$ for each time $t$ curve.} \label{fig:cdfNt}
\end{figure}

Due to previous studies with intraday log-returns \cite{SPY} (see
also chapter \ref{expD}), we assume $\alpha = 1$ for the
simplified CIR model in equation (\ref{eq:CIR}). The parameter
$\theta$ is found from the relation $\theta = \eta \sigma_{N}^{2}$
(\ref{slope}). The remaining parameter $\gamma$ is found by
fitting the empirical $PDF(V_t)$ for time lags $t=1\colon05$ hours
and $t=2\colon10$ hours simultaneously. The regular quality of
such a fit is shown in Figs. \ref{fig:pdfNt} and \ref{fig:cdfNt}.
The theoretical CIR lines are above the data (Fig. \ref{fig:cdfNt}).
Furthermore, the time dependence of the theoretical PDF and CDF
only approximately follow the data. For times below $1$ hour the
probability maximum of the empirical distribution is to the left
of the theoretical distribution and for times above $1$ hour to
the right.

\begin{figure}
\centerline{\epsfig{file=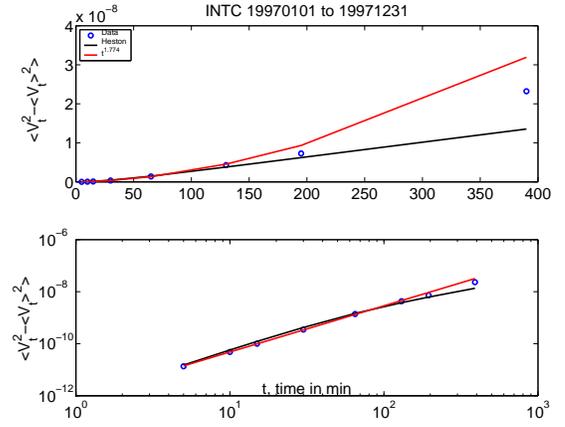,width=0.65\linewidth,angle=-90}}
\caption{Variance of the integrated variance $\langle
V_t^{2}-\langle V_t \rangle^{2} \rangle$ for
  different time lags $t$ for the data (circles) compared to the
  theoretical CIR variance given in equation (\ref{M2cir}) (solid black
  line). For comparison the best power-law fit $\langle V_{t}^{2} - \langle V_{t} \rangle^{2} \rangle \propto t^{1.77}$
  is shown (solid red line).}
\label{fig:N2}
\end{figure}

The results shown in Figs. \ref{fig:pdfNt} and \ref{fig:cdfNt}
indicate that the CIR is only approximately valid. The quality can
be further assessed by constructing the variance of the $V_t$ as a
function of the time lag $t$. Fig. \ref{fig:N2} shows that the
theoretical variance given in equation (\ref{M2cir}) is only
approximately correct. Nevertheless from equation (\ref{M2}), we
know that the variance of $V_t$ corresponds to the kurtosis of
$x_t$. This indicates that even though $V_t$ can not be modelled
well (not even the second moment) the implication of that is only
important to the fourth and higher moments in the log-returns
$x_t$.

\begin{figure}
\centerline{\epsfig{file=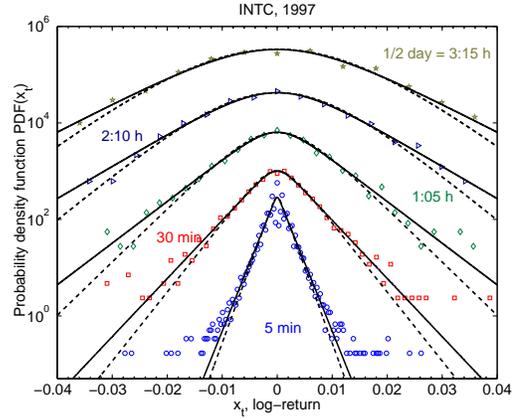,width=0.65\linewidth,angle=-90}}
\caption{Probability distribution function for the log-returns
$x_t$ compared to the Heston model (dashed and solid lines). The
two lines represent a different set of parameters. The solid line
has parameters $\theta$ from Fig. \ref{fig:xN2} and $\gamma$ is
found directly by fitting $x_t$. The dashed lines has $\theta =
\sigma_{N}^{2}\eta$ with $\sigma_{N}^{2}$  from Fig.
\ref{fig:subInt}(solid lines) and Fig. \ref{fig:subRatio} and
$\eta$ from Fig. \ref{fig:Nt}. The parameter $\gamma$ is then
found by fitting the probability density of $V_t$. Curves are offset by factors of 10.}
\label{fig:pdfXt}
\end{figure}

\begin{figure}
\centerline{\epsfig{file=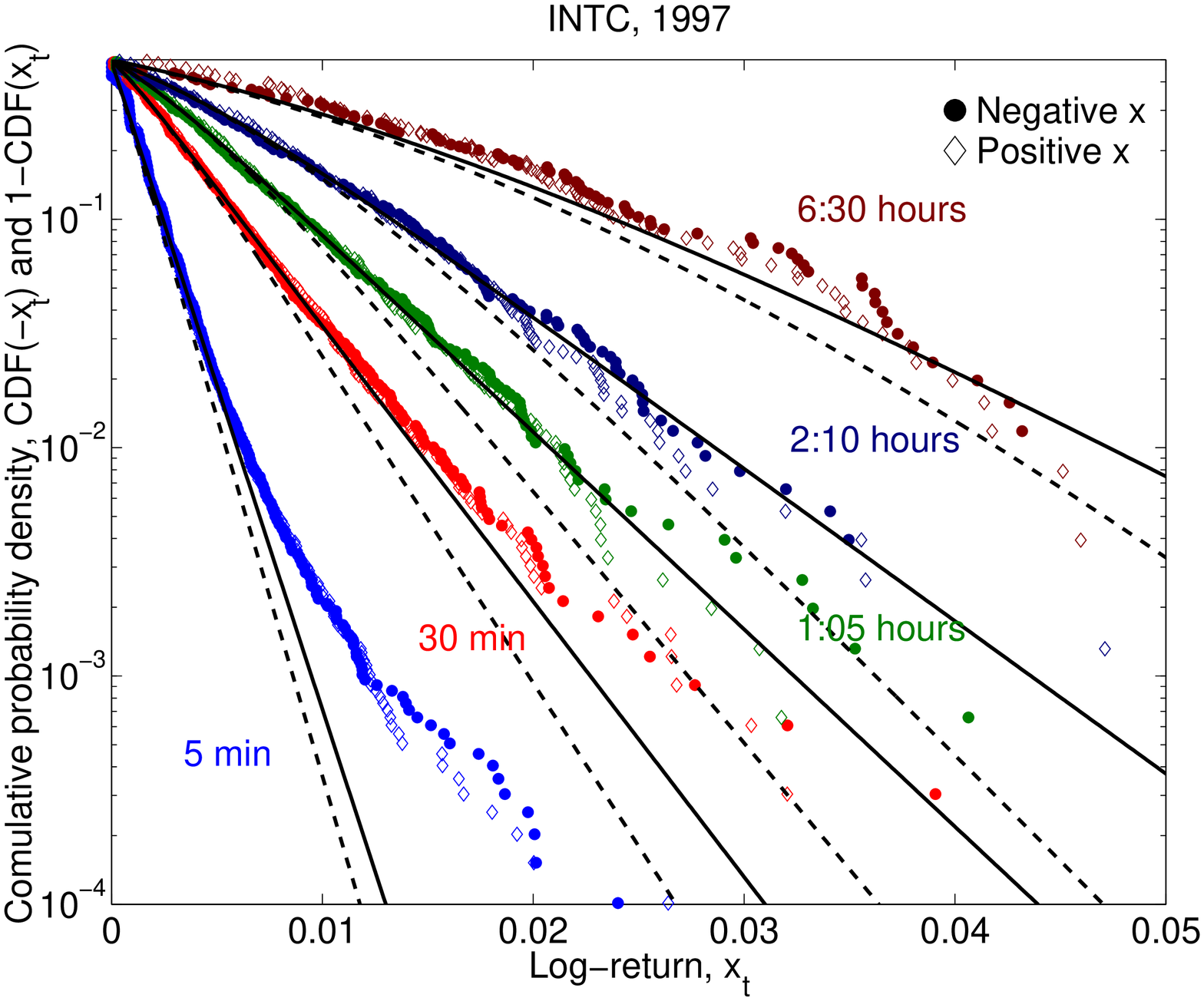,width=0.65\linewidth,angle=-90}}
\caption{Cumulative probability density of $x_t$ compared to the
Heston model. Theoretical lines (dashed and solid) are constructed
by integrating the theoretical probability density functions shown
in Fig. \ref{fig:pdfXt}. The two theoretical lines represent a
different set of parameter. The solid line has parameters $\theta$
from Fig. \ref{fig:xN2} and $\gamma$ is found directly by fitting
$x_t$. The dashed lines has $\theta = \sigma_{N}^{2}\eta$ with
$\sigma_{N}^{2}$  from Fig. \ref{fig:subInt}(solid lines) and
Fig. \ref{fig:subRatio} and $\eta$ from Fig. \ref{fig:Nt}. The
parameter $\gamma$ is then found by fitting the probability
density of $V_t$. Notice that the solid black line clearly
gives a better fit to the data.} \label{fig:cdfXt}
\end{figure}

To verify the quality of the parameters found by fitting the
subordinator, $V_t$, in explaining the log-returns, $x_t$, we present
Figs. \ref{fig:pdfXt} and \ref{fig:cdfXt}. The empirical PDF
(\ref{fig:pdfXt}) and CDF (\ref{fig:cdfXt}) for
$x_t$ show that the corresponding Heston model (dashed black lines), constructed
with parameters found by fitting CIR to the probability density of
$V_t$, is able to
fit only the center of the empirical distributions of $x_t$ ($\approx
80\% - 85\%$) at $t=65,130$ minutes (Fig. \ref{fig:cdfXt}).

To recheck the consistency of the subordination approach, we fit the
empirical PDF of $x_t$ directly with the Heston model (\ref{eq:DY}). We
proceed in similar fashion to the fitting procedure in chapter
\ref{expD}. We assume $\alpha=1$ and take $\theta=8.01 \times
10^{-7}$. The parameter $\theta$ was
found from the relation $\theta=\sigma_{N}^{2} \eta$ (\ref{slope}),
where $\eta$ is found from Fig. \ref{fig:xN2} and $\sigma_{N}^{2}$ is
given such that the subordination in Figs. \ref{fig:subRatio} and
\ref{fig:subInt} is the best possible. Finally, we fit the empirical
PDFs (Fig. \ref{fig:pdfXt}) for the parameter $\gamma$. Therefore,
we are effectively only fitting $\gamma$, since all the other
parameters are the same used in the $V_t$ fit (Fig. \ref{fig:pdfNt}).
We find that the $\gamma$
found from fitting the empirical PDF of $x_t $ directly, is of the
same order of
magnitude as with the one found by fitting the empirical PDF of $V_t$
(0.05 from $x_t$ and 0.06 from $V_t$). This shows, that
the subordination indeed captures most of the information for the
center of the distribution, since fitting $V_t$ or $x_t$ for
$\gamma$ is equivalent.

Notice that the agreement of the theoretical Heston model curves,
constructed with parameters from the $V_t$ fit, is practically
identical to the agreement found in Fig. \ref{fig:subInt}(solid
lines) between the CDF of $x_t$ and the CDF constructed by
subordination using the non-parametric binned probability density
of  $V_t$ as the variance of a Gaussian random walk (\ref{pvt}).
The information content in the number of trades and therefore in
the integrated variance distribution is almost all captured by
CIR, even with a regular fit quality (Fig. \ref{fig:cdfNt}). This
last point implies that even if we had a better fit to the
distribution of $V_t$, the increase in the fitting quality of the
log-returns will not be substantial.

A substantial increase in the fitting quality of the empirical PDF
and CDF of the log-returns in Figs. \ref{fig:pdfXt} and
\ref{fig:cdfXt} is attained if one fits the empirical PDF of
$x_t$ directly with $\theta = 9.53 \times 10^{-7}$ given in Fig.
\ref{fig:xt2}. This amounts to take $\sigma_{N}^{2}$ as given by
Fig. \ref{fig:xN2} and $\eta$ by Fig. \ref{fig:Nt}, such that
relation (\ref{slope}) is still valid. The parameter $\gamma =
0.02$ for the black solid lines in Fig. \ref{fig:cdfXt} is also
considerably different from $\gamma = 0.06$, found by fitting the
empirical PDF of $V_t$ and using $\theta = 8.01 \times 10^{-7}$
such that $\sigma_{N}^{2}$ is the best fit value for the
subordination in Figs. \ref{fig:subInt}(solid line) and
\ref{fig:subRatio}. The substantial increase in the fitting
quality for $x_t$, reemphasizes that the number of trades are only
able to describe the center of the distribution of log-returns
(section \ref{subCheck}).


\subsection{Conclusion}

We have studied the discrete nature of the probability
distribution of absolute returns that arises from the minimal
discrete price change for bid and offers allowed by the stock
exchange. We have shown that such discrete nature implies that
the probability distributions of log-returns for intraday time
lags are only approximately continuous. The continuous
approximation becomes good for returns with time lags longer than
$1$ hour.

We have shown that, using the integrated volatility $V_t = \sigma_{N}^{2}
N_t$ derived from the number of trades $N_t$ as the subordinator
of a driftless Brownian motion (\ref{pvt}), we are able to
describe the center ($\approx 85\%$) of the distribution of
log-returns $x_t$ for time lags $t>1$ hour and smaller than $t<1$
day. The upper limit is restricted by the number of data points we
have, since we are working with only one year of data.

We also have shown that the CIR process is only able to
approximately describe the distribution function for $V_t$.
However, this approximate description is already enough for the
corresponding Heston model to fit the log-returns  $x_t$ with
approximately the maximum quality that the subordination allows
($\approx 80\% - 85\%$).

Finally, a direct fit to the log-returns $x_t$ with the Heston
model results in a considerable increase in the fitting quality.
This reemphasizes that the process of subordination, as implied by
the empirical probability density of $V_t$, is only able to
explain the center of the distribution of returns.


\section{Income distribution}

Attempts to apply the methods of exact sciences, such as physics, to
describe a society have a long history \cite{Ball}.  At the end of the
19th century, Italian physicist, engineer, economist, and sociologist
Vilfredo Pareto suggested that income distribution in a society is
described by a power law \cite{Pareto}.  Modern data indeed confirm
that the upper tail of income distribution follows the Pareto law
\cite{Champernowne,Aoki,Souma,Gallegati,Australia}.  However, the
majority of the population does not belong there, so characterization and
understanding of their income distribution remains an open problem.
Dr\u{a}gulescu and Yakovenko \cite{Yakovenko-money} proposed that the
equilibrium distribution should follow an exponential law analogous to
the Boltzmann-Gibbs distribution of energy in statistical physics.
The first factual evidence for the exponential distribution of income
was found in Ref.\ \cite{Yakovenko-income}.  Coexistence of the
exponential and power-law parts of the distribution was recognized in
Ref.\ \cite{Yakovenko-wealth}.  However, these papers, as well as
Ref.\ \cite{Yakovenko-survey}, studied the data only for a particular
year.  Here we analyze temporal evolution of the personal income
distribution in the USA during 1983--2001.  We show that the US
society has a well-defined two-income-class structure.  The majority of
population (97--99\%) belongs to the lower income class and has a very
stable in time exponential (``thermal'') distribution of income.  The
upper income class (1--3\% of population) has a power-law
(``superthermal'')
distribution, whose parameters significantly change in time with the
rise and fall of the stock market. Using the principle of maximal entropy,
we discuss the concept of equilibrium inequality in a society and
quantitatively show that it applies to the bulk of the population.

\subsection{Data analysis and discussion}

Most
of academic and government literature on income distribution and
inequality \cite{Kakwani,Cowell,Atkinson,Petska} does not attempt to
fit the data by a simple formula.  When fits are performed, usually
the log-normal distribution \cite{Gibrat} is used for the lower part
of the distribution \cite{Souma,Gallegati,Australia}.  Only recently
the exponential distribution started to be recognized in income
studies \cite{Nirei,Mimkes}, and models showing formation of two
classes started to appear \cite{West,Wright}.

Let us introduce the probability density $P(r)$, which gives the
probability $P(r)\,dr$ to have income in the interval $(r,r+dr)$.  The
cumulative probability $C(r)=\int_r^\infty dr'P(r')$ is the
probability to have income above $r$, $C(0)=1$.  By analogy with the
Boltzmann-Gibbs distribution in statistical physics
\cite{Yakovenko-money,Yakovenko-income}, we consider an exponential
function $P(r)\propto\exp(-r/T)$, where $T$ is a parameter analogous
to temperature.  It is equal to the average income $T=\langle
r\rangle=\int_0^\infty dr'r'P(r')$, and we call it the ``income
temperature.''  When $P(r)$ is exponential, $C(r)\propto\exp(-r/T)$ is
also exponential.  Similarly, for the Pareto power law $P(r)\propto
1/r^{\alpha+1}$, $C(r)\propto 1/r^\alpha$ is also a power law.

\begin{figure}
\centerline{
\epsfig{file=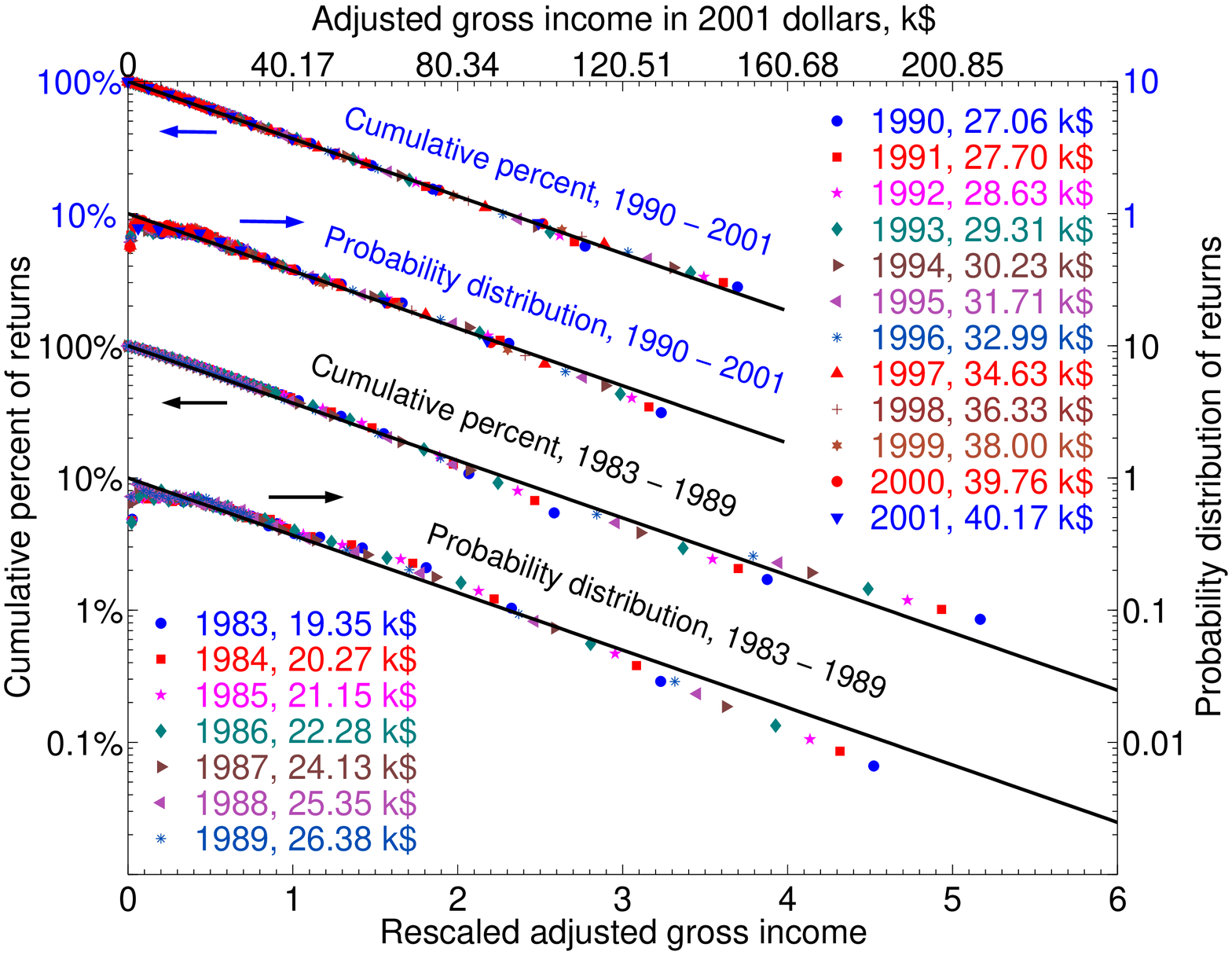,width=0.5\linewidth,angle=-90}}
\caption{
  Cumulative probability $C(r)$ and probability density $P(r)$ plotted
  in the log-linear scale vs.\ $r/T$, the annual personal income $r$
  normalized by the average income $T$ in the exponential part of the
  distribution. The IRS data points are for 1983--2001, and the
  columns of numbers give the values of $T$ for the corresponding
  years.}
\label{fig:LogLin}
centerline{
\epsfig{file=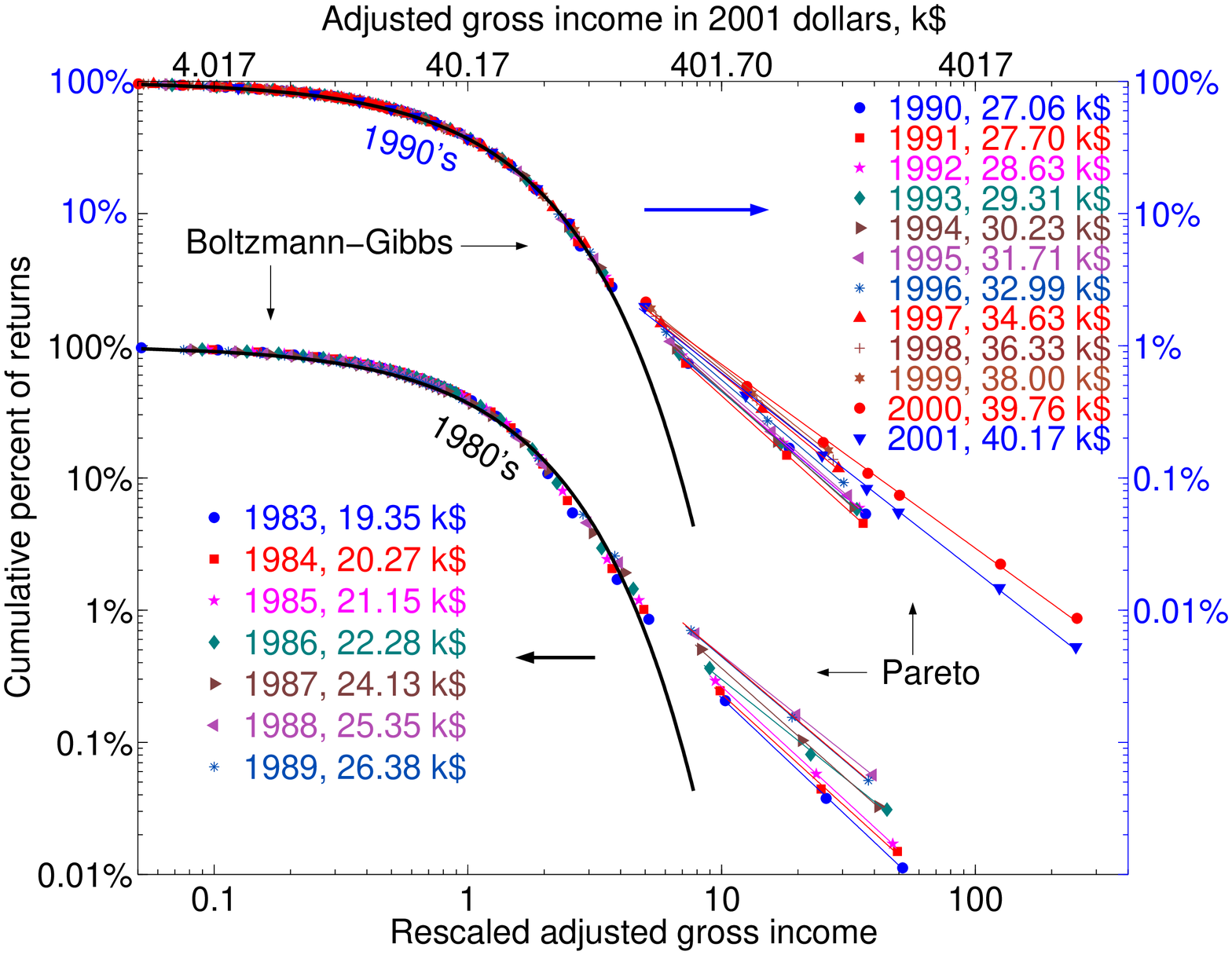,width=0.5\linewidth,angle=-90}}
\caption{
  Log-log plots of the cumulative probability $C(r)$ vs.\ $r/T$ for a
  wider range of income $r$.}
\label{fig:LogLog}
\end{figure}

We analyze the data \cite{IRS} on personal income distribution
compiled by the Internal Revenue Service (IRS) from the tax returns in
the USA for the period 1983--2001 (presently the latest available
year).  The publicly available data are already preprocessed by the
IRS into bins and effectively give the cumulative distribution
function $C(r)$ for certain values of $r$.  First we make the plots of
$\log C(r)$ vs.\ $r$ (the log-linear plots) for each year.  We find
that the plots are straight lines for the lower 97--98\% of
population, thus confirming the exponential law.  From the slopes of
these straight lines, we determine the income temperatures $T$ for
each year.  In Fig.\ \ref{fig:LogLin}, we plot $C(r)$ and $P(r)$ vs.\
$r/T$ (income normalized to temperature) in the log-linear scale.  In
these coordinates, the data sets for different years collapse onto a
single straight line.  (In Fig.\ \ref{fig:LogLin}, the data lines for
1980s and 1990s are shown separately and offset vertically.)  The
columns of numbers in Fig.\ \ref{fig:LogLin} list the values of the
annual income temperature $T$ for the corresponding years, which
changes from 19 k\$ in 1983 to 40 k\$ in 2001.  The upper horizontal
axis in Fig.\ \ref{fig:LogLin} shows income $r$ in k\$ for 2001.

In Fig.\ \ref{fig:LogLog}, we show the same data in the log-log scale
for a wider range of income $r$, up to about $300T$.  Again we observe
that the sets of points for different years collapse onto a single
exponential curve for the lower part of the distribution, when plotted
vs.\ $r/T$.  However, above a certain income $r_*\approx4T$, the
distribution function changes to a power law, as illustrated by the
straight lines in the log-log scale of Fig.\ \ref{fig:LogLog}.  Thus
we observe that income distribution in the USA has a well-defined
two-class structure.  The lower class (the great majority of
population) is characterized by the exponential, Boltzmann-Gibbs
distribution, whereas the upper class (the top few percent of
population) has the power-law, Pareto distribution.  The intersection
point of the exponential and power-law curves determines the income
$r_*$ separating the two classes.  The collapse of data points for
different years in the lower, exponential part of the distribution in
Figs.\ \ref{fig:LogLin} and \ref{fig:LogLog} shows that this part is
very stable in time and, essentially, does not change at all for the
last 20 years, save for a gradual increase of temperature $T$ in
nominal dollars.  We conclude that the majority of population is in
statistical equilibrium, analogous to the thermal equilibrium in
physics.  On the other hand, the points in the upper, power-law part
of the distribution in Fig.\ \ref{fig:LogLog} do not collapse onto a
single line.  This part significantly changes from year to year, so it
is out of statistical equilibrium.  A similar two-part structure in
the energy distribution is often observed in physics, where the lower
part of the distribution is called ``thermal'' and the upper part
``superthermal'' \cite{superthermal}.

\begin{figure}
\centerline{
\epsfig{file=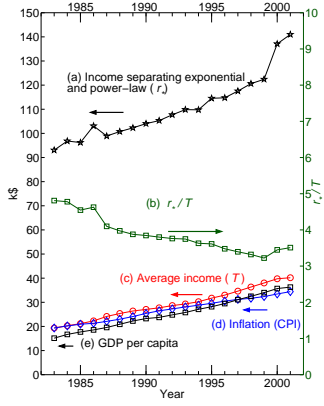,width=0.5\linewidth,angle=0}}
\caption{
  Temporal evolution of various parameters characterizing income
  distribution.}
\label{fig:temperature}
\end{figure}

Temporal evolution of the parameters $T$ and $r_*$ is shown in Fig.\
\ref{fig:temperature}.  We observe that the average income $T$ (in
nominal dollars) was increasing gradually, almost linearly in time,
and doubled in the last twenty years.  In Fig.\ \ref{fig:temperature},
we also show the inflation coefficient (the consumer price index CPI
from Ref.\ \cite{CPI}) compounded on the average income of 1983.  For
the twenty years, the inflation factor is about 1.7, thus most, if not
all, of the nominal increase in $T$ is inflation.  Also shown in Fig.\
\ref{fig:temperature} is the nominal gross domestic product (GDP) per
capita \cite{CPI}, which increases in time similarly to $T$ and CPI.
The ratio $r_*/T$ varies between 4.8 and 3.2 in Fig.\
\ref{fig:temperature}.

\begin{figure}
\centerline{
\epsfig{file=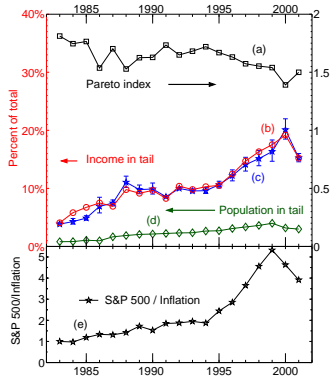,width=0.5\linewidth,angle=0}}
\caption{
  (a) The Pareto index $\alpha$ of the power-law tail
  $C(r)\propto1/r^\alpha$.  (b) The excessive income in the Pareto
  tail, as a fraction $f$ of the total income in the system, obtained
  as the difference between the total income and the income in the
  exponential part of the distribution.  (c) The tail income fraction
  $f$, obtained by integrating the Pareto power law of the tail.  (d)
  The fraction of population belonging to the Pareto tail. (e) The
  stock-market index S\&P 500 divided by the inflation coefficient and
  normalized to 1 in 1983.}
\label{fig:index}
\end{figure}

In Fig.\ \ref{fig:index}, we show how the parameters of the Pareto
tail $C(r)\propto 1/r^\alpha$ change in time.  Curve (a) shows that
the power-law index $\alpha$ varies between 1.8 and 1.4, so the power
law is not universal.  Because a power law decays with $r$ more slowly
than an exponential function, the upper tail contains more income than
we would expect for a thermal distribution, hence we call the tail
``superthermal'' \cite{superthermal}.  The total excessive income in
the upper tail can be determined in two ways: as the integral
$\int_{r_*}^{\infty}dr'r'P(r')$ of the power-law distribution, or as
the difference between the total income in the system and the income
in the exponential part.  Curves (c) and (b) in Fig.\ \ref{fig:index}
show the excessive income in the upper tail, as a fraction $f$ of the
total income in the system, determined by these two methods, which
agree with each other reasonably well.  We observe that $f$ increased
by the factor of 5 between 1983 and 2000, from 4\% to 20\%, but
decreased in 2001 after the crash of the US stock market.  For
comparison, curve (e) in Fig.\ \ref{fig:index} shows the stock market
index S\&P 500 divided by inflation.  It also increased by the factor
of 5.5 between 1983 and 1999, and then dropped after the stock market
crash.  We conclude that the swelling and shrinking of the upper
income tail is correlated with the rise and fall of the stock market.
Similar results were found for the upper income tail in Japan in Ref.\
\cite{Aoki}.  Curve (d) in Fig.\ \ref{fig:index} shows the fraction of
population in the upper tail.  It increased from 1\% in 1983 to 3\% in
1999, but then decreased after the stock market crash.  Notice,
however, that the stock market dynamics had a much weaker effect on
the average income $T$ of the lower, ``thermal'' part of income
distribution shown in Fig.\ \ref{fig:temperature}.


\begin{figure}
\centerline{
\epsfig{file=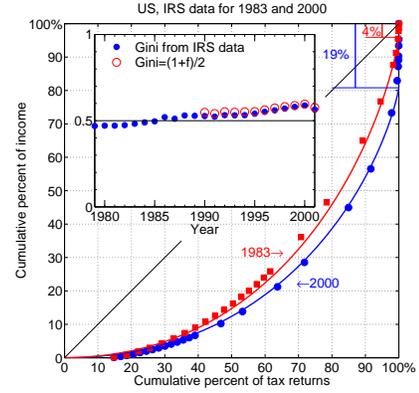,width=0.6\linewidth,angle=-90}}
\caption{
  Main panel: Lorenz plots for income distribution in 1983 and 2000.
  The data points are from the IRS \cite{Petska}, and the theoretical
  curves represent Eq.\ (\ref{eq:Lorenz}) with $f$ from Fig.\
  \ref{fig:index}.  Inset: The closed circles are the IRS data
  \cite{Petska} for the Gini coefficient $G$, and the open circles
  show the theoretical formula $G=(1+f)/2$.}
\label{fig:lorentz}
\end{figure}

For discussion of income inequality, the standard practice is to
construct the so-called Lorenz curve \cite{Kakwani}.  It is defined
parametrically in terms of the two coordinates $x(r)$ and $y(r)$
depending on the parameter $r$, which changes from 0 to $\infty$.  The
horizontal coordinate $x(r)=\int_{0}^{r}dr'P(r')$ is the fraction of
population with income below $r$.  The vertical coordinate
$y(r)=\int_0^rdr'r'P(r')/\int_{0}^{\infty}dr'r'P(r')$ is the total
income of this population, as a fraction of the total income in the
system.  Fig.\ \ref{fig:lorentz} shows the data points for the Lorenz
curves in 1983 and 2000, as computed by the IRS \cite{Petska}.  For a
purely exponential distribution of income $P(r)\propto\exp(-r/T)$, the
formula $y=x+(1-x)\ln(1-x)$ for the Lorenz curve was derived in Ref.\
\cite{Yakovenko-income}.  This formula describes income distribution
reasonably well in the first approximation \cite{Yakovenko-income},
but visible deviations exist.  These deviations can be corrected by
taking into account that the total income in the system is higher than
the income in the exponential part, because of the extra income in the
Pareto tail.  Correcting for this difference in the normalization of
$y$, we find a modified expression \cite{Yakovenko-survey} for the
Lorenz curve
\begin{equation}
  y=(1-f)[x+(1-x)\ln(1-x)]+f\Theta(x-1),
\label{eq:Lorenz}
\end{equation}
where $f$ is the fraction of the total income contained in the Pareto
tail, and $\Theta(x-1)$ is the step function equal to 0 for $x<1$ and
1 for $x\geq1$.  The Lorenz curve (\ref{eq:Lorenz}) experiences a
vertical jump of the height $f$ at $x=1$, which reflects the fact
that, although the fraction of population in the Pareto tail is very
small, their fraction $f$ of the total income is significant.  It does
not matter for Eq.\ (\ref{eq:Lorenz}) whether the extra income in the
upper tail is described by a power law or another slowly decreasing
function $P(r)$.  The Lorenz curves, calculated using Eq.\
(\ref{eq:Lorenz}) with the values of $f$ from Fig.\ \ref{fig:index},
fit the IRS data points very well in Fig.\ \ref{fig:lorentz}.

The deviation of the Lorenz curve from the diagonal in Fig.\
\ref{fig:lorentz} is a certain measure of income inequality.  Indeed,
if everybody had the same income, the Lorenz curve would be the
diagonal, because the fraction of income would be proportional to the
fraction of population.  The standard measure of income inequality is
the so-called Gini coefficient $0\leq G\leq1$, which is defined as the
area between the Lorenz curve and the diagonal, divided by the area of
the triangle beneath the diagonal \cite{Kakwani}.  It was calculated
in Ref.\ \cite{Yakovenko-income} that $G=1/2$ for a purely exponential
distribution.  Temporal evolution of the Gini coefficient, as
determined by the IRS \cite{Petska}, is shown in the inset of Fig.\
\ref{fig:lorentz}.  In the first approximation, $G$ is quite close to
the theoretically calculated value 1/2.  The agreement can be improved
by taking into account the Pareto tail, which gives $G=(1+f)/2$ for
Eq.\ (\ref{eq:Lorenz}).  The inset in Fig.\ \ref{fig:lorentz} shows
that this formula very well fits the IRS data for the 1990s with the
values of $f$ taken from Fig.\ \ref{fig:index}.  We observe that
income inequality was increasing for the last 20 years, because of
swelling of the Pareto tail, but started to decrease in 2001 after the
stock market crash.  The deviation of $G$ below 1/2 in the 1980s
cannot be captured by our formula.  The data points for the Lorenz
curve in 1983 lie slightly above the theoretical curve in Fig.\
\ref{fig:lorentz}, which accounts for $G<1/2$.

Thus far we discussed the distribution of individual income.  An
interesting related question is the distribution of family income
$P_2(r)$.  If both spouses are earners, and their incomes are
distributed exponentially as $P_1(r)\propto\exp(-r/T)$\footnote{Even
  thought the income of women is generally lower that men, this seems
  not to make a difference in temperature significant enough to be
  noticed.}, then
\begin{equation}
  P_2(r)=\int_0^r dr' P_1(r')P_1(r-r')\propto r\exp(-r/T).
\label{eq:family}
\end{equation}
Eq.\ (\ref{eq:family}) is in a good agreement with the family income
distribution data from the US Census Bureau \cite{Yakovenko-income}.
In Eq.\ (\ref{eq:family}), we assumed that incomes of spouses are
uncorrelated.  This assumption was verified by comparison with the
data in Ref.\ \cite{Yakovenko-survey}.  The Gini coefficient for
family income distribution (\ref{eq:family}) was found to be
$G=3/8=37.5\%$ \cite{Yakovenko-income}, in agreement with the data.
Moreover, the calculated value 37.5\% is close to the average $G$ for
the developed capitalist countries of North America and Western
Europe, as determined by the World Bank \cite{Yakovenko-survey}.

On the basis of the analysis presented above, we propose a concept of
the \emph{equilibrium inequality} in a society, characterized by
$G=1/2$ for individual income and $G=3/8$ for family income.  It is a
consequence of the exponential Boltzmann-Gibbs distribution in thermal
equilibrium, which maximizes the entropy $S=\int dr\,P(r)\,\ln P(r)$
of a distribution $P(r)$ under the constraint of the conservation law
$\langle r\rangle=\int_0^\infty dr\,P(r)\,r=\rm const$.  Thus, any
deviation of income distribution from the exponential one, to either
less inequality or more inequality, reduces entropy and is not
favorable by the second law of thermodynamics.  Such deviations may be
possible only due to non-equilibrium effects.  The presented data show
that the great majority of the US population is in thermal
equilibrium.

Finally, we briefly discuss how the two-class structure of income
distribution can be rationalized on the basis of a kinetic approach,
which deals with temporal evolution of the probability distribution
$P(r,t)$.  Let us consider a diffusion model, where income $r$ changes
by $\Delta r$ over a period of time $\Delta t$.  Then, temporal
evolution of $P(r,t)$ is described by the Fokker-Planck equation
\cite{Kinetics}
\begin{equation}
   \frac{\partial P}{\partial t}=\frac{\partial}{\partial r}
   \left(AP + \frac{\partial}{\partial r}(BP)\right), \quad
   A=-{\langle\Delta r\rangle \over \Delta t}, \quad
   B={\langle(\Delta r)^2\rangle \over 2\Delta t}.
\label{eq:diffusion}
\end{equation}
For the lower part of the distribution, it is reasonable to assume
that $\Delta r$ is independent of $r$.  In this case, the coefficients
$A$ and $B$ are constants.  Then, the stationary solution
$\partial_tP=0$ of Eq.\ (\ref{eq:diffusion}) gives the exponential
distribution \cite{Yakovenko-money} $P(r)\propto\exp(-r/T)$ with
$T=B/A$.  Notice that a meaningful solution requires that $A>0$, i.e.\
$\langle\Delta r\rangle<0$ in Eq.\ (\ref{eq:diffusion}).  On the other
hand, for the upper tail of income distribution, it is reasonable to
expect that $\Delta r\propto r$ (the Gibrat law \cite{Gibrat}), so
$A=ar$ and $B=br^2$.  Then, the stationary solution $\partial_tP=0$ of
Eq.\ (\ref{eq:diffusion}) gives the power-law distribution
$P(r)\propto1/r^{\alpha+1}$ with $\alpha=1+a/b$.  The former process
is additive diffusion, where income changes by certain amounts,
whereas the latter process is multiplicative diffusion, where income
changes by certain percentages.  The lower class income comes from
wages and salaries, so the additive process is appropriate, whereas
the upper class income comes from investments, capital gains, etc.,\
where the multiplicative process is applicable.  Ref.\ \cite{Aoki}
quantitatively studied income kinetics using tax data for the upper
class in Japan and found that it is indeed governed by a
multiplicative process.  The data on income mobility in the USA are
not readily available publicly, but are accessible to the Statistics
of Income Research Division of the IRS.  Such data would allow to
verify the conjectures about income kinetics.

The exponential probability distribution $P(r)\propto\exp(-r/T)$ is a
monotonous function of $r$ with the most probable income $r=0$.  The
probability densities shown in Fig.\ \ref{fig:LogLin} agree reasonably
well with this simple exponential law.  However, a number of other
studies found a nonmonotonous $P(r)$ with a maximum at $r\neq0$ and
$P(0)=0$.  These data were fitted by the log-normal
\cite{Souma,Gallegati,Australia} or the gamma distribution
\cite{Mimkes,West,Ferrero}.  The origin of the discrepancy in the
low-income data between our work and other papers is not completely
clear at this moment.  The following factors may possibly play a role.
First, one should be careful to distinguish between personal income
and group income, such as family and household income.  As Eq.\
(\ref{eq:family}) shows, the latter is given by the gamma distribution
even when the personal income distribution is exponential.  Very often
statistical data are given for households and mix individual and group
income distributions (see more discussion in Ref.\
\cite{Yakovenko-income}).  Second, the data from tax agencies and
census bureaus may differ.  The former data are obtained from tax
declarations of all the taxable population, whereas the latter data from
questionnaire surveys of a limited sample of population.  These two
methodologies may produce different results, particularly for low
incomes.  Third, it is necessary to distinguish between distributions
of money \cite{Yakovenko-money,Ferrero,Chakrabarti}, wealth
\cite{West,wealth}, and income.  They are, presumably, closely
related, but may be different in some respects.  Fourth, the low-income
probability density may be different in the USA and in other countries
because of different Social Security or more general policies.  All these questions
require careful investigation in future work.  We can only say that
the data sets analyzed in this paper and our previous papers are well
described by a simple exponential function for the whole lower class.
This does not exclude a possibility that other functions can also fit
the data \cite{Dragulescu}.  However, the exponential law has only one
fitting parameter $T$, whereas log-normal, gamma, and other
distributions have two or more fitting parameters, so they are less
parsimonious.


\end{document}